\documentclass[11pt]{article}
\usepackage[utf8]{inputenc}
\usepackage{mathtools}
\usepackage{hyperref}
\usepackage{dsfont}
\usepackage{listings}
\usepackage{amsmath,amssymb}
\usepackage[superscript,biblabel]{cite}
\usepackage{authblk}
\usepackage[margin=1.3in]{geometry}
\usepackage[labelfont=bf]{caption}
\PassOptionsToPackage{hyphens}{url}\usepackage{hyperref}
\DeclareMathOperator{\logit}{logit}
\DeclareMathOperator{\expit}{expit}

\title{Marginalization of Regression-Adjusted Treatment Effects in Indirect Comparisons with Limited Patient-Level Data}
\author[1,2]{Antonio Remiro-Azócar}
\author[1,3,4]{Anna Heath}
\author[1]{Gianluca Baio}
\affil[1]{Department of Statistical Science, University College London}
\affil[2]{Quantitative Research, Statistical Outcomes Research \& Analytics (SORA)}
\affil[3]{Child Health Evaluative Sciences, The Hospital for Sick Children}
\affil[4]{Dalla Lana School of Public Health, University of Toronto}
\date{v0.8, \today}

\begin{document}

\maketitle

\begin{abstract}
Population adjustment methods such as matching-adjusted indirect comparison (MAIC) are increasingly used to compare marginal treatment effects when there are cross-trial differences in effect modifiers and limited patient-level data. MAIC is based on propensity score weighting, which is sensitive to poor covariate overlap and cannot extrapolate beyond the observed covariate space. Current outcome regression-based alternatives can extrapolate but target a conditional treatment effect that is incompatible in the indirect comparison. When adjusting for covariates, one must integrate or average the conditional estimate over the population of interest to recover a compatible marginal treatment effect. We propose a marginalization method based on parametric G-computation that can be easily applied where the outcome regression is a generalized linear model or a Cox model. In addition, we introduce a novel general-purpose method based on the ideas underlying multiple imputation, which we term multiple imputation marginalization (MIM) and is applicable to a wide range of models, including parametric survival models. The approaches view the covariate adjustment regression as a nuisance model and separate its estimation from the evaluation of the marginal treatment effect of interest. Both methods can accommodate a Bayesian statistical framework, which naturally integrates the analysis into a probabilistic framework. A simulation study provides proof-of-principle for the methods and benchmarks their performance against MAIC and the conventional outcome regression. The simulations are based on scenarios with binary outcomes and continuous covariates, with the log-odds ratio as the measure of effect. The marginalized outcome regression approaches achieve more precise and more accurate estimates than MAIC, particularly when covariate overlap is poor, and yield unbiased marginal treatment effect estimates under no failures of assumptions. Furthermore, the marginalized regression-adjusted estimates provide greater precision and accuracy than the conditional estimates produced by the conventional outcome regression, which are systematically biased because the log-odds ratio is a non-collapsible measure of effect.
\textbf{Keywords:} Health technology assessment; indirect treatment comparison; causal inference; marginal treatment effect; outcome regression; standardization
\end{abstract}

\clearpage

\begin{center}
	\tableofcontents
	\clearpage
	\listoffigures
\end{center}
\pagebreak

\renewcommand{\thefootnote}{\alph{footnote}}

\section{Introduction}

The development of novel pharmaceuticals requires several stages, which include regulatory evaluation and, in several jurisdictions, health technology assessment (HTA).\cite{vreman2020decision} To obtain regulatory approval, a new technology must demonstrate efficacy and randomized controlled trials (RCTs) are the gold standard for this purpose, due to their potential in limiting bias.\cite{temple2000placebo} Evidence supporting regulatory approval is often provided by a two-arm RCT, typically comparing the new technology to placebo or standard of care. Then, in certain jurisdictions, HTA addresses whether the health care technology should be publicly funded by the health care system. For HTA, manufacturers must convince payers that their product offers the best ``value for money'' of all available options in the market. This demands more than a demonstration of efficacy\cite{paul2001fourth} and will often require the comparison of treatments that have not been trialed against each other.\cite{sutton2008use}

In the absence of head-to-head trials, indirect treatment comparisons (ITCs) are at the top of the hierarchy of evidence to inform treatment and reimbursement decisions, and are very prevalent in HTA.\cite{dias2013evidence} Standard ITCs use indirect evidence obtained from RCTs through a common comparator arm.\cite{dias2013evidence, bucher1997results} These techniques are compatible with both individual patient data (IPD) and aggregate-level data (ALD). However, they are biased when the distribution of effect measure modifiers differs across trial populations, meaning that relative treatment effects are not constant.\cite{phillippo2018methods} 

% Standard propensity score methods\cite{austin2011introduction} can adjust for these differences in a pairwise comparison but require patient-level data for all studies.\cite{faria2015nice} 

Often in HTA, there are: (1) no head-to-head trials comparing the interventions of interest; (2) IPD available from the manufacturer's own trial but only published ALD for the comparator(s); and (3) imbalances in effect measure modifiers across studies. Several ``pairwise'' methods, labeled \textit{population-adjusted indirect comparisons}, have been introduced to estimate relative treatment effects in this scenario. These include matching-adjusted indirect comparison (MAIC),\cite{signorovitch2010comparative} based on inverse propensity score weighting, and simulated treatment comparison (STC),\cite{caro2010no} based on outcome modeling/regression. There is a simpler alternative, crude direct post-stratification (also known as non-parametric standardization, subclassification or direct adjustment)\cite{miettinen1972standardization}, but this fails if any of the covariates are continuous or where there are several covariates for which one must account,\cite{stuart2011use} in which case it is also inefficient. Very recently, a relevant outcome modeling-based approach called multilevel network meta-regression (ML-NMR) has been developed.\cite{phillippo2020multilevel, phillippo2019calibration} This incorporates larger networks of treatments and studies. The focus of this article is on the pairwise approaches but ML-NMR is considered in the discussion. 

Recommendations on the use of MAIC and STC in HTA have been provided, defining the relevant terminology and evaluating the theoretical validity of these methods.\cite{phillippo2018methods, phillippo2016nice} However, this guidance is only provisional as further research must: (1) examine these methods through comprehensive simulation studies; and (2) develop novel methods for population adjustment.\cite{phillippo2018methods, phillippo2016nice} In addition, recommendations have highlighted the importance of embedding the methods within a Bayesian framework, which allows for the principled propagation of uncertainty to the wider health economic model,\cite{baio2012bayesian} and is particularly appealing for ``probabilistic sensitivity analysis''.\cite{claxton2005probabilistic} This is a required component in the normative framework of HTA bodies such as NICE,\cite{baio2012bayesian, claxton2005probabilistic} used to characterize the impact of the uncertainty in the model inputs on decision-making.

Recently, several simulation studies have been conducted to assess population-adjusted indirect comparisons.\cite{remiro2020methods, cheng2019statistical, hatswell2020effects, phillippo2020assessing} Remiro-Az\'ocar et al. perform a simulation study benchmarking the performance of the typical use of MAIC and STC against the standard ITC for the Cox model and survival outcomes.\cite{remiro2020methods} In this study, MAIC yields unbiased and relatively accurate treatment effect estimates under no failures of assumptions, but the robust sandwich variance may underestimate standard errors where effective sample sizes are small. In the simulation scenarios, there is some degree of overlap between the studies' covariate distributions. Nevertheless, it is well known that weighting methods like MAIC are highly sensitive to poor overlap, are not asymptotically efficient, and incapable of extrapolation.\cite{phillippo2020assessing, stuart2010matching, lee2011weight, hirano2001estimation} With poor overlap, extreme weights may produce unstable treatment effect estimates with high variance. A related problem in finite samples is that feasible weighting solutions may not exist\cite{jackson2020alternative} due to separation problems where samples sizes are small and the number of covariates is large.\cite{van2011targeted, neugebauer2005prefer}

Outcome regression approaches such as STC are appealing as these tend to be more efficient than weighting, providing more stable estimators and allowing for model extrapolation.\cite{robins1992estimating} We view extrapolation as an advantage because poor overlap, with small effective sample sizes and large percentage reductions in effective sample size, is a pervasive issue in HTA.\cite{phillippo2019population} While extrapolation can also be viewed as a disadvantage if it is not valid, in our case it expands the range of scenarios in which population adjustment can be used.

The aforementioned simulation study\cite{remiro2020methods} demonstrates that the typical usage of STC, as described by HTA guidance and recommendations,\cite{phillippo2016nice} produces systematically biased estimates of the marginal treatment effect, with inappropriate coverage rates, because it targets a conditional estimand instead. With the Cox model and survival outcomes, there is bias because the conditional (log) hazard ratio is non-collapsible. In addition, the conditional estimand cannot be combined in any indirect treatment comparison or compared between studies because non-collapsible conditional estimands vary across different covariate adjustment sets. This is a recurring problem in meta-analysis.\cite{hauck1998should, daniel2020making}

The crucial element that has been missing from the typical usage of STC is the marginalization of treatment effect estimates. When adjusting for covariates, one must integrate or average the conditional estimate over the joint covariate distribution to recover a marginal treatment effect that is compatible in the indirect comparison. We propose a simple marginalization method based on parametric G-computation\cite{robins1986new, robins1987graphical} or model-based standardization,\cite{moore2009covariate, austin2010absolute, rosenblum2010simple, zhang2008estimating} often applied in observational studies in epidemiology and medical research where treatment assignment is non-random. In meta-analysis, Vo et al.\cite{vo2019novel, vo2021assessing} have used parametric G-computation to transport RCT results to a specific target population. We extend these approaches to population-adjusted indirect comparisons with limited patient-level data. In addition, we introduce a novel general-purpose method based on the ideas underlying multiple imputation,\cite{rubin2004multiple} which we term \textit{multiple imputation marginalization} (MIM) and is applicable to a wide range of models, including parametric survival models. 

Both parametric G-computation and multiple imputation marginalization can be viewed as extensions to the conventional STC, with all methods making use of effectively the same outcome model. The novel methodologies are outcome regression approaches, thereby capable of extrapolation, that target marginal treatment effects. They do so by separating the covariate adjustment regression model from the evaluation of the marginal treatment effect of interest. The conditional parameters of the regression are viewed as nuisance parameters, not directly relevant to the research question. The methods can be implemented in a Bayesian statistical framework, which explicitly accounts for relevant sources of uncertainty, allows for the incorporation of prior evidence (e.g.~expert opinion), and naturally integrates the analysis into a probabilistic framework, typically required for HTA.\cite{baio2012bayesian} 

In this paper, we carry out a simulation study to benchmark the performance of the novel methods against MAIC and the conventional STC. The simulations provide proof-of-principle and are based on scenarios with binary outcomes and continuous covariates, with the log-odds ratio as the measure of effect. 
% The methods are evaluated in 162 scenarios that vary the trial sample size, effect-modifying strength of covariates, prognostic effect of covariates, covariate overlap/imbalance and the level of correlation in the covariates. 
The marginalized outcome regression approaches achieve greater precision and accuracy than MAIC and are unbiased under no failures of assumptions. Furthermore, the marginalized regression-adjusted estimates provide greater precision than the conditional estimates produced by the conventional version of STC. While this precision comparison is irrelevant, because it is made for estimators of different estimands, it supports previous research on non-collapsible measures of effect.\cite{moore2009covariate, daniel2020making}  

In Section \ref{sec2}, we present the context and data requirements for population-adjusted indirect comparisons. Section \ref{sec3} provides a detailed description of the outcome regression methodologies. Section \ref{sec4} outlines a simulation study, which  evaluates the statistical properties of different approaches to outcome regression with respect to MAIC. Section \ref{sec5} describes the results from the simulation study. An extended discussion of our findings is presented in Section \ref{sec6}. 

\section{Context}\label{sec2}

Consider an active treatment $A$, which needs to be compared to another active treatment $B$ for the purposes of reimbursement. Treatment $A$ is new and being tested for cost-effectiveness, while treatment $B$ is typically an established intervention, already on the market. Both treatments have been evaluated in a RCT against a common comparator $C$, e.g.~standard of care or placebo, but not against each other. Indirect comparisons are performed to estimate the relative treatment effect for a specific outcome. The objective is to perform the analysis that would be conducted in a hypothetical head-to-head RCT between $A$ and $B$, which indirect treatment comparisons seek to emulate. 

The RCT is widely considered the gold standard design to evaluate treatments\cite{glenny2005indirect} due to its internal validity.\cite{temple2000placebo} Appropriate randomization guarantees covariate balance on expectation, so that the treatment groups are comparable and confounding is limited. Therefore, assuming no structural issues (e.g.~no dropout, measurement error, etc.), RCTs allow for unbiased estimation of the relative treatment effect within the study. 

RCTs have different types of potential target average estimands of interest: \textit{marginal} or \textit{population-average} effects, calibrated at the population level, and \textit{conditional} effects, calibrated at the individual level. The former are typically, but not necessarily, estimated by an ``unadjusted'' analysis. This may be a simple comparison of the expected outcomes for each group or a univariable regression including only the main treatment effect. Conditional treatment effects are typically estimated by an ``adjusted'' analysis (e.g.~a multivariable regression of outcome on treatment and covariates), accounting for prognostic variables that are pre-specified in the protocol or analysis plan, such as prior medical/treatment history, demographics and physiological status. A recurring theme throughout this article is that the terms ``conditional and adjusted (likewise marginal and unadjusted) should not be used interchangeably'' because marginal need not mean unadjusted and covariate-adjusted analyses may also target marginal estimands.\cite{daniel2020making,remiro2021target}

The marginal effect would be the average effect, at the population level (conditional on the entire population distribution of covariates), of moving all individuals in the trial population between two hypothetical worlds: from one where everyone receives treatment $B$ to one where everyone receives treatment $A$.\cite{austin2011introduction, imbens2004nonparametric, hernan2020causal} The conditional effect corresponds to the average treatment effect at the unit level, conditional on the effects of the covariates that have also been included in the model. This would be the average effect of switching the treatment of an individual in the trial population from $B$ to $A$, fully conditioned on the average combination of subject-level covariates, or the average effect across sub-populations of subjects who share the same covariates. Population-adjusted indirect comparisons are used to inform reimbursement decisions in HTA at the population level. Therefore, marginal treatment effect estimates are required.\cite{remiro2021marginalization}

The indirect comparison between treatments $A$ and $B$ is typically carried out in the ``linear predictor'' scale;\cite{dias2013evidence, bucher1997results} namely, using additive effects for a given linear predictor, e.g.~log-odds ratio for binary outcomes or log hazard ratio for survival outcomes. Indirect treatment comparisons can be ``anchored'' or ``unanchored''. Anchored comparisons make use of a connected treatment network. In this case, this is available through a common comparator $C$. Unanchored comparisons use disconnected treatment networks or single-arm trials and require much stronger assumptions than their anchored counterparts.\cite{phillippo2018methods} The use of unanchored comparisons where there is connected evidence is discouraged and often labeled as problematic.\cite{phillippo2018methods, phillippo2016nice} This is because it does not respect within-study randomization and is not protected from imbalances in any covariates that are prognostic of outcome (almost invariably, a larger set of covariates than the set of effect measure modifiers). Hence, our focus is on anchored comparisons. 

In the standard anchored scenario, a manufacturer submitting evidence to HTA bodies has access to IPD from its own trial that compares its treatment $A$ against the standard health technology $C$. The disclosure of proprietary, confidential IPD from industry-sponsored clinical trials is rare. Hence, individual-level data on baseline covariates, treatment and outcomes for the competitor's trial, evaluating the relative efficacy or effectiveness of intervention $B$ vs.~$C$, are regularly unavailable, for both the submitting company and the national HTA agency assessing the evidence. For this study, only summary outcome measures and marginal moments of the covariates, e.g.~means with standard deviations for continuous variables or proportions for binary and categorical variables, as found in so-called ``Table 1'' of clinical trial publications, are available. We consider, without loss of generality, that IPD are available for a study comparing treatments $A$ and $C$ (denoted $AC$) and published ALD are available for a study comparing interventions $B$ and $C$ ($BC$).

Standard ITCs such as the Bucher method\cite{bucher1997results} assume that there are no differences across trials in \textit{effect measure modifiers}. A variable is an effect measure modifier, \textit{effect modifier} for short, if the relative effect of a particular intervention, as measured on a specific scale, varies at different levels of the variable. For instance, if women react differently to a drug therapy than men on the log-odds ratio scale, then gender modifies the effect of the drug on such scale. Within the biostatistics literature, effect modification is usually referred to as heterogeneity or interaction, because effect modifiers are considered to alter the effect of treatment by interacting with it on a specific scale,\cite{vanderweele2009concerning} and are typically detected by examining statistical interactions.\cite{rothman1980concepts}

Consider that $Z$ denotes a treatment indicator. Active treatment $A$ is denoted $Z=1$, active treatment $B$ is denoted $Z=2$, and the common comparator $C$ is denoted $Z=0$. In addition, $S$ denotes a specific study. The $AC$ study, comparing treatments $A$ and $C$ is denoted $S=1$. The $BC$ study is denoted $S=2$. The true relative treatment effect between $Z$ and $Z'$ in study population $S$ is indicated by $\Delta_{ZZ'}^{(S)}$ and is estimated by $\hat{\Delta}_{ZZ'}^{(S)}$.

In standard ITCs, one assumes that the $A$ vs.~$C$ treatment effect $\Delta_{10}^{(1)}$ in the $AC$ population is equal to $\Delta_{10}^{(2)}$, the effect that would have have occurred in the $BC$ population. Note that the Bucher method and most conventional network meta-analysis methods do not explicitly specify a target population of policy interest (whether this is $AC$, $BC$ or otherwise).\cite{manski2019meta} Hence, they cannot account for differences in covariates across study populations. The Bucher method is only valid when either: (1) the $A$ vs.~$C$ treatment effect is homogeneous, such that there is no effect modification; or (2) the distributions of the effect modifiers are the same in both studies. 

If the $A$ vs.~$C$ treatment effect is heterogeneous and the effect modifiers are not equidistributed across trials, relative treatment effects are no longer constant across the trial populations, except in the pathological case where the bias induced by different effect modifiers is in opposite directions and cancels out. Hence, the assumptions of the Bucher method are broken. In this scenario, standard ITC methods are liable to produce biased and overprecise estimates of the treatment effect.\cite{song2003validity} These features are undesirable, particularly from the economic modeling point of view, as they impact negatively on the probabilistic sensitivity analysis.

Conversely, MAIC and STC target the $A$ vs.~$C$ treatment effect that would be observed in the $BC$ population, thereby performing an adjusted indirect comparison in such population. MAIC and STC implicitly assume that the target population is the $BC$ population. The estimate of the adjusted $A$ vs.~$B$ treatment effect is:
\begin{equation}
\hat{\Delta}_{12}^{(2)} = \hat{\Delta}_{10}^{(2)} - \hat{\Delta}_{20}^{(2)},
\label{eqn1}
\end{equation}
where $\hat{\Delta}_{10}^{(2)}$ is the estimated relative treatment effect of $A$ vs $C$ (mapped to the $BC$ population), and $\hat{\Delta}_{20}^{(2)}$ is the estimated marginal treatment effect of $B$ vs.~$C$ (in the $BC$ population). The estimate $\hat{\Delta}_{20}^{(2)}$ and an estimate of its variance may be directly published or derived non-parametrically from crude aggregate outcomes made available in the literature. The majority of RCT publications will report an estimate targeting a marginal treatment effect, derived from a simple regression of outcome on a single independent variable, treatment assignment. In addition, the estimate $\hat{\Delta}_{12}^{(2)}$ should target a marginal treatment effect for reimbursement decisions at the population level. Therefore, $\hat{\Delta}_{10}^{(2)}$ should target a marginal treatment effect that is compatible with $\hat{\Delta}_{20}^{(2)}$.\cite{remiro2020conflating}

As the relative effects, $\hat{\Delta}_{10}^{(2)}$ and $\hat{\Delta}_{20}^{(2)}$, are specific to separate studies, the within-trial randomization of the originally assigned patient groups is preserved. Because the estimates are based on different study samples (IPD are unavailable for $BC$), the within-trial relative effects are assumed statistically independent of each other. Hence, their variances are simply summed to estimate the variance of the $A$ vs.~$B$ treatment effect. One can also take a Bayesian approach to estimating the indirect treatment comparison, in which case variances would be derived empirically from draws of the posterior density. We believe that a Bayesian analysis is helpful because simulation from the posterior distribution provides a framework for probabilistic decision-making, directly allowing for both statistical estimation and inference, and for principled uncertainty propagation.\cite{dias2013evidence} 

A reference intervention is required to define the effect modifiers. In the methods considered in this article, we are selecting the effect modifiers of treatment $A$ with respect to $C$ (as opposed to the treatment effect modifiers of $B$ vs.~$C$). This is because we have to adjust for these in order to perform the indirect comparison in the $BC$ population, implicitly assumed to be the target population. If one had access to IPD for the $BC$ study and only published ALD for the $AC$ study, one would have to adjust for the factors modifying the effect of treatment $B$ with respect to $C$, in order to perform the comparison in the $AC$ population.

In some contexts, a distinction is made between sample-average and population-average marginal effects.\cite{stuart2011use, cole2010generalizing, kern2016assessing, hartman2015sample} Typically, another implicit assumption made by population-adjusted indirect comparisons is that the marginal treatment effects estimated in the $BC$ trial sample (as described by its published covariate summaries for $\hat{\Delta}_{10}^{(2)}$), match those that would be estimated in the study's target population. This implies that either: (1) the sample on which inferences are made is the trial's target population; or (2) it is a simple random sample representative of such population, ignoring sampling variability. 

\subsection{The need for outcome regression approaches}\label{subsec21}

At present, matching-adjusted indirect comparison (MAIC)\cite{signorovitch2010comparative} is the most commonly used population-adjusted indirect comparison method.\cite{phillippo2019population} ``Matching-adjusted'' is a misnomer, as the indirect comparison is actually ``weighting-adjusted'', with the population adjustment based on propensity score weighting.\cite{rosenbaum1987model} A logistic regression is used to model the trial assignment odds conditional on a selected set of baseline covariates. The weights estimated by the model represent the ``trial selection'' odds, namely, the odds of being enrolled in the $BC$ trial. These are balancing scores that, when applied to the $AC$ IPD, form a pseudo-population that has balanced covariate moments with respect to the $BC$ population. The weights are often applied to a weighted simple regression to estimate the marginal treatment effect for $A$ vs.~$C$ in the $BC$ population. However, MAIC does not explicitly require an outcome model. The development of outcome regression methods, which estimate an outcome-generating mechanism given treatment and the baseline covariates, is appealing for several reasons:

(1) \textbf{Statistical precision and efficiency.} Outcome regression tends to give more precise estimates than weighting. Weighting is particularly inefficient and unstable where covariate overlap is poor and effective sample sizes are small,\cite{neugebauer2005prefer, van2003unified, ertefaie2010comparing, daniel2013methods, lunceford2004stratification} as it is sensitive to inordinate influence by extreme weights. Outcome regression can extrapolate the association between outcome and covariates where overlap is insufficient, while weighting methods cannot extrapolate beyond the observed covariate space in the $AC$ IPD. Valid extrapolation, using the linearity assumption or other appropriate assumptions about the input space, requires accurately capturing the true covariate-outcome relationships. 

(2) \textbf{Different modeling assumptions.} While MAIC relies on a correctly specified model for the conditional odds of trial assignment given the covariates, outcome regression methods rely on a correctly specified model for the conditional expectation of the outcome given treatment and the covariates. In my experience, identifying the variables that affect outcome is more straightforward than identifying the factors that drive trial assignment in the context of population-adjusted indirect comparisons. This is not typically the case in the standard use of propensity score weighting in observational studies, where one identifies the factors that drive treatment (as opposed to trial) assignment. Nevertheless, in our scenario, the factors driving selection into different RCTs are often arbitrary.\cite{shenoy2015elderly, khan2020participation} Researchers may benefit from the use of distinct modeling approaches with different assumptions, as these can yield different results, especially if there is a violation of assumptions. 
    
(3) \textbf{Flexibility.} Researchers could use augmented or doubly robust methods\cite{hernan2020causal, bang2005doubly, vansteelandt2011invited, dahabreh2019generalizing} that combine the model for the expectation of the outcome with the trial assignment odds model. These are attractive due to their increased robustness to model misspecification: consistent estimation only requires the correct specification of either of the two models, not necessarily both.\cite{bang2005doubly, kang2007demystifying} Even with the reduced  misspecification risk, they tend to have improved precision and efficiency with respect to the standard weighting estimators.\cite{tan2007comment}

\vspace{0.4cm}

\subsection{Some assumptions}\label{subsec22}

Besides the differences in (typically parametric) model specification, weighting methods such as MAIC and outcome regression methods such as those discussed in this article mostly require the same set of assumptions. An in-depth non-technical description of these is detailed in \hyperref[SA]{Supplementary Appendix A} and a short discussion is presented in subsection \ref{subsec62}. The assumptions include: 

\begin{enumerate}
    \item Internal validity of the $AC$ and $BC$ trials, e.g.~appropriate randomization and sufficient sample sizes so that the treatment groups are comparable, no interference, negligible measurement error or missing data, the absence of non-compliance, etc.
    \item Consistency under parallel studies such that both trials have identical control treatments, sufficiently similar study designs and outcome measure definitions, and have been conducted in care settings with a high degree of similarity. 
    \item Accounting for all effect modifiers of treatment $A$ vs.~$C$ in the adjustment. This assumption is called the conditional constancy of the $A$ vs.~$C$ marginal treatment effect,\cite{phillippo2016nice} and requires that a sufficiently rich set of baseline covariates has been measured for the $AC$ study and is available in the $BC$ study publication.\footnote{In the anchored scenario, we are interested in a comparison of \textit{relative} outcomes or effects, not \textit{absolute} outcomes. Hence, an anchored comparison only requires conditioning on the effect modifiers of the $A$ vs.~$C$ treatment effect. This assumption is named the \textit{conditional constancy of relative effects},\cite{phillippo2018methods, phillippo2016nice} i.e., given the selected effect-modifying covariates, the marginal $A$ vs.~$C$ treatment effect is constant across the $AC$ and $BC$ populations. There are other formulations of this assumption,\cite{stuart2011use, cole2010generalizing, kern2016assessing, pearl2014external, hartman2015sample} such as trial assignment/selection being conditionally ignorable, unconfounded or exchangeable for such treatment effect, i.e., conditionally independent of the treatment effect, given the selected effect modifiers. One can consider that being in population $AC$ or population $BC$ does not carry over any information about the marginal $A$ vs $C$ treatment effect, once we condition on the treatment effect modifiers. This means that after adjusting for these effect modifiers, treatment effect heterogeneity and trial assignment are conditionally independent. Another advantage of outcome regression with respect to weighting is that, by being less sensitive to overlap issues, it allows for the inclusion of larger numbers of effect modifiers. This makes it easier to satisfy the conditional constancy of relative effects.}
    \item Overlap between the covariate distributions in $AC$ and $BC$. More specifically, that the ranges of the selected covariates in the $AC$ trial cover their respective moments in the $BC$ population. The overlap assumption (often referred to as ``positivity'') can be overcome in outcome regression if one is willing to rely on model extrapolation, assuming correct model specification.\cite{hernan2020causal} 
    \item Correct specification of the $BC$ population. Namely, that it is appropriately represented by the information available to the analyst, that does not have access to patient-level data from the $BC$ study. 
    % \item A final assumption, the shared effect modifier assumption (described in subsection \ref{subsec43}), is required to transport the treatment effect estimate for $A$ vs.~$B$ from the $BC$ population to any other target population. Otherwise, one has to assume that the $BC$ population is the target for the analysis to be valid.
\end{enumerate}
Most assumptions are causal and untestable, with their justification typically requiring prior substantive knowledge.\cite{remiro2020principled} Nevertheless, we shall assume that they hold throughout the article. We discuss potential failures of assumptions and their consequences in subsection \ref{subsec62}, in the context of the simulation study, and in \hyperref[SA]{Supplementary Appendix A}.

\section{Methodology}\label{sec3}

\subsection{Data structure}\label{subsec31}

For the $AC$ trial IPD, let $\mathcal{D}_{AC} = ({\boldsymbol{x}, \boldsymbol{z}, \boldsymbol{y}})$. Here, $\boldsymbol{x}$ is a matrix of baseline characteristics (covariates), e.g.~age, gender, comorbidities, baseline severity, of size $N \times K$, where $N$ is the number of subjects in the trial and $K$ is the number of available covariates. For each subject $n=1,\dots,N$, a row vector $\boldsymbol{x}_n$ of $K$ covariates is recorded. Each baseline characteristic can be classed as a prognostic variable (a covariate that affects outcome), an effect modifier, both or none. For simplicity in the notation, it is assumed that all available baseline characteristics are prognostic of the outcome and that a subset of these, $\boldsymbol{x}^{\boldsymbol{(EM)}} \subseteq \boldsymbol{x}$, is selected as effect modifiers on the linear predictor scale, with a row vector $\boldsymbol{x}^{(EM)}_n$ recorded for each subject. We let $\boldsymbol{y} = (y_1, y_2, \dots, y_N)$ represent a vector of outcomes, e.g.~a time-to-event or binary indicator for some clinical measurement; and $\boldsymbol{z} = (z_1, z_2, \dots, z_N)$ is a treatment indicator ($z_n=1$ if subject $n$ is under treatment $A$ and $z_n=0$ if under $C$). For simplicity, we shall assume that there are no missing values in $\mathcal{D}_{AC}$. As outlined in the \hyperref[sec6]{Discussion}, 
the outcome regression methodologies can be readily adapted to address this issue, particularly under a Bayesian implementation, but this is an area for future research.

We let $\mathcal{D}_{BC}=[\boldsymbol{\theta}, \boldsymbol{\rho}, \hat{\Delta}_{20}^{(2)}, \hat{V}(\hat{\Delta}_{20}^{(2)})]$ denote the information available for the $BC$ study. No individual-level information on covariates, treatment or outcomes is available. Here, $\boldsymbol{\theta}$ represents a vector of published covariate summaries, e.g.~proportions or means. For ease of exposition, we shall assume that these are available for all $K$ covariates (otherwise, one would take the intersection of the available covariates), and that the selected effect modifiers are also available such that $\boldsymbol{\theta}^{\boldsymbol{(EM)}} \subseteq \boldsymbol{\theta}$. An estimate $\hat{\Delta}_{20}^{(2)}$ of the $B$ vs.~$C$ treatment effect in the $BC$ population, and an estimate of its variance $\hat{V}(\hat{\Delta}_{20}^{(2)})$, either published directly or derived from crude aggregate outcomes in the literature, are also available. Note that these are not used in the adjustment mechanism but are ultimately required to perform inference for the indirect comparison in the $BC$ population. 

Finally, we let the symbol $\boldsymbol{\rho}$ stand for the dependence structure of the $BC$ covariates. Under certain assumptions about representativeness, this can be retrieved from the $AC$ trial, e.g.~through the observed pairwise correlations, or from external data sources such as registries. This information, together with the published covariate summary statistics, is required to characterize the joint covariate distribution of the $BC$ population. A pseudo-population of $N^*$ subjects is simulated from this joint distribution, such that $\boldsymbol{x}^*$ denotes a matrix of baseline covariates of dimensions $N^* \times K$, with a row vector $\boldsymbol{x}_i^*$ of $K$ covariates simulated for each subject $i=1,\dots N^*$. Notice that the value of $N^*$ does not necessarily have to correspond to the actual sample size of the $BC$ study; however, the simulated cohort must be sufficiently large so that the sampling distribution is stabilized, minimizing sampling variability. Again, a subset of the simulated covariates, $\boldsymbol{x}^{*\boldsymbol{(EM)}} \subseteq \boldsymbol{x^*}$, makes up the treatment effect modifiers on the linear predictor scale, with a row vector $\boldsymbol{x}_i^{*\boldsymbol{(EM)}}$ for each subject $i=1,\dots,N^*$. In this article, the asterisk superscript represents unobserved quantities that have been constructed in the $BC$ population. 

The outcome regression approaches discussed in this article estimate treatment effects with respect to a hypothetical pseudo-population for the $BC$ study. Before outlining the specific outcome regression methods, we explain how to generate values for the individual-level covariates $\boldsymbol{x^*}$ for the $BC$ population using Monte Carlo simulation. 

\subsection{Individual-level covariate simulation}\label{subsec32}

Ideally, the $BC$ population should be characterized by the full joint distribution of covariates. However, the restriction of limited IPD makes it unlikely that the joint distribution of the $BC$ covariates is available. Where there are not many covariates and these are binary, this is sometimes available as a cross-tabulation. However, most of the time we need to approximate the joint distribution appropriately. This is important to avoid bias arising from the incomplete specification of the $BC$ population. The published summary values $\boldsymbol{\theta}$ and the correlation structure $\boldsymbol{\rho}$ are combined, making certain parametric assumptions about the marginal distributional forms, to infer the joint distribution of the $BC$ covariates and construct an appropriate pseudo-population for inferences. The proposed approaches allow the analyst to bring in some prior knowledge or evidence to inform the potential distributions of the covariates. However, it is worth noting that we cannot give a general recipe for this step, which requires context-specific knowledge that is likely not available from the observed data in the trials. 

Firstly, the marginal distributions for each covariate are specified. The mean and, if applicable, the standard deviation of the marginals are sourced from the $BC$ report to match the published summary statistics. As the true marginal distributional forms are not known, some parametric assumptions are required. For instance, if it is reasonable to assume that sampling variability for a continuous covariate can be described using a normal distribution, and the covariate's mean and standard deviation are published in the $BC$ report, we can assume that it is marginally normally distributed. Hence, we can also select the family for the marginal distribution using the theoretical validity of the candidate distributions alongside the IPD. For example, the marginal distribution of duration of prior treatment at baseline could be modeled as a log-normal or Gamma distribution as these distributions are right-skewed and bounded to the left by zero. Truncated distributions can be used to resemble the inclusion/exclusion criteria for continuous covariates in the $BC$ trial, e.g.~age limits, and avoid deterministic overlap violations.  

Secondly, the correlations between covariates are specified. We suggest two possible data-generating model structures for this purpose: (1) simulating the covariates from a multivariate Gaussian copula;\cite{phillippo2020multilevel, nelsen2007introduction} or (2) factorizing the joint distribution of the covariates into the product of marginal and conditional distributions.
%; for instance, if the covariates are $\boldsymbol{x}=(x_1,x_2)$ indicating age and sex, respectively, we may write $p(x_1,x_2)=p(x_2)p(x_1\mid x_2)$, where the marginal distribution over sexes and the conditional distribution for age (groups) by sexes could be informed using census data. 
The former approach is perhaps more general-purpose. The latter is more flexible, defining separate models for each variable, but its specification can be daunting where there are many covariates and interdependencies are complex. 

Any multivariate joint distribution can be decomposed in terms of univariate marginal distribution functions and a dependence structure.\cite{sklar1959fonctions} A Gaussian copula ``couples'' the marginal distribution functions for each covariate to a multivariate Gaussian distribution function. The main appeal of a copula is that the correlation structure of the covariates and the marginal distribution for each covariate can be modeled separately. We may use the pairwise correlation structure observed in the $AC$ patient-level data as the dependence structure, while keeping the marginal distributions inferred from the $BC$ summary values and the IPD. Note that the term ``Gaussian'' does not refer to the marginal distributions of the covariates but to the correlation structure. While the Gaussian copula is sufficiently flexible for most modeling purposes, more complex copula types (e.g.~Clayton, Gumbel, Frank) may provide different and more customizable correlation structures.\cite{nelsen2007introduction}

Alternatively, we can account for the correlations by factorizing the joint distribution of covariates in terms of marginal and conditional densities. This strategy is common in implementations of sequential conditional algorithms for parametric multiple imputation.\cite{royston2004multiple, buuren2010mice} For instance, consider two baseline characteristics: $age$, which is a continuous variable, and the presence of a comorbidity $c$, which is dichotomous. We can factorize the joint distribution of the covariates such that $p(age, c)=p(c \mid age)p(age)$. 

\begin{figure}[!htb]
\center{\includegraphics[width=0.74\textwidth]{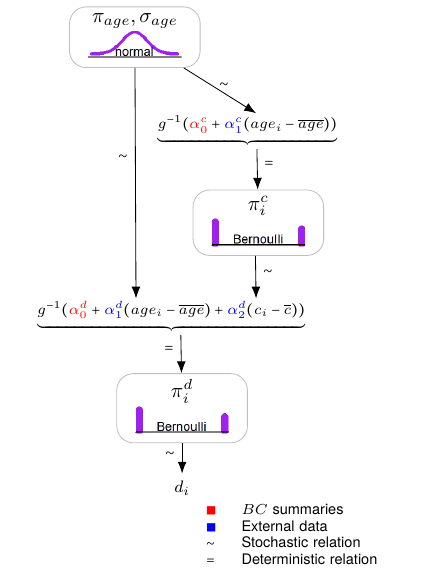}}
\caption[]
{
An example of individual-level Monte Carlo covariate simulation where the joint distribution of three baseline characteristics, $age$, comorbidity $c$ and comorbidity $d$, is factorized into the product of marginal and conditional distributions, such that $p(age, c, d)=p(d \mid c, age)p(c \mid age)p(age)$. The joint distribution is valid because the conditional distributions defining the covariates are compatible: we start with a marginal distribution for age and construct the joint distribution by modeling each additional covariate, one-by-one, conditionally on the covariates that have already been simulated. This diagram adopts the convention of Kruschke.\cite{kruschke2014doing}}
\label{fig1}
\end{figure}

In this scenario, we draw $age_i$ for subject $i$ from a suitable marginal distribution, e.g.~a normal, with the mean and standard deviation sourced from the published $BC$ summaries or official life tables. The mean $\pi_i^{c}$ of $c$ (the conditional proportion of the comorbidity) given the age, can be modeled through a regression: $\pi_i^{c} = g^{-1}(\alpha_0^{c} + \alpha_1^{c}(age_i - \overline{age}))$, with $c_i \sim \textnormal{Bernoulli}(\pi_i^c)$ where $g(\cdot)$ is an appropriate link function. Here, the coefficients $\alpha_0^{c}$ and $\alpha_1^{c}$ represent respectively the overall proportion of comorbidity $c$ in the $BC$ population (marginalizing out the age), and the correlation level between comorbidity $c$ and (the centered version of) age. The former coefficient can be directly sourced from the published $BC$ summaries, whereas the latter could be derived from pairwise correlations observed in the $AC$ IPD or from external sources, e.g.~clinical expert opinion, registries or administrative data, applying the selection criteria of the $BC$ trial to subset the data. Figure \ref{fig1} provides an example of a similar probabilistic structure with three covariates: $age$ and the presence of two comorbidities, $c$ and $d$. In this example, the distribution of the covariates is factorized such that $p(age, c, d)=p(d \mid c, age)p(c \mid age)p(age)$. 

\clearpage

It is important to acknowledge that this ``covariate simulation'' step arises due to a suboptimal scenario, where patient-level data on covariates are unavailable for the $BC$ study. Ideally, this should be freely available or, at least, disclosed by the sponsor company. Raw patient-level data are always the preferred input for statistical inference, allowing for the testing of assumptions.\cite{chan2014increasing} The underlying reasons for unavailable IPD are diverse and span across a range of issues. Perhaps the most sensitive of these is privacy, with the General Data Protection Regulation\cite{voigt2017eu} ratified by the European Union in 2018 recognizing data concerning health as a special category of data with specific protection safeguards and disclosure regulations.

We note that, if the main hindrance to the availability of IPD is privacy, the manufacturer itself could facilitate statistical inference by using the IPD to create fully artificial covariate datasets.\cite{nowok2016synthpop} The release of such datasets would not involve a violation of privacy or confidentiality and would avoid the need for the ``covariate simulation'' step. Alternatively, Bonofiglio et al.\cite{bonofigliorecovery} have recently proposed a framework where access to covariate correlation summaries is made possible through distributed computing. It is unclear whether access to such framework would be granted to a competitor submitting evidence for reimbursement to HTA bodies, albeit the summaries could be reported in clinical trial publications.

\subsection{Conventional outcome regression}\label{subsec33}

In simulated treatment comparison (STC), IPD from the $AC$ trial are used to fit a regression model describing the observed outcomes $\boldsymbol{y}$ in terms of the relevant baseline characteristics $\boldsymbol{x}$ and the treatment variable $\boldsymbol{z}$. 

STC has different formulations.\cite{phillippo2018methods, caro2010no, phillippo2016nice, ishak2015simulation} In the conventional version described by the NICE Decision Support Unit Technical Support Document 18,\cite{phillippo2018methods, phillippo2016nice} the individual-level covariate simulation step described in subsection \ref{subsec32} is not performed. The covariates are centered at the published mean values $\boldsymbol{\theta}$ from the $BC$ population. Under a generalized linear modeling framework, the following regression model is fitted to the observed $AC$ IPD:
\begin{equation}
g(\mu_n) = \beta_0 +  \left(\boldsymbol{x}_n  - \boldsymbol{\theta}  \right)\boldsymbol{\beta_1} + \left[\beta_z +  \left(\boldsymbol{x}_n^{\boldsymbol{(EM)}} - \boldsymbol{\theta}^{\boldsymbol{(EM)}} \right) \boldsymbol{\beta_2}\right]\mathds{1}(z_n=1),
\label{eqn2}
\end{equation}
where, for a generic subject $n$, $\mu_n$ is the expected outcome on the natural scale, e.g.~the probability scale for binary outcomes, $g(\cdot)$ is an invertible canonical link function, $\beta_0$ is the intercept, $\boldsymbol{\beta_1}$ is a vector of $K$ regression coefficients for the prognostic variables, $\boldsymbol{\beta_2}$ is a vector of interaction coefficients for the effect modifiers (modifying the $A$ vs.~$C$ treatment effect) and $\beta_z$ is the $A$ vs.~$C$ treatment coefficient. For binary outcomes in logistic regression, one uses the $\logit(\mu_n) = \ln[\mu_n/(1-\mu_n)]$ link function, but other choices are possible in practice, e.g.~the identity link for standard linear regression with continuous-valued outcomes, or the log link for Poisson regression with count outcomes. In the population adjustment literature, covariates are sometimes centered separately for active treatment and common comparator arms.\cite{petto2019alternative, belger2015inclusion, aouni2020matching} We do not recommend this approach because it may break randomization, distorting the balance between treatment arms $A$ and $C$ on covariates that are not accounted for. If these covariates are prognostic of outcome, this would compromise the internal validity of the within-study treatment effect estimate for $A$ vs.~$C$. 

The regression in Equation \ref{eqn2} models the conditional outcome mean given treatment and the centered baseline covariates. Because the IPD prognostic variables and the effect modifiers are centered at the published mean values from the $BC$ population, the estimated $\hat{\beta}_z$ is directly interpreted as the $A$ vs.~$C$ treatment effect in the $BC$ population or, more specifically, in a pseudopopulation with the $BC$ covariate means and the $AC$ correlation structure. Typically, analysts set $\hat{\Delta}_{10}^{(2)}=\hat{\beta}_z$ in Equation \ref{eqn1}, inputting this coefficient into the health economic decision model.\cite{grimm2019nivolumab, ren2019pembrolizumab} For uncertainty quantification purposes, the variance of said treatment effect is obtained from the standard error estimate of the treatment coefficient in the fitted model.\cite{phillippo2018methods, phillippo2016nice}

An important issue with this approach is that the treatment parameter $\hat{\beta}_z$, extracted from the fitted model, has a conditional interpretation at the individual level, because it is conditional on the baseline covariates included as predictors in the multivariable regression.\cite{austin2011introduction, remiro2020methods} However, we require that $\hat{\Delta}_{10}^{(2)}$ estimates a marginal treatment effect, for reimbursement decisions at the population level. In addition, we require that $\hat{\Delta}_{10}^{(2)}$ is compatible with the published marginal effect for $B$ vs.~$C$, $\hat{\Delta}_{20}^{(2)}$, for comparability in the indirect treatment comparison in Equation \ref{eqn1}. Even if the published estimate $\hat{\Delta}_{20}^{(2)}$ targets a conditional estimand, this cannot be combined in the indirect treatment comparison because it likely has been adjusted using a different covariate set and model specification than $\hat{\beta}_z$.\cite{daniel2020making} An indirect comparison of conditional treatment effects cannot be performed because we do not have access to the $BC$ study patient-level data and cannot derive a compatible conditional estimate for $B$ vs.~$C$, using the same outcome regression specification. In any case, a comparison of conditional treatment effects is not of interest when making decisions at the population level in HTA.\cite{remiro2020conflating} Therefore, the treatment coefficient $\hat{\beta}_z$ does not target the estimand of interest. 

With non-collapsible effect measures such as the odds ratio in logistic regression, marginal and conditional estimands for non-null effects do not coincide,\cite{janes2010quantifying} even with covariate balance and in the absence of confounding.\cite{greenland1987interpretation, greenland1999confounding} Targeting the wrong estimand may induce systematic bias, as observed in a recent simulation study.\cite{remiro2020methods}  Most applications of population-adjusted indirect comparisons are in oncology\cite{remiro2020methods,phillippo2019population} and are concerned with non-collapsible measures of treatment effect such as (log) hazard ratios\cite{austin2011introduction, austin2014use, greenland1987interpretation} or (log) odds ratios.\cite{austin2011introduction, austin2014use, greenland1987interpretation,greenland1999confounding} With both collapsible and non-collapsible measures of effect, maximum-likelihood estimators targeting distinct estimands will have different standard errors. Therefore, marginal and conditional estimates quantify parametric uncertainty differently, and conflating these will lead to the incorrect propagation of uncertainty to the wider health economic decision model, which will be problematic for probabilistic sensitivity analyses. 

\subsection{Marginalization via parametric G-computation}\label{subsec34}

The crucial element that has been missing from the typical application of outcome regression is the marginalization of the $A$ vs. $C$ treatment effect estimate. When adjusting for covariates, one must integrate or average the conditional estimate over the joint $BC$ covariate distribution to recover a marginal treatment effect that is compatible in the indirect comparison. Parametric G-computation\cite{robins1986new, robins1987graphical, keil2018bayesian, snowden2011implementation, wang2017g} is an established method for marginalizing regression-adjusted conditional estimates. The literature on population-adjusted indirect comparisons has been developed separately to G-computation, despite the close relationships between the methodologies. We build a new link between the two in the next paragraphs. 

G-computation in this context consists of: (1) predicting the conditional outcome expectations under treatments $A$ and $C$ for each subject in the $BC$ population; (2) averaging the predictions to produce marginal outcome means on the natural scale; and (3) back-transforming the averages to the linear predictor scale, contrasting the linear predictions to estimate the marginal $A$ vs.~$C$ treatment effect in the $BC$ population. This marginal effect is compatible in the indirect treatment comparison. This procedure is a form of standardization, a technique which has been performed for decades in epidemiology, e.g.~when computing standardized mortality ratios.\cite{vansteelandt2011invited} Parametric G-computation is often called model-based standardization\cite{moore2009covariate, austin2010absolute, rosenblum2010simple} because a parametric model is used to predict the conditional outcome expectations under each treatment. When the covariates and outcome are discrete, the estimation of the conditional expectations could be non-parametric, in which case G-computation is numerically identical to crude direct post-stratification.\cite{miettinen1972standardization} 

G-computation marginalizes the conditional estimates by separating the regression modeling outlined in subsection \ref{subsec33} from the estimation of the marginal treatment effect for $A$ vs.~$C$. Firstly, a regression model of the observed outcome $\boldsymbol{y}$ on the covariates $\boldsymbol{x}$ and treatment $\boldsymbol{z}$ is fitted to the $AC$ IPD:
\begin{equation}
g(\mu_n) = \beta_0 +  \boldsymbol{x}_n\boldsymbol{\beta_1} + \left(\beta_z +  \boldsymbol{x}_n^{\boldsymbol{(EM)}} \boldsymbol{\beta_2}\right)\mathds{1}(z_n=1).
\label{eqn3}
\end{equation}
In the context of G-computation, this regression model is often called the ``Q-model''. Contrary to Equation \ref{eqn2}, it is not centered on the mean $BC$ covariates for reasons we shall explain shortly.

Having fitted the Q-model, the regression coefficients are treated as nuisance parameters. The parameters are applied to the simulated covariates $\boldsymbol{x^*}$ to predict hypothetical outcomes for each subject under both possible treatments. Namely, a pair of predicted outcomes, also called \textit{potential} outcomes,\cite{imbens2015causal} under $A$ and under $C$, is generated for each subject. 

Parametric G-computation typically relies on maximum-likelihood estimation to fit the regression model in Equation \ref{eqn3}. In this case, the methodology proceeds as follows. We denote the maximum-likelihood estimate of the regression parameters as  $\boldsymbol{\hat{\beta}} = (\hat{\beta}_0, \boldsymbol{\hat{\beta}_1}, \boldsymbol{\hat{\beta}_2}, \hat{\beta}_z)$. Leaving the simulated covariates $\boldsymbol{x^*}$ at their set values, we fix the treatment values, indicated by a vector $\boldsymbol{z^*} = (z^*_1, z^*_2, \dots, z^*_{N^*})$, for all $N^*$. By plugging treatment $A$ into the maximum-likelihood regression fit for each simulated individual, we predict the marginal outcome mean, on the natural scale, when all subjects are under treatment $A$:

\begin{align}
\hat{\mu}_1(\boldsymbol{x^*}) 
&= \int_{\boldsymbol{x^*}} g^{-1}(\hat{\beta}_0 +  \boldsymbol{x^*}\boldsymbol{\hat{\beta}_1} + \hat{\beta}_z +  \boldsymbol{x^{*(EM)}}\boldsymbol{\hat{\beta}_2}) p(\boldsymbol{x^*}) d\boldsymbol{x^*} \label{eqn4}\\
&\approx
\frac{1}{N^*}\sum_{i=1}^{N^*} g^{-1}  (\hat{\beta}_0 +  \boldsymbol{x}_i^*\boldsymbol{\hat{\beta}_1} + \hat{\beta}_z +  \boldsymbol{x}_i^{*\boldsymbol{(EM)}} \boldsymbol{\hat{\beta}_2}).
\label{eqn5}
\end{align}
Equation \ref{eqn4} follows from the law of total expectation, such that the (marginal) expected outcome is equal to the expected value of the conditional expected outcome, given the predictors. The joint probability density function for the $BC$ covariates is denoted $p(\boldsymbol{x}^*)$. This could be replaced by a probability mass function if the covariates are discrete or by a mixture density if there is a combination of discrete and continuous covariates. Replacing the integral by the summation in Equation \ref{eqn5} follows from using the empirical joint distribution of the simulated covariates as a non-parametric estimator of the density $p(\boldsymbol{x}^*)$.\cite{daniel2020making} 

Similar to above, by plugging treatment $C$ into the regression fit for every simulated observation, we predict the marginal outcome mean in the hypothetical scenario in which all units are under treatment $C$:

\begin{align}
\hat{\mu}_0(\boldsymbol{x^*}) 
&= \int_{\boldsymbol{x^*}} g^{-1}(\hat{\beta}_0 +  \boldsymbol{x^*}\boldsymbol{\hat{\beta}_1}) p(\boldsymbol{x^*}) d\boldsymbol{x^*} \label{eqn6} \\
&\approx \frac{1}{N^*}\sum_{i=1}^{N^*} g^{-1}(\hat{\beta}_0 +  \boldsymbol{x}_i^*\boldsymbol{\hat{\beta}_1}). 
\label{eqn7} 
\end{align}
To estimate the marginal or population-average treatment effect for $A$ vs.~$C$ in the linear predictor scale, one back-transforms to this scale the average predictions, taken over all subjects on the natural outcome scale, and calculates the difference between the average linear predictions:

\begin{equation}
\hat{\Delta}_{10}^{(2)} = g(\hat{\mu}_1) - g(\hat{\mu}_0),
\label{eqn8}
\end{equation}
where we have removed the dependence on $\boldsymbol{x^*}$ for simplicity in the notation. If the outcome model in Equation \ref{eqn3} is correctly specified, the estimators of the marginal outcome means on the natural scale $\hat{\mu}_1 \rightarrow \mu_1$ and $\hat{\mu}_0 \rightarrow \mu_0$, i.e., are consistent with respect to convergence to their true value, and so is the marginal treatment effect estimate $\hat{\Delta}_{10}^{(2)} \rightarrow \Delta_{10}^{(2)}$. 

% This is provided that the $BC$ population is correctly specified and other assumptions for valid population adjustment in $BC$ are met (see \hyperref[SA]{Supplementary Appendix A})

For illustrative purposes, consider a logistic regression for binary outcomes. In this case, $\hat{\mu}_1$ is the average of the individual probabilities predicted by the regression when all participants are assigned to treatment $A$. Similarly, $\hat{\mu}_0$ is the average probability when everyone is assigned to treatment $C$. The inverse link function $g^{-1}(\cdot)$ would be the inverse logit function $\expit(\cdot)=\exp(\cdot)/[1+\exp(\cdot)]$, and the average predictions in the probability scale could be substituted into Equation \ref{eqn8} and transformed to the log-odds ratio scale, using the logit link function. More interpretable summary measures of the marginal contrast, e.g.~odds ratios, relative risks or risk differences, can also be produced by manipulating the average natural outcome means differently than in Equation \ref{eqn8}, mapping these to other scales. For instance, a marginal odds ratio can be estimated as $\exp[g(\hat{\mu}_1)]/\exp[g(\hat{\mu}_0)] = \frac{\hat{\mu}_1/(1-\hat{\mu}_1)}{\hat{\mu}_0/(1-\hat{\mu}_0)}$, where $g(\cdot)$ denotes the logit link function. The standard scale commonly used for performing indirect treatment comparisons is the log-odds ratio scale\cite{dias2013evidence, bucher1997results, phillippo2018methods} and this linear predictor scale is used to define effect modification, which is scale-specific.\cite{phillippo2016nice} Hence, we assume that the marginal log-odds ratio is the relative effect measure of interest. 

Note that the estimated absolute outcomes $\hat{\mu}_1$ and $\hat{\mu}_0$, e.g.~the average outcome probabilities under each treatment in the case of logistic regression, are sometimes desirable in health economic models without any further processing.\cite{phillippo2020target} In addition, these could be useful in unanchored comparisons, where there is no common comparator group included in the analysis, e.g.~if the competitor trial is an RCT without a common control or a single-arm trial evaluating the effect of treatment $B$ alone. In the unanchored case, absolute outcome means are compared across studies as opposed to relative effects. However, unanchored comparisons make very strong assumptions which are largely considered impossible to meet (absolute effects are conditionally constant as opposed to relative effects being conditionally constant).\cite{phillippo2018methods, phillippo2016nice}

\subsubsection{Cox proportional hazards regression}\label{subsec343}

The most popular outcome types in applications of population-adjusted indirect comparisons are survival or time-to-event outcomes (e.g.~overall or progression-free survival), and the most prevalent measure of effect is the (log) hazard ratio.\cite{phillippo2019population} Therefore, developing G-computation approaches where the nuisance model is a Cox proportional hazards regression is important and useful to practitioners. In this setting, $\hat{\Delta}_{10}^{(2)}$ and $\hat{\Delta}_{20}^{(2)}$ should target marginal log hazard ratios for indirect treatment comparisons in the linear predictor scale. Something to bear in mind is that, even if Cox models are very frequently used in evidence synthesis for time-to-event data, health economic modelers typically use parametric survival models for extrapolation purposes. In subsection \ref{subsec36}, we develop a novel general-purpose methodology that can be used in scenarios where the outcome regression of interest is a parametric survival model. 

The G-computation formulae for the Cox regression are provided by Stitelman et al.\cite{stitelman2011targeted} Consider that a Cox proportional hazards model has first been fitted, conditional on covariates which follow the functional form in the linear predictor of Equation \ref{eqn3}. For the generalized linear model, we were interested in the average outcome predictions in the natural scale. With Cox regression, the average survival probabilities are of interest. 

We proceed similarly as in Equations \ref{eqn4}-\ref{eqn8}. Leaving the simulated covariates $\boldsymbol{x^*}$ at their set values, we fix the value of treatment at $z^*_i$ for all $i=1,\dots, N^*$. By plugging treatment $A$ into the Cox regression fit for each simulated unit, we compute the expected marginal survival probability when all subjects are under treatment $A$:
\begin{align}
% \hat{P}(T_1 > t) &= \int \hat{S}^{(1)}(t \mid \boldsymbol{x^*})f(\boldsymbol{x^*})d(\boldsymbol{x^*}) 
% \label{eqn8}
% \\
\hat{P}(T_1 > t)  &= \frac{1}{N^*} \sum_{i=1}^{N^*} \hat{S}_i^{(1)}(t \mid \boldsymbol{x}_i^*) 
\label{eqn9} \\
&=
\frac{1}{N^*} \sum_{i=1}^{N^*}
\exp[-\hat{H}_0(t))]^{\exp(\hat{\beta}_0 +  \boldsymbol{x}_i^*\boldsymbol{\hat{\beta}_1} + \hat{\beta}_z +  \boldsymbol{x}_i^{*\boldsymbol{(EM)}} \boldsymbol{\hat{\beta}_2})}.
\label{eqn10}
\end{align}
Above, $t$ denotes a particular time point and $T_1$ denotes a potential event time under treatment $A$, such that $\hat{P}(T_1 > t)$ is the mean treatment-specific probability of surviving beyond $t$. In Equation \ref{eqn9}, $\hat{S}_i^{(1)}(t \mid \boldsymbol{x}_i^*)$  denotes an estimate of the survival probability under treatment $A$ at time $t$ for simulated subject $i$ with covariates $\boldsymbol{x}_i^*$. Equation \ref{eqn10} follows from expressing the survival function in terms of $\hat{H}_0(t)$, an estimate of the baseline cumulative hazard function at time $t$, exponentiated and raised to the power of the exponentiated linear predictor term. Estimates of the baseline cumulative hazard are easily obtained from Cox regressions fitted with the standard survival analysis software packages. 

Similarly, the expected marginal survival probability when all simulated subjects are under treatment $C$ is given by:  
\begin{align}
\hat{P}(T_0 > t)  &= \frac{1}{N^*} \sum_{i=1}^{N^*} \hat{S}_i^{(0)}(t \mid \boldsymbol{x}_i^*) 
\label{eqn11} \\
&=
\frac{1}{N^*} \sum_{i=1}^{N^*}
\exp[-\hat{H}_0(t)]^{\exp(\hat{\beta}_0 +  \boldsymbol{x}_i^*\boldsymbol{\hat{\beta}_1})},
\label{eqn12}
\end{align}
where $T_0$ denotes a potential event time under treatment $C$, and $\hat{S}_i^{(0)}(t \mid \boldsymbol{x}_i^*)$  denotes the estimated survival probability under treatment $C$ at time $t$ for subject $i$ with simulated covariates $\boldsymbol{x}_i^*$. The marginal hazard at time $t$ for treatment $z^* \in \{0,1\}$ can be expressed as the negative logarithm of the survival probability, $-\ln[\hat{P}(T_{z^*} > t)]$. Therefore, the estimate of the marginal log hazard ratio for $A$ vs.~$C$ in the $BC$ population at time $t$ is:
\begin{equation}
\hat{\Delta}_{10,t}^{(2)} =  \ln \{ -\ln [\hat{P}(T_1 > t)] \} - \ln \{ -\ln [ \hat{P} (T_0 > t)]\},
\label{eqn13}
\end{equation}
where $\hat{P}(T_1 > t)$ and $\hat{P} (T_0 > t)$ are obtained using Equations \ref{eqn9}-\ref{eqn10} and Equations \ref{eqn11}-\ref{eqn12}, respectively. 

The Cox regression assumes that the true marginal log hazard ratio is independent of time due to the proportional hazards assumption. However, as pointed out by Varadhan et al.,\cite{varadhan2016cross} the estimate $\hat{\Delta}_{10,t}^{(2)}$ in Equation \ref{eqn13} may vary across different values of $t$. We have to set $t$ to a specific time point, or alternatively, to estimate the marginal hazard ratio over a set of time points and display the estimates graphically. When selecting a value of $t$, bear in mind that, in Equation \ref{eqn13}, the marginal log hazard ratio estimate is undefined at $t$ for which $\hat{P}(T_{z^*} > t)=1$ for treatment $z^* \in \{0,1\}$.\cite{stitelman2011targeted} A simulation procedure for marginalizing estimates of conditional hazard ratios has recently been proposed by Daniel et al.\cite{daniel2020making} This approach should avoid these issues by averaging the marginal log hazard ratio over a set time frame, but adapting the methodology to the current setting is beyond the scope of this article. 

One can manipulate the expected marginal survival probabilities differently than in Equation \ref{eqn13} to produce estimates of the marginal risk difference (the additive difference in survival probabilities) or the marginal log relative risk at a particular time point.\cite{stitelman2011targeted} These effect measures are more easily interpreted. However, indirect treatment comparisons with survival outcomes are typically performed in the log hazard ratio scale,\cite{dias2013evidence} and this linear predictor scale is used to define effect modification, which is scale-specific.\cite{phillippo2016nice} Therefore, the marginal log hazard ratio is the relative effect measure of interest. 

\subsubsection{Model fitting and selection}\label{subsec341}

Because the regression in Equation \ref{eqn3} will be our working model from now onward, we briefly discuss some good practices for model fitting and model selection. Time and care should be taken to perform these exercises and fit an appropriate regression.

The inclusion of all imbalanced effect modifiers in Equation \ref{eqn3} is required for unbiased estimation of both the marginal and conditional $A$ vs.~$C$ treatment effects in the $BC$ population.\cite{zhang2016new} A strong fit of the regression model, evaluated by model checking criteria such as the residual deviance and information criteria, may increase precision. Hence, we could select the model with the lowest information criterion conditional on including all effect modifiers.\cite{zhang2016new} Model checking criteria should not guide causal decisions on effect modifier status, which should be defined prior to fitting the outcome model. As effect-modifying covariates are likely to be good predictors of outcome, the inclusion of appropriate effect modifiers should provide an acceptable fit. In addition, note that any model comparison criteria will only provide information about the observed $AC$ data and therefore tell just part of the story.\cite{gabrio2019full} We have no information on the fit of the selected model to the $BC$ patient-level data. 

At this point, the readers may be wondering why the outcome regressions fitted in subsections \ref{subsec33} and \ref{subsec34} are different. In the conventional outcome regression described in \ref{subsec33} and by the NICE technical support document,\cite{phillippo2016nice} the IPD covariates are centered by plugging in the mean $BC$ covariate values. In the Q-model required for G-computation, outlined in \ref{subsec34}, the covariates are not centered and the regression fit is used to make predictions for the simulated covariates. The underlying reason for this has been described for generalized linear models with non-linear link functions, such as logistic or Poisson regression.\cite{bartlett2018covariate, qu2015estimation, lane1982analysis} On the natural scale, averaging the individual outcome predictions made at the centered covariates of the sample does not consistently estimate the marginal mean response for the centered sample. In the words of Bartlett,\cite{bartlett2018covariate} ``prediction at the mean'' value of the baseline covariates for a treatment group does not result in the ``marginal mean'' under such treatment. Similarly, in the words of Qu and Luo,\cite{qu2015estimation} the ``mean at mean covariates'' of the study sample is generally not equivalent to the marginal response over the subjects in the sample. The former results in a conditional estimate whereas the latter produces a marginal population-level estimate, of interest in our scenario. 

We have postulated a single outcome model for all subjects in the $AC$ IPD, which includes the necessary treatment-covariate interaction terms to capture effect modification over the covariates. Nevertheless, another possible strategy is to fit two outcome models separately for each treatment group in the randomized trial, i.e., to fit one regression to the patients under treatment $A$ and then another regression among the patients under $C$, then predicting the conditional outcome expectations and averaging these out on the entire simulated pseudo-population. This is perhaps a more ``objective'' approach to covariate adjustment, as the model-fitting is performed independently of reference to a conditional treatment effect (in this case, the fitted regressions do not have a treatment coefficient), but obviates the estimation of treatment-by-covariate interactions.\cite{lunceford2004stratification, dahabreh2020extending} Throughout the article, we consider the nuisance model in Equation \ref{eqn3} to be a parametric regression. Alternatively, non-parametric estimators of the conditional expectation may be less susceptible to model misspecification. We discuss the potential application of these methods in subsection \ref{subsec62}.  

\subsubsection{Variance estimation}\label{subsec342}

From a frequentist perspective, it is not easy to derive analytically a closed-form expression for the standard error of the marginal $A$ vs.~$C$ treatment effect. Deriving the asymptotic distribution is not straightforward because, typically, the estimate is a non-linear function of each of the components of $\boldsymbol{\hat{\beta}}$. When using maximum-likelihood estimation to fit the outcome model, standard errors and interval estimates can be obtained using resampling-based methods such as the traditional non-parametric bootstrap\cite{efron1986bootstrap} or the m-out-of-n bootstrap.\cite{varadhan2016cross} In our bootstrap implementation, we only resample the IPD of the $AC$ trial due to patient-level data limitations for the $BC$ study. The standard error would be estimated as the sample standard deviation of the resampled marginal treatment effect estimates. Assuming that the sample size $N$ is reasonably large, we can appeal to the asymptotic normality of the marginal treatment effect and construct Wald-type normal distribution-based confidence intervals. Alternatively, one can construct interval estimates using the relevant quantiles of the bootstrapped treatment effect estimates, without necessarily assuming normality. This avoids relying on the adequacy of the asymptotic normal approximation, an approximation which will be inappropriate where the true model likelihood is distinctly non-normal,\cite{rubin1987logit} and may allow for the more principled propagation of uncertainty.

An alternative to bootstrapping for statistical inference is to simulate the parameters of the multivariable regression in Equation \ref{eqn3} from the asymptotic multivariate normal distribution with means set to the maximum-likelihood estimator and with the corresponding variance-covariance matrix, iterate over Equations \ref{eqn4}-\ref{eqn8} and compute the sample variance. This parametric simulation approach is less computationally intensive than bootstrap resampling. It has the same reliance on random numbers and may offer similar performance.\cite{aalen1997markov} It is equivalent to approximating the posterior distribution of the regression parameters, assuming constant non-informative priors and a large enough sample size. Again, this large-sample formulation relies on the adequacy of the asymptotic normal approximation. 

\subsection{Bayesian parametric G-computation}\label{subsec35}

A Bayesian approach to parametric G-computation may be beneficial for several reasons. Firstly, the maximum-likelihood estimates of the outcome regression coefficients may be unstable where the sample size $N$ of the $AC$ IPD is small, the data are sparse or the covariates are highly correlated, e.g.~due to finite-sample bias or variance inflation. This leads to poor frequentist properties in terms of precision. A Bayesian approach with default shrinkage priors, i.e., priors specifying a low likelihood of a very large effect, can reduce variance, stabilize the estimates and improve their accuracy in these cases.\cite{keil2018bayesian} 

Secondly, we can use external data and/or contextual information on the prognostic effect and effect-modifying strength of covariates, e.g.~from covariate model parameters reported in the literature, to construct informative prior distributions for $\boldsymbol{\beta_1}$ and $\boldsymbol{\beta_2}$, respectively, and skeptical priors (i.e., priors with mean zero, where the variance is chosen so that the probability of a large effect is relatively low) for the conditional treatment effect $\beta_z$, if necessary. Where meaningful prior knowledge cannot be leveraged, one can specify generic default priors instead. For instance, it is unlikely in practice that conditional odds ratios are outside the range $0.1-10$. Therefore, we could use a null-centered normal prior with standard deviation 1.15, which is equivalent to just over 95\% of the prior mass being between 0.1 and 10. As mentioned earlier, this ``weakly informative'' contextual knowledge may result in shrinkage that improves accuracy with respect to maximum-likelihood estimators.\cite{keil2018bayesian}  Finally, it is simpler to account naturally for issues in the $AC$ IPD such as missing data and measurement error within a Bayesian formulation.\cite{keil2014autism, josefsson2021bayesian}

In the generalized linear modeling context, consider that we use Bayesian methods to fit the outcome regression model in Equation \ref{eqn3}. The difference between Bayesian G-computation and its maximum-likelihood counterpart is in the estimated distribution of the predicted outcomes. The Bayesian approach also marginalizes, integrates or standardizes over the joint posterior distribution of the conditional nuisance parameters of the outcome regression, as well as the joint covariate distribution $p(\boldsymbol{x^*})$. Following Keil et al.,\cite{keil2018bayesian} Rubin\cite{rubin1978bayesian} and Saarela et al.,\cite{saarela2015predictive} we draw a vector of size $N^*$ of predicted outcomes $\boldsymbol{y}^*_{z^*}$ under each set intervention $z^* \in \{0,1\}$ from its posterior predictive distribution under the specific treatment. This is defined as $p(\boldsymbol{y}^*_{z^*} \mid \mathcal{D}_{AC})=\int_{\boldsymbol{\beta}} p(\boldsymbol{y}^*_{z^*} \mid \boldsymbol{\beta}) p(\boldsymbol{\beta} \mid \mathcal{D}_{AC}) d\boldsymbol{\beta}$, where $p(\boldsymbol{\beta} \mid \mathcal{D}_{AC})$ is the posterior distribution of the outcome regression coefficients $\boldsymbol{\beta}$, which encode the predictor-outcome relationships observed in the $AC$ trial IPD. This\cite{keil2018bayesian} is given by:
\begin{align}
p(\boldsymbol{y}^*_{z^*} \mid \mathcal{D}_{AC}) &= \int_{\boldsymbol{x^*}} p(\boldsymbol{y}^* \mid \boldsymbol{z^*}, \boldsymbol{x^*}, \mathcal{D}_{AC}) p(\boldsymbol{x^*} \mid \mathcal{D}_{AC}) d\boldsymbol{x^*} 
\label{eqn14}
\\
&= \int_{\boldsymbol{x^*}} \int_{\boldsymbol{\beta}} p(\boldsymbol{y^*} \mid \boldsymbol{z^*}, \boldsymbol{x^*},\boldsymbol{\beta}) p(\boldsymbol{x^*} \mid \boldsymbol{\beta}) p (\boldsymbol{\beta} \mid \mathcal{D}_{AC}) d \boldsymbol{\beta}  d \boldsymbol{x^*}.
\label{eqn15}
\end{align}
As noted by Keil et al.,\cite{keil2018bayesian} the posterior predictive distribution $p(\boldsymbol{y}^*_{z^*} \mid \mathcal{D}_{AC})$ is a function only of the observed data $\mathcal{D}_{AC}$, the joint probability density function $p(\boldsymbol{x^*})$ of the simulated $BC$ pseudo-population, which is independent of $\boldsymbol{\beta}$, the set treatment values $\boldsymbol{z^*}$, and the prior distribution $p(\boldsymbol{\beta})$ of the regression coefficients.

In practice, the integrals in Equations \ref{eqn14} and \ref{eqn15} can be approximated numerically, using full Bayesian estimation via Markov chain Monte Carlo (MCMC) sampling. This is carried out as follows. As per the maximum-likelihood procedure, we leave the simulated covariates at their set values and fix the value of treatment to create two datasets: one where all simulated subjects are under treatment $A$ and another where all simulated subjects are under treatment $C$. The outcome regression model in Equation \ref{eqn3} is fitted to the original $AC$ IPD with the treatment actually received. From this model, conditional parameter estimates are drawn from their posterior distribution $p(\boldsymbol{\beta} \mid \mathcal{D}_{AC})$, given the observed patient-level data and some suitably defined prior $p(\boldsymbol{\beta})$. 

It is relatively straightforward to integrate the model-fitting and outcome prediction within a single Bayesian computation module using efficient simulation-based sampling methods such as MCMC. Assuming convergence of the MCMC algorithm, we form realizations of the parameters $ \{ \hat{\boldsymbol{\beta}}^{(l)} = (\hat{\beta}_0^{(l)}, \boldsymbol{\hat{\beta}}^{(l)}_{\boldsymbol{1}},
\boldsymbol{\hat{\beta}}^{(l)}_{\boldsymbol{2}}, \hat{\beta}^{(l)}_z,): l=1,2,\dots, L \}$, where $L$ is the number of MCMC draws after convergence and $l$ indexes each specific draw. Again, these conditional coefficients are nuisance parameters, not of direct interest in our scenario. Nevertheless, the samples are used to extract draws of the conditional expectations for each simulated subject $i$ (the posterior draws of the linear predictor transformed by the inverse link function) from their posterior distribution. The $l$-th draw of the conditional expectation for simulated subject $i$ set to treatment $A$ is:
\begin{equation}
\hat{\mu}_{1,i}^{(l)}=g^{-1}(\hat{\beta}_0^{(l)} +  \boldsymbol{x}_i^*\boldsymbol{\hat{\beta}}^{(l)}_{\boldsymbol{1}} + \hat{\beta}^{(l)}_z +  \boldsymbol{x}_i^{*\boldsymbol{(EM)}} \boldsymbol{\hat{\beta}}^{(l)}_{\boldsymbol{2}}).
\label{eqn16}
\end{equation}
Similarly, the $l$-th draw of the conditional expectation for simulated subject $i$ under treatment $C$ is:
\begin{equation}
\hat{\mu}_{0,i}^{(l)}=g^{-1}(\hat{\beta}_0^{(l)} +  \boldsymbol{x}_i^*\boldsymbol{\hat{\beta}}^{(l)}_{\boldsymbol{1}}). 
\label{eqn17}
\end{equation}

The conditional expectations drawn from Equations \ref{eqn16} and \ref{eqn17} are used to impute the individual-level outcomes $\{ y_{1,i}^{*(l)} : i=1,\dots,N^*; l=1,2,\dots, L\}$ under treatment $A$ and $\{ y_{0,i}^{*(l)} : i=1,\dots,N^*; l=1,2,\dots, L\}$ under treatment $C$, as independent draws from their posterior predictive distribution at each iteration of the MCMC chain. For instance, if the outcome model is a normal linear regression with a Gaussian likelihood, one multiplies the simulated covariates and the set treatment $z^*_i$ for each subject $i$ by the $l$-th random draw of the posterior distribution of the regression coefficients, given the observed IPD and some suitably defined prior, to form draws of the conditional expectation $\hat{\mu}^{(l)}_{z^*,i}$ (which is equivalent to the linear predictor because the link function is the identity link in linear regression). Then each predicted outcome $y_{z^*,i}^{*(l)}$ would be drawn from a normal distribution with mean equal to $\hat{\mu}^{(l)}_{z^*,i}$ and standard deviation equal to the corresponding posterior draw of the error standard deviation. With a logistic regression as the outcome model, one would impute values of a binary response $y_{z^*,i}^{*(l)}$ by random sampling from a Bernoulli distribution with mean equal to the expected conditional probability $\hat{\mu}^{(l)}_{z^*,i}$.

Producing draws from the posterior predictive distribution of outcomes is fairly simple using dedicated Bayesian software such as \texttt{BUGS},\cite{lunn2012bugs} \texttt{JAGS}\cite{plummer2003jags} or \texttt{Stan},\cite{carpenter2017stan} where the outcome regression and prediction can be implemented simultaneously in the same module. Over the $L$ MCMC draws, these programs typically return a $L \times N^*$ matrix of simulations from the posterior predictive distribution of outcomes. The $l$-th row of this matrix is a vector of outcome predictions of size $N^*$ using the corresponding draw of the regression coefficients from their posterior distribution. We can estimate the marginal treatment effect for $A$ vs.~$C$ in the $BC$ population by: (1) averaging out the imputed outcome predictions in each draw over the simulated subjects, i.e., over the columns, to produce the marginal outcome means on the natural scale; and (2) taking the difference in the sample means under each treatment in a suitably transformed scale. Namely, for the $l$-th draw, the $A$ vs.~$C$ marginal treatment effect is:
\begin{equation}
\hat{\Delta}_{10}^{(2,l)} = g \Bigg (\frac{1}{N^*} \sum_{i=1}^{N^*} y^{*(l)}_{1,i} \Bigg ) - g \Bigg (\frac{1}{N^*} \sum_{i=1}^{N^*} y^{*(l)}_{0,i} \Bigg ).
\label{eqn18}
\end{equation}

The average, variance and interval estimates of the marginal treatment effect can be derived empirically from draws of the posterior density, i.e., by taking the sample mean, variance and the relevant percentiles over the $L$ draws, which approximate the posterior distribution of the marginal treatment effect. The computational expense of the Bayesian approach to G-computation is expected to be similar to that of the maximum-likelihood version, given that the latter typically requires bootstrapping for uncertainty quantification. Computational cost can be reduced by adopting approximate Bayesian inference methods such as integrated nested Laplace approximation (INLA)\cite{rue2009approximate} instead of MCMC sampling to draw from the posterior predictive distribution of outcomes. 

Note that Equation \ref{eqn18} is the Bayesian version of Equation \ref{eqn8}. Other parameters of interest can be obtained, e.g.~the risk difference by using the identity link function in this equation, but these are typically not of direct relevance in our scenario. Again, where the contrast between two different interventions is not of primary interest, the absolute outcome draws from their posterior predictive distribution under each treatment may be relevant. The average, variance and interval estimates of the absolute outcomes can be derived empirically over the $L$ draws. An argument in favor of a Bayesian approach is that, once the simulations have been conducted, one can obtain a full characterization of uncertainty on any scale of interest.  

In the Cox regression scenario, parametric Bayesian G-computation would follow a similar approach, and would involve drawing the marginal survival probabilities under each treatment from their posterior predictive distribution. Implementing Bayesian parametric G-computation in the Cox regression scenario is a research priority. 

\subsection{Multiple imputation marginalization}\label{subsec36}

We now develop a general-purpose marginalization procedure labeled multiple imputation marginalization (MIM) because it contains many similarities to multiple imputation. This procedure might be useful where the effect measure of interest cannot be readily summarized in terms of predicted outcomes and G-computation cannot be easily applied. An example scenario where this is the case is when the outcome model is a parametric survival regression. Parametric survival distributions, e.g.~exponential, Weibull, Gompertz, log-logistic, log-normal and generalized gamma, are commonly used in health economic evaluations to extrapolate published Kaplan-Meier survival curves from the clinical trial follow-up period to a lifetime horizon.\cite{latimer2013survival, bagust2014survival, grieve2013extrapolation, vickers2019evaluation, baio2020survhe} As well as permitting survival extrapolation, these may allow for non-proportional and time-varying hazards. The area under the extrapolation is used to estimate the mean survival benefit of an intervention in cost-effectiveness analyses, typically in terms of life years or quality-adjusted life years. Parametric survival models are particularly of interest in oncology health technology appraisals.

Consider that the outcome model of interest is a parametric survival model. In this scenario, an anchored regression-adjusted indirect comparison would be conducted as follows: (1) a univariable parametric survival regression of outcome on treatment group is fitted to the $BC$ trial data (the subject-level data is typically reconstructed from digitized Kaplan-Meier curves, e.g.~using the algorithm by Guyot et al.\cite{guyot2012enhanced}); (2) a multivariable covariate-adjusted parametric survival model (of the same family as the model in Step 1) is fitted to the $AC$ trial data with treatment group as a covariate; (3) the coefficients of the covariate-adjusted regression are marginalized to derive a marginal treatment effect for $A$ vs. $C$ in the $BC$ population (with the location or rate coefficient and, potentially, ancillary coefficients such as shapes being treated as nuisance parameters); and (4) this relative treatment effect is applied to the survival curve of common comparator $C$ in the $BC$ study to yield a survival curve for treatment $A$ in the $BC$ population. To our knowledge, the marginalization procedure in Step 3 cannot be easily conducted in this scenario (and in many others) using parametric G-computation. This motivates the development of a general-purpose framework such as MIM. 

Conceptually, MIM splits the population adjustment into two separate stages: (1) the generation (\textit{synthesis}) of synthetic datasets; and (2) the \textit{analysis} of the generated datasets. The synthesis is completely separated from the analysis --- only after the synthesis has been completed is the marginal effect of treatment on the outcome estimated. This is analogous to the separation between design and analysis in propensity score methods, between imputation and analysis in multiple imputation, or between fitting (and predicting outcomes with) the Q-model and estimating the marginal treatment effect in G-computation. 

\begin{figure}[!htb]
\center{\includegraphics[width=0.84\textwidth]{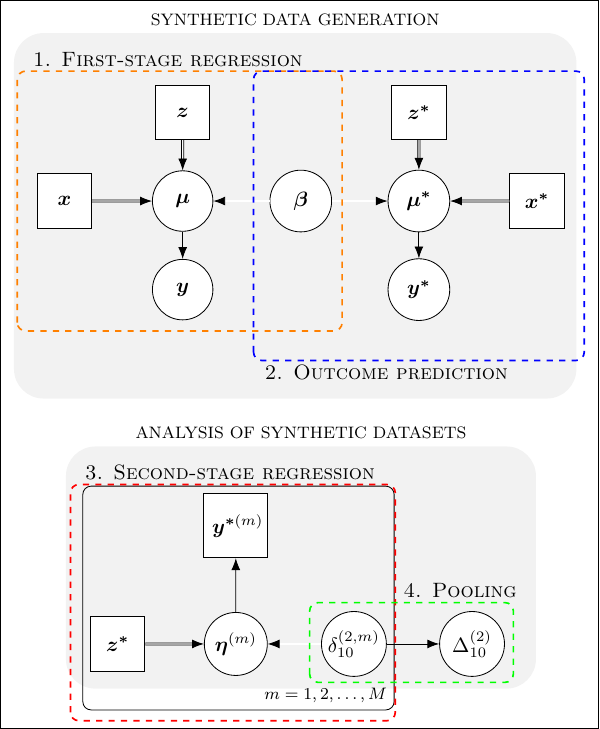}}
\caption{A Bayesian directed acyclic graph representing multiple imputation marginalization (MIM) and accounting for its two main stages: (1) synthetic data generation; and (2) the analysis of synthetic datasets. Square nodes represent constant variables, circular nodes indicate stochastic variables, single arrows denote stochastic dependence, double arrows indicate logical relationships and the plate notation indicates repeated analyses. The difference between MIM and Bayesian G-computation is that MIM requires specifying a marginal structural model for each synthesis, the second-stage regression, in the analysis stage. The results of these regressions are then pooled across all syntheses.\label{fig2}}
\end{figure}

Similarly to Bayesian G-computation, MIM sits naturally within a Bayesian framework in integrating different sources of evidence to fully characterize probabilistic relationships among a set of relevant variables, using a simulation approach. A more detailed explanation of each module is provided below. Figure \ref{fig2} displays a Bayesian directed acyclic graph (DAG) summarizing the general MIM structure and the links between the modules. In this graphical representation, the nodes represent the variables of the model (constants are denoted as squares and stochastic nodes are circular); single arrows indicate probabilistic relationships and double arrows indicate logical functions. The plate notation indicates repeated analyses. We return to Figure \ref{fig2} and provide further explanations for the notation throughout this section. For consistency with the rest of the article, MIM is presented within a generalized linear modeling formulation. Nevertheless, its formal integration in a unified survival analysis framework for HTA, which contains many particularities, is a necessary and important piece of currently ongoing research. 

\subsubsection{Generation of synthetic datasets: a missing data problem}\label{subsec361}

The first stage, synthetic data generation, consists of two steps. Initially, the \textit{first-stage regression} builds a model to capture the relationship between the outcome $\boldsymbol{y}$ and the covariates $\boldsymbol{x}$ and treatment $\boldsymbol{z}$ in the observed IPD. In the \textit{outcome prediction} step, we generate predicted outcomes $\boldsymbol{y^*}$ for $A$ and $C$ in the $BC$ population by drawing from the posterior predictive distribution of outcomes, given the observed predictor-outcome relationships in the $AC$ trial IPD, the simulated covariates $\boldsymbol{x^*}$ and the set treatment.

These steps are identical to those described for the Bayesian G-computation procedure in subsection \ref{subsec35}. Interestingly, Bayesian G-computation follows closely the basic principles of multiple imputation.\cite{rubin2004multiple} This is a simulation technique where missing data points are replaced with a set of plausible values conditional on some pre-specified imputation mechanism. Multiple imputation can be regarded as a fundamentally Bayesian operation,\cite{rubin2004multiple, meng1994multiple, gabrio2019full} as the imputed outcomes are drawn from the posterior predictive distribution of observed outcomes. In addition, our problem can be conceptualized as a missing data problem, where the individual-level outcomes for treatments $A$ and $C$ in the $BC$ population are treated as systematically missing data under a complete case analysis.\cite{leon2003semiparametric} Namely, we only observe the outcomes for the subjects in the $AC$ trial, with the outcomes experienced in the $BC$ population ``missing''. The imputation mechanism would be the statistical model in Equation \ref{eqn3} relating the outcomes $\boldsymbol{y}$ to the predictors $(\boldsymbol{x}, \boldsymbol{z})$. This dependence structure is estimated using the original IPD and used to construct the posterior predictive distribution of outcomes. 

\clearpage

Practically, we may frame Bayesian G-computation as conducting $L$ hypothetical trials comparing $A$ vs.~$C$ in the $BC$ population. Extending the parallel with the missing data literature, the outcome-generation process in these trials is based on the assumption of a missing-at-random mechanism. Namely, the missing relative outcomes for $A$ vs.~$C$ in the $BC$ population are conditionally exchangeable with those observed in the $AC$ population (conditioning on the predictors that have been adjusted for). Therefore, this missing-at-random assumption is analogous to the conditional constancy of relative effects mentioned in subsection \ref{subsec22}, which is untestable using the available data alone. 

There is one conceptual difference between the synthesis stage of MIM and parametric G-computation, which arises in order to facilitate the presentation of MIM. Instead of considering two datasets with $N^*$ subjects each (one under treatment $A$ and the other under treatment $C$), each synthesis considers a single dataset with $N^*$ individuals that maintains the original treatment allocation ratio of the $AC$ trial. Treatment in all synthetic datasets will be fixed to $\boldsymbol{z^*}=(z^*_1,\dots,z^*_{N^*})$, a vector of size $N^*$. In practice, this different conceptualization will not make a difference provided that $N^*$ is reasonably large. This is because, in the synthetic samples, we ``enforce'' the randomization of individuals into $A$ and $C$ by simulating the covariates for active treatment and control arms combined. 

As per Bayesian G-computation, the synthesis stage can be performed using MCMC. Iterating over the $L$ converged draws of the MCMC algorithm, one generates $M \leq L$ synthetic datasets, $\mathcal{D}_{AC}^{*} = \{ \mathcal{D}_{AC}^{*(m)}: m=1,2\dots, M \}$, where $\mathcal{D}_{AC}^{*(m)}=(\boldsymbol{x}^{\boldsymbol{*}}, \boldsymbol{z^*}, \boldsymbol{y}^{\boldsymbol{*}(m)})$. Here, $\boldsymbol{x}^{\boldsymbol{*}}$ is a $N^* \times K$ matrix of individual-level $BC$ covariates, drawn from their approximate joint distribution as per subsection \ref{subsec32}, and $\boldsymbol{z^*}$ is the assigned treatment in the syntheses, as previously described. Each $\boldsymbol{y}^{\boldsymbol{*}(m)}$ is a vector of predicted outcomes of size $N^*$. We fill in $\boldsymbol{y}^{\boldsymbol{*}(m)}$ by drawing from its posterior predictive distribution. In line with the multiple imputation framework, these draws are repeated independently $M$ times to create $M$ completed syntheses, with the posterior samples making up the imputed datasets. In standard multiple imputation, it is not uncommon to release as little as 5 imputed datasets.\cite{rubin2004multiple, schafer1997analysis} However, MIM is likely to require a larger value of $M$ as it imputes an entire dataset as opposed to a relatively small proportion of missing values, i.e., the ``fraction of missing information'' is 1.

Within a survival analysis framework, the fitted first-stage regression would be used to predict survival times in the simulated pseudo-population for the $BC$ study. One would assume that censoring is non-informative, which is an assumption made, in any case, by the Cox proportional hazards regression and the standard parametric survival models. Namely, one would not attempt to simulate censoring according to any given distribution, or to mimic any particular censoring pattern (essentially, assuming that all times are uncensored in the simulated data structure). 

\subsubsection{Analysis of synthetic datasets}\label{subsec362}

In the second stage, the analysis of synthetic datasets, we seek inferences about the marginal $A$ vs.~$C$ treatment effect in the $BC$ population, $\Delta_{10}^{(2)}$, given the synthesized outcomes. This will ultimately be compared with the treatment effect for $B$ vs.~$C$, to produce a marginal treatment effect for $A$ vs.~$B$ in the $BC$ population. The analysis stage consists of another two steps. In the \textit{second-stage regression} step, we regress each predicted outcome $\boldsymbol{y}^{\boldsymbol{*}(m)}$ on the treatment indicator $\boldsymbol{z^*}$ alone, to generate estimates of the marginal $A$ vs.~$C$ treatment effect in each synthesis. This second-stage regression is effectively a marginal structural model of outcome on treatment,\cite{sato2003marginal} and adds some computational expense with respect to Bayesian parametric G-computation. In the \textit{pooling} step, the treatment effect estimates and their variances are combined across all $M$ syntheses to produce an estimate of the average marginal treatment effect in the $BC$ population. 

In a typical problem involving multiple imputation, the imputation (i.e., synthesis) and analysis stages may be performed simultaneously in a joint model.\cite{gabrio2019full} However, this is problematic in MIM as the dependent variable $\boldsymbol{y}^{\boldsymbol{*}}$ of the analysis is completely synthesized. Consider the Bayesian DAG in Figure \ref{fig2}. In a joint model, the predicted outcomes are a \textit{collider} variable and block the only path between the first and the second module, i.e., information from the directed arrows ``collides'' at the node. Due to these non-trivial joint modeling issues, we have considered the data synthesis and analysis stages as separate units in a two-stage modular framework. The analysis stage conditions on the response variable predicted by the synthesis stage, treating it as observed data.

\paragraph{Second-stage regression}

We fit $M$ second-stage regressions of predicted outcomes $\boldsymbol{y}^{\boldsymbol{*}}$ on the treatment $\boldsymbol{z^*}$. Identical analyses are performed on each $\boldsymbol{y}^{*(m)}$ ($\boldsymbol{z^*}$ is fixed), such that for $m=1,2, \dots, M$:
\begin{equation}
g(\eta_i^{(m)}) = \delta_0^{(m)} + \delta_{10}^{(m)} z^*_i,
\label{eqn19}
\end{equation}
where $\eta_i^{(m)}$ is the expected outcome on the natural scale of subject $i$ in the $m$-th synthesis, the coefficient $\delta_0^{(m)}$ is an intercept term and $\delta_{10}^{(m)}$ denotes the marginal $A$ vs.~$C$ treatment effect in the $m$-th synthesis. 
There is some non-trivial computational complexity to performing a Bayesian fit in this step. This would embed a nested simulation scheme. Namely, if we draw $M$ samples 
$\{ \boldsymbol{y}^{\boldsymbol{*}(m)} : m = 1,2,\dots M \}$ in the synthesis stage, a further number of samples, say $R$, of the treatment effect $\{ \hat{\delta}_{10}^{(m,r)}: m = 1,2\dots M; r = 1, 2,\dots R \}$ would be drawn for each of these realizations separately. This structure is likely to be unfeasible in terms of running time and we choose to prioritize computational efficiency. 

Using maximum-likelihood estimation, we generate a point estimate $\hat{\delta}_{10}^{(m)}$ of the \allowbreak{marginal} treatment effect and a measure of its variance $\hat{v}^{(m)}$ in each synthesis $\boldsymbol{y}^{\boldsymbol{*}(m)}$. 
The model is relatively simple as we have enforced randomization in the trial by simulating covariates for both arms jointly. Hence, this step emulates the unadjusted analysis of an RCT. A marginal treatment effect estimate is produced because a simple regression of outcome on treatment alone is performed (covariate adjustment was performed by the first-stage regression, i.e., the Q-model), with the fitted coefficient $\hat{\delta}_{10}^{(m)}$ estimating a relative effect between subjects that, on expectation, have the same distribution of covariates.\cite{remiro2020methods} Assigned treatment was already included as a predictor in the first-stage regression. Hence, the second-stage regression is more restrictive and therefore ``congenial'' (i.e., compatible  for unbiased estimation) with the synthesis stage.\cite{meng1994multiple}

\paragraph{Pooling}

We now combine the $M$ point estimates of the $A$ vs.~$C$ treatment effect and their variances to generate a posterior distribution for the $A$ vs.~$C$ marginal treatment effect, in the $BC$ population. Due to the two-stage structure of MIM, it is necessary to pool the estimates across the analyses to estimate this effect. The analysis of a single synthesis accounts for two sources of uncertainty: (1) the uncertainty in the regression coefficients used to generate the predicted outcomes; and (2) prediction error or random individual variation. However, it will produce attenuated measures of variability for the average $A$ vs.~$C$ treatment effect. We must account for a third source of variation to produce valid statistical inference: the uncertainty due to the data being synthesized. This is incorporated by pooling across multiple syntheses, a question shared with the domain of statistical disclosure limitation.\cite{raghunathan2003multiple, rubin1993statistical, reiter2002satisfying, reiter2005releasing, si2011comparison, reiter2007multiple, raab2016practical}

In statistical disclosure limitation, data agencies mitigate the risk of identity disclosure by releasing multiple \textit{fully synthetic} datasets, i.e., datasets that only contain simulated values, in lieu of the original confidential data of real survey respondents. Raghunathan et al.\cite{raghunathan2003multiple} describe full synthesis as a two-step process: (1) construct multiple synthetic populations by repeatedly drawing from the posterior predictive distribution, conditional on a model fitted to the original data; and (2) draw random samples from each synthetic population and release these synthetic samples to the public. In practice, as indicated by Reiter and Raghunathan,\cite{reiter2007multiple} it is not a requirement to generate the populations, but only to generate values for the synthetic samples. Once the samples are released, the analyst seeks inferences based on the synthetic data alone.  

MIM is analogous to this problem, albeit there are some differences. In MIM, the analyst also acts as the synthesizer of data, and there is no ``original data'' on outcomes as such -- the $AC$ trial has not been conducted in the $BC$ population. In any case, values for the samples are generated in the synthesis stage by repeatedly drawing from the posterior predictive distribution of outcomes. This is conditional on the predictor-outcome relationships indexed by the model fitted to the $AC$ IPD, and on the simulated covariates.

The target for inference in this step is the marginal $A$ vs.~$C$ treatment effect conditional on the synthetic outcomes (and treatment), i.e., we seek to construct the posterior distribution $p(\Delta_{10}^{(2)} \mid \boldsymbol{y}^{\boldsymbol{*}}, \boldsymbol{z^*})$. Following Raab et al.,\cite{raab2016practical} we view each $\boldsymbol{y}^{\boldsymbol{*}(m)}$ as a random sample from $p(\boldsymbol{y}^{\boldsymbol{*}} \mid \boldsymbol{\hat{\beta}}^{(m)}, \boldsymbol{x}^{\boldsymbol{*}}, \boldsymbol{z^*})$, where $\boldsymbol{\hat{\beta}}^{(m)}$ is sampled from its posterior $p(\boldsymbol{\beta} \mid \mathcal{D}_{AC})$. Hence, the ``true'' marginal treatment effect $\delta_{10}^{(m)}$ for the $m$-th synthesis, corresponding to $\boldsymbol{\hat{\beta}}^{(m)}$, can be defined as a function of this sample. In each second-stage regression in Equation \ref{eqn19}, this is the treatment effect estimated by $\hat{\delta}_{10}^{(m)}$. 

Therefore, following Raghunathan et al.,\cite{raghunathan2003multiple} the estimators $\{ \hat{\delta}_{10}^{(m)}, \hat{v}^{(m)}; m=1,2,\dots,M \}$ from the second-stage regressions are treated as ``data'', and are used to construct an approximation to the posterior density $p(\Delta_{10}^{(2)} \mid \boldsymbol{y}^{\boldsymbol{*}}, \boldsymbol{z^*})$. This density is assumed to be approximately normal and is parametrized by its first two moments: the mean $\mu_{\Delta}$, and the variance $\sigma_{\Delta}^2$. To derive the conditional distribution $p(\mu_\Delta, \sigma_\Delta^2 \mid 
\boldsymbol{y}^{\boldsymbol{*}}, \boldsymbol{z^*})$ of these moments given the syntheses, the estimators $\{ \hat{\delta}_{10}^{(m)}, \hat{v}^{(m)}; m=1,2,\dots,M \}$, where $\hat{v}^{(m)}$ is the the point estimate of the variance in the $m$-th second-stage regression, are treated as sufficient summaries of the syntheses, and $\mu_{\Delta}$ and $\sigma_{\Delta}^2$ are treated as parameters. Then, the posterior distribution $p(\Delta_{10}^{(2)} \mid \boldsymbol{y}^{\boldsymbol{*}}, \boldsymbol{z^*})$ is constructed as:
\begin{equation}
p(\Delta_{10}^{(2)} \mid \boldsymbol{y}^{\boldsymbol{*}}, \boldsymbol{z^*}) = \int_{\mu_\Delta, \sigma_\Delta^2} p(\Delta_{10}^{(2)}
\mid \mu_\Delta, \sigma_\Delta^2) 
p(\mu_\Delta, \sigma_\Delta^2 \mid 
\boldsymbol{y}^{\boldsymbol{*}}, \boldsymbol{z^*})
d(\mu_\Delta, \sigma_\Delta^2).
\label{eqn20}
\end{equation}
% $\mu_{\Delta} = \sum_{m=1}^M \delta_{AC}^{*(m)} /M$, and the variance $\mathcal{B} = \sum_{m=1}^{M}(\delta_{AC}^{*(m)} - \bar{\Delta}_{AC}^*)^2/(M-1)$. 

We have two options to approximate the posterior distribution. The first involves direct Monte Carlo simulation and the second uses a simple normal approximation. In analogy with the theory of multiple imputation,\cite{rubin2004multiple} both approaches require the following quantities for inference:
\begin{align}
\bar{\delta}_{10} &= \sum_{m=1}^M
\hat{\delta}_{10}^{(m)}/M, 
\label{eqn21}
\\
\bar{v} &= \sum_{m=1}^M \hat{v}^{(m)}/M, 
\label{eqn22}
\\
b &= \sum_{m=1}^{M}(\hat{\delta}_{10}^{(m)} - \bar{\delta}_{10})^2/(M-1),
\label{eqn23}
\end{align}
where $\bar{\delta}_{10}$ is the average of the treatment effect point estimates across the $M$ syntheses, $\bar{v}$ is the average of the point estimates of the variance (the ``within'' variance), and $b$ is the sample variance of the point estimates (the ``between'' variance). These quantities are computed using the point estimates from the second-stage regressions. 

\paragraph{Pooling via posterior simulation}

After deriving the quantities in Equations \ref{eqn21}, \ref{eqn22} and \ref{eqn23}, the posterior in Equation \ref{eqn20} is approximated by direct Monte Carlo simulation. Firstly, one draws $\mu_\Delta$ and $\sigma_\Delta^2$ from their posterior distributions, conditional on the syntheses. These distributions are derived by Raghunathan et al.\cite{raghunathan2003multiple} We draw values of $\mu_\Delta$ from a normal distribution:
\begin{equation}
p(\mu_\Delta \mid \boldsymbol{y}^{\boldsymbol{*}}, \boldsymbol{z^*}) \sim \textnormal{N}(\bar{\delta}_{10}, \bar{v}/M),
\label{eqn24}
\end{equation}
We draw values of $\sigma_\Delta^2$ from a chi-squared distribution with $M-1$ degrees of freedom:
\begin{equation}
p\left((M-1)b/(\sigma_\Delta^2 + \bar{v}) \mid \boldsymbol{y}^{\boldsymbol{*}}, \boldsymbol{z^*}\right)
\sim \chi^2_{M-1}.
\label{eqn25}
\end{equation}
Given draws of $\mu_\Delta$ and $\sigma_\Delta^2$, we draw values of $\Delta_{10}^{(2)}$ from a $t$-distribution with $M-1$ degrees of freedom:\cite{raghunathan2003multiple}
\begin{equation}
p(\Delta_{10}^{(2)} \mid \mu_\Delta, \sigma_\Delta^2) \sim t_{M-1}\left(\mu_\Delta, (1 + 1/M) \sigma_\Delta^2\right),
\label{eqn26}
\end{equation}
where the $\sigma_\Delta^2/M$ term in the variance is necessary as an adjustment for there being a finite number of syntheses; as $M \rightarrow \infty$, the variance tends to $\sigma_\Delta^2$.

By performing a large number of simulations, we are estimating the posterior distribution in Equation \ref{eqn20} by approximating the integral of the posterior in Equation \ref{eqn26} with respect to the posteriors in Equations \ref{eqn24} and \ref{eqn25}.\cite{raghunathan2003multiple} Hence, the resulting draws of $\Delta_{10}^{(2)}$ are samples from the posterior distribution $p(\Delta_{10}^{(2)} \mid \boldsymbol{y}^{\boldsymbol{*}}, \boldsymbol{z^*})$ in Equation \ref{eqn20}. We can take the expectation over the posterior draws to produce a point estimate $\hat{\Delta}_{10}^{(2)}$ of the marginal $A$ vs.~$C$ treatment effect, in the $BC$ population. An estimate of its variance $\hat{V}(\hat{\Delta}_{10}^{(2)})$ can be directly computed from the draws of the posterior density. Uncertainty measures such as 95\% interval estimates can be calculated from the corresponding empirical quantiles.

The posterior distributions in Equations \ref{eqn24}, \ref{eqn25} and \ref{eqn26} have been derived under certain normality assumptions, which are adequate for reasonably large sample sizes, where the relevant sample sizes are both the size of the $AC$ trial and the size $N^*$ of the synthetic datasets. Another assumption is that priors for the parameters in this step are diffuse, i.e., non-informative in the range where the posteriors have support from the data.\cite{raghunathan2003multiple}

\paragraph{Pooling via combining rules}

A simple alternative to direct Monte Carlo simulation is to use a basic normal approximation to the posterior density in Equation \ref{eqn20}, such that the sampling distribution in Equation \ref{eqn26} is a Normal as opposed to a $t$-distribution. The posterior mean is the average of the treatment effect point estimates across the $M$ syntheses. The simple combining rule for the variance arises from using $b - \bar{v}$ to estimate $\sigma^2_\Delta$, which is equivalent to setting $\sigma^2_\Delta$ at its approximate posterior mean in Equation \ref{eqn25}.\cite{si2011comparison} Again, the $b/M$ term is necessary as an adjustment for there being a finite number of syntheses. 

Consequently, point estimates for the $A$ vs.~$C$ treatment effect and its variance can be derived using the following plug-in estimators:
\begin{equation}
\hat{\Delta}^{(2)}_{10}=\bar{\delta}_{10},
\label{eqn27}
\end{equation}
\begin{equation}
\hat{V}\left(\hat{\Delta}^{(2)}_{10}\right)  =(1+1/M)b - \bar{v}.
\label{eqn28}
\end{equation}
Interval estimates can be approximated using a normal distribution, e.g.~taking $\pm 1.96$ times the square root of the variance computed in Equation \ref{eqn28}.\cite{raghunathan2003multiple} A heavier-tailed $t$-distribution with $\nu_f = (M-1)(1+\bar{v}/[(1+1/M)b)]^2$ degrees of freedom has also been proposed, as normal distributions may produce excessively narrow intervals and undercoverage when $M$ is more modest.\cite{reiter2002satisfying} Note that the combining rules in Equations \ref{eqn27} and \ref{eqn28} are only appropriate for reasonably large $M$. The choice of $M$ is discussed in subsection \ref{subsec38}.

\subsection{Indirect treatment comparison}\label{subsec37}

The estimated marginal treatment effect for $A$ vs.~$C$ is typically compared with that for $B$ vs.~$C$ to estimate the marginal treatment effect for $A$ vs.~$B$ in the $BC$ population. This is the indirect treatment comparison in the $BC$ population performed in Equation \ref{eqn1}. 

There is some flexibility in this step. Bayesian G-computation and the indirect comparison can be performed in one step under an MCMC approach. Similarly, so can the MIM pooling stage and the indirect comparison. In these cases, the estimation of $\Delta_{20}^{(2)}$ would be integrated within the estimation or simulation of the posterior of $\Delta_{10}^{(2)}$, under suitable priors, and a posterior distribution for $\Delta_{12}^{(2)}$ would be generated. This would require inputting as data the available aggregate outcomes for each treatment group in the published $BC$ study, or reconstructing subject-level data from these outcomes. For binary outcomes, event counts from the cells of a $2\times2$ contingency table would be required to estimate probabilities of the binary outcome as the incidence proportion for each treatment (dividing the number of subjects with the binary outcome in a treatment group by the total number of subjects in the group), to then estimate a marginal log-odds ratio for $B$ vs.~$C$. For survival outcomes, one can input patient-level data (with outcome times and event indicators for each subject) reconstructed from digitized Kaplan-Meier curves, e.g.~using the algorithm by Guyot et al.\cite{guyot2012enhanced} 

The advantage of this approach is that it directly generates a full posterior distribution for $\Delta_{12}^{(2)}$. Hence, its output is perfectly compatible with a probabilistic cost-effectiveness model. Samples of the posterior are directly incorporated into the decision analysis, so that the relevant economic measures can be evaluated for each sample without further distributional assumptions.\cite{dias2013evidence} If necessary, we can take the expectation over the draws of the posterior density to produce a point estimate $\hat{\Delta}_{12}^{(2)}$ of the marginal $A$ vs.~$B$ treatment effect, in the $BC$ population. Variance and interval estimates are derived empirically from the draws. 

Alternatively, we can perform the G-computation and indirect comparison, or the MIM pooling and indirect comparison, in two steps. Irrespective of the selected inferential framework, point estimates $\hat{\Delta}_{10}^{(2)}$ and $\hat{\Delta}_{20}^{(2)}$ can be directly substituted in Equation \ref{eqn1}. As the associated variance estimates $\hat{V}(\hat{\Delta}_{10}^{(2)})$ and $\hat{V}(\hat{\Delta}_{20}^{(2)})$ are statistically independent, these are summed to estimate the variance of the $A$ vs.~$B$ treatment effect:
\begin{equation}
\hat{V}(\hat{\Delta}_{12}^{(2)}) = \hat{V}(\hat{\Delta}_{10}^{(2)}) + \hat{V}(\hat{\Delta}_{20}^{(2)}).
\label{eqn38}
\end{equation}
With relatively large $M$ and sample sizes, interval estimates can be constructed using normal distributions, $\hat{\Delta}_{12}^{(2)} \pm 1.96 \sqrt{\hat{V}(\hat{\Delta}_{12}^{(2)})}$. This two-step strategy is simpler and easier to apply but sub-optimal in terms of integration with probabilistic sensitivity analysis, although one could perform forward Monte Carlo simulation from a normal distribution with mean $\hat{\Delta}_{12}^{(2)}$ and variance $\hat{V}(\hat{\Delta}_{12}^{(2)})$. Ultimately, it is the distribution of $\Delta_{12}^{(2)}$ that is relevant for HTA purposes. 

\subsection{Number of resamples or synthetic datasets}\label{subsec38}

When performing parametric G-computation with maximum-likelihood estimation, the choice of the number of bootstrap resamples is important. Similarly, when performing Bayesian parametric G-computation, the number $L$ of converged MCMC draws is important, as is the number $M \leq L$ of syntheses in MIM. Given recent advances in computing power, we encourage setting these values as large as possible, in order to maximize the precision and accuracy of the treatment effect estimator, and to minimize the Monte Carlo error in the estimate. A sensible strategy is to increase the number of bootstrap resamples/syntheses until repeated analyses across different random seeds give similar results, within a specified degree of accuracy.

In MIM, MCMC simulation is used in the synthesis stage. The value of $M$ is likely to be a fraction of the total number of iterations/posterior samples required for convergence. As computation time is driven by the synthesis stage, increasing $M$ provides more precise and efficient estimation\cite{reiter2002satisfying, reiter2003inference} of the treatment effect at little cost in the analysis stage. In the context of statistical disclosure limitation, it is not uncommon to set the number of syntheses as low as $M=10$.\cite{raghunathan2020synthetic} This is because the original survey data may involve several hundreds of subjects and variables. Releasing a large number of syntheses with a large number of subjects may not be practical, placing undue demands on the analyst, e.g.~in terms of storage costs and processing needs. MIM has been developed with much smaller numbers of subjects and covariates in mind, in a context in which the data synthesizer and analyst are the same entity. 

For MIM, an inconvenience of the expressions in Equation \ref{eqn25} and Equation \ref{eqn28} is that these may produce negative variances. When the posterior in Equation \ref{eqn25} generates a negative value of $\sigma^2_\Delta$, i.e., when $\frac{(M-1)b}{\chi^*} < v$ (where $\chi^*$ is the draw from the posterior in Equation \ref{eqn25}), the variance of the posterior distribution in Equation \ref{eqn26} is negative. Similarly, Equation \ref{eqn28} produces a negative variance when $(1+1/M)b < \bar{v}$. This is because the formulations have been derived using method-of-moments approximations, where estimates are not necessarily constrained to fall in the parameter space. Negative variances are unlikely to occur if $M$ and the size of the synthetic datasets are relatively large. This is due to lower variability in $\sigma^2_\Delta$ and $\hat{V}\left(\hat{\Delta}_{10}^{(2)}\right)$:\cite{reiter2005releasing}  $\bar{v}$ decreases with larger syntheses and $b$ is less variable with larger $M$.\cite{reiter2002satisfying} Reasonable values of $M$ are likely to depend on the specific setting, e.g.~the size of the $AC$ trial and the properties of the outcome model type, and we discuss these in the context of the simulation study in subsection \ref{subsec445}. 

\section{Simulation study}\label{sec4}

\subsection{Aims}\label{subsec41}

The objectives of the simulation study are to benchmark the performance of the novel ``marginalized'' outcome regression methods, and compare it with that of MAIC and the conventional version of STC across a range of scenarios that may be encountered in practice. We evaluate each estimator on the basis of the following finite-sample frequentist characteristics:\cite{morris2019using} (1) unbiasedness; (2) variance unbiasedness; (3) randomization validity;\footnote{In a sufficiently large number of repetitions, $(100\times (1-\alpha))$\% interval estimates based on normal distributions should contain the true value $(100\times(1-\alpha))$\% of the time, for a nominal significance level $\alpha$.} and (4) precision. The selected performance measures assess these criteria specifically (see subsection \ref{subsec45}). The simulation study is reported following the ADEMP (Aims, Data-generating mechanisms, Estimands, Methods, Performance measures) structure.\cite{morris2019using} All simulations and analyses were performed using \texttt{R} software version 3.6.3.\cite{team2013r} The design of the simulation study is similar to that presented by Remiro-Az\'ocar et al.,\cite{remiro2020methods} but features binary outcomes instead of survival outcomes, with a logistic regression outcome model as opposed to a Cox model, and a different treatment allocation ratio in the trials.\footnote{The files required to run the simulations are available at \url{http://github.com/remiroazocar/marginalized_indirect_comparisons_simstudy}.} Example \texttt{R} code implementing the methods on a simulated example is provided in \hyperref[SB]{Supplementary Appendix B}. 

\subsection{Data-generating mechanisms}\label{subsec42}

We consider binary outcomes using the log-odds ratio as the measure of effect. Most applications of population-adjusted indirect comparisons in HTA are in oncology,\cite{phillippo2019population} where the binary outcome may be response to treatment or the occurrence of an adverse event. 

For trials $AC$ and $BC$, outcome $y_n$ for subject $n$ is simulated from a Bernoulli distribution with probabilities of success generated from logistic regression, such that:
\begin{equation*}
\logit[p(y_n \mid \boldsymbol{x}_n, z_n)] = \beta_0 + \boldsymbol{x}_n \boldsymbol{\beta_1} + (\beta_z +  \boldsymbol{x}_n^{\boldsymbol{(EM)}} \boldsymbol{\beta_2})\mathds{1}(z_n = 1),
\end{equation*}
using the notation of the $AC$ trial data. Four correlated continuous covariates $\boldsymbol{x}_n$ are generated per subject by simulating from a multivariate normal distribution with pre-specified variable means and covariance matrix.\cite{ripley2009stochastic} Two of the covariates are purely prognostic variables; the other two ($\boldsymbol{x}_n^{\boldsymbol{(EM)}}$) are effect modifiers, modifying the effect of both treatments $A$ and $B$ versus $C$ on the log-odds ratio scale, and prognostic variables. 

The strength of the association between the prognostic variables and the outcome is set to $\beta_{1,k} = -\ln (0.5)$, where $k$ indexes a given covariate. This regression coefficient fixes the conditional odds ratio for the effect of each prognostic variable on the odds of outcome at 2, indicating a strong prognostic effect. The strength of interaction of the effect modifiers is set to $\beta_{2,k} = - \ln (0.67)$, where $k$ indexes a given effect modifier.  This fixes the conditional odds ratio for the interaction effect on the odds of the outcome at approximately 1.5. The shared effect modifier assumption\cite{phillippo2016nice} holds in the simulation study by design. That is, both active treatments have the same effect modifiers with respect to the common comparator and identical interaction coefficients for each. Pairwise Pearson correlation coefficients between the covariates are set to 0.2, indicating a moderate level of positive correlation. 

% This parameter has a material impact on the marginal $A$ vs.~$B$ treatment effect. Hence, population adjustment is necessary in order to remove the bias induced by covariate imbalances. 

The binary outcome represents the occurrence of an adverse event. Each active intervention has a very strong conditional treatment effect $\beta_z = \ln(0.17)$ at baseline (when the effect modifiers are zero) versus the common comparator. Such relative effect is associated with a ``major'' reduction of serious adverse events in a classification of extent categories by the German national HTA agency.\cite{skipka2016methodological} The covariates may represent comorbidities, which are associated with greater rates of the adverse event and, in the case of the effect modifiers, which interact with treatment to render it less effective. The intercept $\beta_0 = -0.62$ is set to fix the baseline event percentage at 35\% (under treatment $C$, when the values of the covariates are zero). 

The number of subjects in the $BC$ trial is 600, under a 2:1 active treatment vs.~control allocation ratio. For the $BC$ trial, the individual-level covariates and outcomes are aggregated to obtain summaries. The continuous covariates are summarized as means and standard deviations, which would be available to the analyst in the published study in a table of baseline characteristics in the RCT publication. The binary outcomes are summarized as overall event counts, e.g.~from the cells of a $2\times2$ contingency table. Typically, the published study only provides this aggregate information to the analyst. 

The simulation study investigates two factors in an arrangement with nine scenarios, thus exploring the interaction between these factors. The simulation scenarios are defined by the values of the following parameters:

\begin{itemize}
\item The number of subjects in the $AC$ trial, $N \in \{200, 400, 600\}$ under a 2:1 active intervention vs.~control allocation ratio. The sample sizes correspond to typical values for a Phase III RCT\cite{stanley2007design} and for trials included in applications of MAIC submitted to HTA authorities.\cite{phillippo2019population}
\item The degree of covariate imbalance.\footnote{In the simulation study, covariate \textit{balance} is a proxy for covariate \textit{overlap}. Imbalance refers to the difference in covariate distributions across studies, as measured by the difference in (standardized) average values. Imbalances in effect measure modifiers across studies induce bias in the standard indirect comparison and motivate the use of population adjustment. Overlap describes how similar the covariate ranges are  across studies --- there is complete overlap if the ranges are identical. Lack of complete overlap hinders the use of population adjustment.} For both trials, each covariate $k$ follows a normal marginal distribution with mean $\mu_k$ and standard deviation $\sigma_k$, such that $x_{i,k} \sim \textnormal{Normal}(\mu_k, \sigma_k^2)$ for subject $i$. For the $BC$ trial, we fix $\mu_k=0.6$. For the $AC$ trial, we vary the means of the marginal normal distributions such that $\mu_k \in \{ 0.45, 0.3, 0.15\}$. The standard deviation of each marginal distribution is fixed at $\sigma_k=0.4$ for both trials. This setup corresponds to standardized differences\cite{flury1986standard} or Cohen effect size indices\cite{cohen2013statistical} (the difference in means in units of the pooled standard deviation) of 0.375, 0.75 and 1.125, respectively. This yields strong, moderate and poor covariate overlap; with overlap between the univariate marginal distributions of 85\%, 71\% and 57\%, respectively, when $N=600$. To compute the overlap percentages, we have followed a derivation by Cohen\cite{cohen2013statistical} for normally-distributed populations with equal size and equal variance. Note that the percentage overlap between the joint covariate distributions of each study is substantially lower. The strong, moderate and poor covariate overlap scenarios correspond to average percentage reductions in effective sample size of 22\%, 60\% and 85\%, respectively. These percentage reductions are representative of the range encountered in NICE technology appraisals.\cite{remiro2020methods, phillippo2019population}
\end{itemize}

\subsection{Estimands}\label{subsec43}

The estimand of interest is the marginal log-odds ratio for $A$ vs.~$B$ in the $BC$ population. The treatment coefficient $\beta_z = \ln(0.17)$ is the same for both $A$ vs.~$C$ and $B$ vs.~$C$, and the shared effect modifier assumption holds in the simulation study. Therefore, the true conditional treatment effect for $A$ vs.~$B$ in the $BC$ population is zero. As the true subject-level conditional effects are zero for all units, the true marginal log-odds ratio in the $BC$ population is zero ($\Delta_{12}^{(2)}=0$). This implies a null hypothesis-like simulation setup of no treatment effect for $A$ vs.~$B$, and marginal and conditional estimands in the $BC$ population coincide by design. 

Note that the true marginal effect for $A$ vs.~$B$ in the $BC$ population is a composite of that for $A$ vs.~$C$ and that for $B$ vs.~$C$, both of which are non-null. These are the same and cancel out. For reference, the true marginal log-odds ratio in the $BC$ population for the active treatments vs.~the common comparator ($\Delta_{10}^{(2)}$ and $\Delta_{20}^{(2)}$) is -1.15. This has been calculated by simulating two potential cohorts of 500,000 subjects, with the $BC$ covariate distribution and the outcome-generating mechanism in subsection \ref{subsec42}. One cohort is under the active treatment and the other is under the common comparator. The number of simulated subjects is sufficiently large to minimize sampling variability. The two cohorts are concatenated and a simple logistic regression is fitted, regressing the simulated binary outcomes on an indicator variable for treatment assignment. The treatment coefficient estimates the average difference in the potential outcomes on the log-odds ratio scale, and serves as the log of the true marginal odds ratio for the two interventions under consideration. This is because the outcomes have been generated according to the true data-generating mechanism, where the true conditional effects are explicit, and which uses the correct conditional model by definition. Due to the non-collapsibility of the odds ratio, this simulation-based approach is necessary to determine the true marginal effect for $A$ vs. $C$ and $B$ vs. $C$, and has been adopted in previous research involving non-collapsible effect measures \cite{austin2013performance, austin2016variance, lesko2017bias}. 

All methods compared in the simulation study perform the same unadjusted analysis (i.e., a simple regression of outcome on treatment) to estimate the marginal treatment effect of $B$ versus $C$. Because the $BC$ study is a relatively large RCT, this comparison should be unbiased with respect to the true marginal log-odds ratio in $BC$. Therefore, any bias in the $A$ vs. $B$ comparison should arise from bias in the $A$ vs. $C$ comparison, for which marginal and conditional relative treatment effects are non-null. 

% The shared effect modifier assumption\cite{phillippo2016nice} holds in the simulation study by design. That is, both active treatments have identical effect modifiers and interaction coefficients for each effect modifier. Therefore, the $A$ vs.~$B$ marginal treatment effect can be generalized to any given target population, because effect modifiers are guaranteed to cancel out (the marginal effect for $A$ vs.~$B$ is conditionally constant across all populations). If the shared modifier assumption does not hold, the true marginal treatment effect for $A$ vs.~$B$ in the $BC$ population will not be applicable in any target population (one has to assume that the target population is $BC$ for the analysis to be valid), and the marginal and conditional estimands for $A$ vs.~$B$ will likely not coincide where the measure of effect is non-collapsible.

\subsection{Methods}\label{subsec44}

\subsubsection{Matching-adjusted indirect comparison}\label{subsec441}

Matching-adjusted indirect comparison (MAIC) is implemented using the original method of moments formulation presented by Signorovitch et al.\cite{signorovitch2010comparative, phillippo2016nice, remiro2020methods,  phillippo2020equivalence} To avoid further reductions in effective sample size and precision, only the effect modifiers are included in the weighting model. A weighted logistic regression is fitted to the $AC$ IPD and standard errors for the $A$ vs.~$C$ marginal treatment effect are computed by resampling via the ordinary non-parametric bootstrap with replacement,\cite{efron1994introduction, sikirica2013comparative} with 1,000 resamples of each simulated dataset. Note that the standard version of MAIC\cite{signorovitch2010comparative, phillippo2016nice, remiro2020methods,  phillippo2020equivalence} uses a robust sandwich estimator for variance estimation\cite{white1980heteroskedasticity, huber1967behavior, hill2006interval} that accounts for the heteroskedasticity or correlation induced by the weighting. Nevertheless, this has understated variability under small effective sample sizes in previous simulation studies,\cite{remiro2020methods, kauermann2001note, fay2001small} and most software implementations of the estimator treat the weights as fixed quantities. The bootstrap approach should account for the uncertainty in estimating the weights from the data. The average marginal log-odds ratio for $A$ vs.~$C$ is calculated as the mean across the 1,000 bootstrap resamples. Its corresponding standard error is the sample standard deviation across the resamples. 

In our implementation of MAIC, we only balance the covariate means and balance these for active treatment and control arms combined. Other approaches have been proposed, such as balancing the covariates separately for active treatment and common comparator arms,\cite{petto2019alternative, belger2015inclusion, aouni2020matching} or balancing terms of higher order than means, e.g.~by including squared covariates in the weight estimation to balance variances. The former approach is discouraged because it may break randomization in the IPD, distorting the balance between treatment arms $A$ and $C$ on covariates that are not accounted for in the weighting, and potentially compromising the internal validity of the within-study estimate. The latter approach may increase finite-sample bias\cite{windmeijer2005finite} and has performed poorly in recent simulation studies, in terms of both bias and precision, where covariate variances differ across studies.\cite{petto2019alternative, hatswell2020effects, phillippo2020assessing, phillippo2020target} 

Given the often arbitrary factors driving selection into different trials, the mechanism for generating the simulation study data in subsection \ref{subsec42} does not specify a trial assignment model. Nevertheless, the logistic regression model for estimating the weights is considered the ``best-case'' model because it selects the ``right'' subset of covariates as effect modifiers. The estimated weights are adequate for bias removal because the balancing property\cite{dehejia1999causal, waernbaum2010propensity, zhao2008sensitivity, rubin2000combining} holds with respect to the effect modifier means. Namely, conditional on the weights, the effect modifier means are balanced between the two trials, and one can potentially achieve unbiased estimation of treatment effects in the $BC$ population due to conditional exchangeability over trial assignment.

In a test simulation scenario with $N=200$, bootstrapped MAIC has a running time of approximately 2.7 seconds per simulated dataset, using an Intel Core i7-8650 CPU (1.90 GHz) processor. Computation time increases linearly with the number of bootstrap resamples. 

\subsubsection{Conventional simulated treatment comparison}\label{subsec442}

The conventional version of simulated treatment comparison (STC), as described by HTA guidance and recommendations,\cite{phillippo2016nice} is implemented. A covariate-adjusted logistic regression is fitted to the IPD using maximum-likelihood estimation. The outcome regression is correctly specified.\footnote{It is more burdensome to specify an outcome regression model than a propensity score model (for the MAIC weights). The former requires specifying both prognostic and interaction terms, whereas the latter only requires the specification of effect modifiers. In practice, one cannot typically ascertain which covariates are purely prognostic variables and which covariates are effect modifiers. Exploratory simulations show that the relative precision and accuracy of MAIC deteriorate, with respect to outcome regression, if we treat all four covariates as effect modifiers. On the other hand, the relative precision and accuracy of outcome regression deteriorate if the terms corresponding to the purely prognostic covariates are not included in the outcome model. Nevertheless, the other terms in the regression already account for a considerable portion of the variability of the outcome, and relative effects are accurately estimated in any case. These alternative setups do not alter the conclusions of the simulation study.} All covariates are accounted for in the regression but only the treatment effect modifiers are centered at their mean $BC$ values, and interaction terms are only included for the effect modifiers. The log-odds ratio estimate for $A$ vs.~$C$ is the treatment coefficient of the centered multivariable regression, with its standard error quantifying the standard deviation of the treatment effect.

In a test simulation scenario with $N=200$, the conventional STC has a running time of 0.02 seconds per simulated dataset.

\subsubsection{Maximum-likelihood parametric G-computation}\label{subsec443}

We consider two implementations of parametric G-computation. In the first implementation, we use maximum-likelihood estimation to fit the multivariable outcome regression. The Q-model is correctly specified. We construct the joint distribution of the four $BC$ covariates by simulating these from a multivariate Gaussian copula. This uses normally-distributed marginals with the $BC$ means and standard deviations, and the pairwise linear correlations of the $AC$ IPD. $N^*=1000$ subject profiles are simulated for the $BC$ pseudo-population, a value high enough to minimize sampling variability and provide an adequate degree of precision. Outcomes in the $BC$ population are predicted by plugging the simulated covariates into the maximum-likelihood fit. The procedure is resampled using the ordinary non-parametric bootstrap with replacement, with 1,000 resamples of each simulated dataset. Increasing further the number of resamples produces minimal gains in estimation precision and accuracy, with the Monte Carlo error across different random seeds remaining relatively insensitive to these increases. The average marginal log-odds ratio for $A$ vs.~$C$ is calculated as the mean across the 1,000 bootstrap resamples. Its corresponding standard error is the sample standard deviation across the resamples.  

In a test simulation scenario with $N=200$, parametric G-computation using maximum-likelihood estimation has a running time of approximately 3.5 seconds per replicate. Computation time increases linearly with the number of bootstrap resamples. 

\subsubsection{Bayesian parametric G-computation}\label{subsec444}

In the second implementation of parametric G-computation, we use MCMC simulation to fit the outcome regression. This is implemented using the package \texttt{rstanarm},\cite{goodrich2018rstanarm} a high-level appendage to the \texttt{rstan} package,\cite{team2016rstan} the \texttt{R} interface for \texttt{Stan}.\cite{carpenter2017stan} Again, the Q-model is correctly specified. The joint distribution of the $BC$ covariates is constructed by simulating $N^*=1000$ subjects from a multivariate Gaussian copula, with normally-distributed marginals with the $BC$ means and standard deviations, and the pairwise linear correlations of the $AC$ IPD. Predicted outcomes for the simulated covariates are drawn from their posterior predictive distribution.

We use the default independent ``weakly informative'' priors for the logistic regression intercept and predictor coefficients, i.e., the likelihood dominates under a reasonably large amount of data and the prior strongly influences the posterior if the data are weak.\cite{gelman2008weakly} These are normally-distributed priors centered at mean 0. The scale of the normal prior distribution for the intercept is 1. The scale parameter of the normal priors for the other coefficients is 2.5, rescaled in terms of the standard deviation of the predictor in question. This places most of the prior mass in the range of plausible effects, discarding coefficient values that are implausibly strong, e.g.~log-odds ratios over 3 (corresponding, approximately, to odds ratios over 20). This provides some regularization and helps stabilize computation. Alternative prior specifications are considered to check that we are not incorporating any unintended information into the models through the priors. Results are robust to the definitions of the prior distributions. 

We run two Markov chains with 4,000 total iterations per chain. These include 2,000 warmup/burn-in iterations for each chain that are not used for posterior inference. This gives a total of 4,000 iterations for performing the analysis. Approximate mixing of the chains was attained, with all within-chain relative to between-chain statistics (R-hat) below 1.1.\cite{gelman2013bayesian} Satisfactory convergence was confirmed by the inspection of trace plots and the assessment of diagnostics such as the effective sample size and the Gelman-Rubin convergence diagnostic (potential scale reduction factor).\cite{gelman2013bayesian} The average marginal treatment effect for $A$ vs.~$C$ is estimated taking the sample mean of the marginal log-odds ratio across the 4,000 MCMC iterations. The corresponding standard error is estimated using the sample standard deviation of the posterior draws of the marginal log-odds ratio. 

In a test simulation scenario with $N=200$, Bayesian parametric G-computation has a running time of approximately 4.2 seconds per replicate. Computation time increases linearly with the total number of MCMC iterations. 

\subsubsection{Multiple imputation marginalization}\label{subsec445}

In the synthesis stage (first-stage regression and outcome prediction), we follow the MCMC procedure outlined for Bayesian G-computation to generate the syntheses, using identical prior specifications. We run two MCMC chains using \texttt{rstanarm} with 4,000 iterations per chain, where the burn-in is of 2,000 iterations. The convergence and mixing of the chains are satisfactory. The first-stage logistic regression is correctly specified. The joint distribution of the $BC$ covariates is constructed by simulating from a multivariate Gaussian copula, with normally-distributed marginals with the $BC$ means and standard deviations, and the pairwise linear correlations of the $AC$ IPD. Predicted outcomes for the simulated covariates are drawn from their posterior predictive distribution as for Bayesian G-computation.  

The MCMC chains are thinned every 4 iterations to use $M=4000/4=1000$ syntheses in the analysis stage. Each synthesis is of size $N^*=1000$, while keeping the same treatment allocation ratio of the original $AC$ trial. We consider the selected value of $M$ to provide an adequate degree of precision. In a test simulation scenario ($N=200$), $M=1000$ is high enough to minimize the Monte Carlo noise in the treatment effect estimate, such that the Monte Carlo error across different random seeds is small with respect to the uncertainty in the estimator (estimates are approximately within 0.01 across seeds).  

We consider a two-step approach to pooling and the indirect comparison. In this formulation, the combining rules in Equations \ref{eqn27} and \ref{eqn28} are used to pool the point estimates of the second stage regressions and to estimate the marginal log-odds ratio for $A$ vs.~$C$ in the $BC$ population. Under $M=1000$, variance estimates for the marginal $A$ vs.~$C$ treatment effect are never negative under any scenario. 

In a test simulation scenario with $N=200$, MIM has a running time of approximately 7.9 seconds per replicate. Assuming that the total number of MCMC iterations is fixed, computation time increases linearly with the number of syntheses $M$.

\subsubsection{Indirect treatment comparison}\label{subsec446}

For all methods, the marginal log-odds ratio for $B$ vs.~$C$ is estimated directly from the event counts, and its standard error is computed using the delta method.\cite{bland2000odds} The marginal log-odds ratio estimate for $A$ vs.~$B$ and its standard error are obtained by combining the within-study point estimates, as per subsection \ref{subsec37} (using Equation \ref{eqn1} to compare point estimates and Equation \ref{eqn38} to sum the point estimates of the variance). Wald-type 95\% interval estimates are constructed for the marginal $A$ vs.~$B$ treatment effect using normal distributions. 

In Bayesian G-computation, we have used a two-step approach for: (1) the population-adjusted analysis of the $AC$ trial (estimation of the marginal effect for $A$ vs.~$C$); and (2) the indirect treatment comparison (estimation of the marginal effect for $A$ vs.~$B$). We also consider integrating the two in one stage, using MCMC sampling. In this case, for estimation of the marginal log-odds ratio for $B$ vs.~$C$, the true underlying event rates/proportions for the treatments are given non-informative Jeffreys $\textnormal{Beta}(0.5,0.5)$ priors. The number of events in each arm is sampled from two independent Binomial likelihoods, parametrized by the aforementioned event probabilities and the total number of subjects in each arm. Means and variances for the marginal $A$ vs.~$B$ treatment effect are obtained empirically from the posterior samples, with interval estimates calculated from the quantiles of the posterior distribution. 

Similarly, in MIM, we have used a two-step approach for: (1) pooling (estimation of the average marginal effect for $A$ vs.~$C$); and (2) the indirect treatment comparison (estimation of the marginal effect for $A$ vs.~$B$). We also consider using posterior simulation (Equations \ref{eqn24}-\ref{eqn26}) to pool the point estimates from the second-stage regressions, integrating the pooling and the indirect comparison within a single Bayesian computation module. In this case, we apply MCMC sampling using the aforementioned prior specifications for the event rates in the $BC$ study. 

While the Bayesian inferential frameworks might be convenient for parametric G-computation and for MIM in the context of probabilistic sensitivity analysis, the selected inferential framework has little bearing on computation time and on the results of this simulation study, both in terms of a single case study and of the long-run frequentist statistical properties of the methods. Integrating the indirect treatment comparison step within a Bayesian module leads to virtually identical performance measures than the two-step approaches. Therefore, results are not reported.

\subsection{Performance measures}\label{subsec45}

We generate and analyze 2,000 Monte Carlo replicates of trial data per simulation scenario. Recall that in our implementations of MAIC, G-computation (both versions) and MIM, a large number of bootstrap resamples, MCMC draws or syntheses are performed for each of the 2,000 replicates. For instance, the analysis for one simulation scenario using Bayesian G-computation contains 4,000 MCMC draws (after burn-in) times 2,000 simulation replicates, which equals a total of 8 million posterior draws. Based on the method and simulation scenario with the highest long-run variability (MAIC with $N=200$ and poor covariate overlap), we consider the degree of precision provided by the Monte Carlo standard errors under 2,000 replicates to be acceptable in relation to the size of the effects.\footnote{Conservatively, we assume that $\textnormal{SD}(\hat{\Delta}_{12}^{(2)}) \leq 1.71$ and that the variance across simulations of the estimated treatment effect is always less than 2.92. Given that the MCSE of the bias is equal to $\sqrt{\textnormal{Var}(\hat{\Delta}_{12}^{(2)})/N_{sim}}$, where $N_{sim}=2000$ is the number of simulations, it is at most 0.038 under 2,000 simulations. We consider the degree of precision provided by the MCSE of the bias to be acceptable in relation to the size of the effects. If the empirical coverage rate of the methods is 95\%, $N_{sim}=2000$ implies that the MCSE of the coverage is $\left (\sqrt{(95 \times 5)/2000} \right)\%=0.49\%$, with the worst-case MCSE being $1.12\%$ under 50\% coverage. We also consider this degree of precision to be acceptable. Hence, the simulation study is conducted under $N_{sim}=2000$.} 

We evaluate the performance of the outcome regression methods and MAIC on the basis of the following criteria: (1) bias; (2) variability ratio; (3) empirical coverage rate of the interval estimates; (4) empirical standard error (ESE); and (5) mean square error (MSE). These criteria are explicitly defined in a previous simulation study by the authors.\cite{remiro2020methods} 

With respect to the simulation study aims in subsection \ref{subsec41}, the bias in the estimated treatment effect assesses aim 1. This is equivalent to the average estimated treatment effect across simulations because the true treatment effect $\Delta_{12}^{(2)} = 0$. The variability ratio evaluates aim 2. This represents the ratio of the average model standard error and the sample standard deviation of the treatment effect estimates (the empirical standard error).\cite{leyrat2014propensity} Variability ratios greater than (or lesser than) one indicate that model standard errors overestimate (or underestimate) the variability of the treatment effect estimate. It is worth noting that this metric assumes that the correct estimand and corresponding variance are being targeted. A variability ratio of one is of little use if this is not the case, e.g.~if both the model standard errors and the empirical standard errors are taken over estimates targeting the wrong estimand. Coverage targets aim 3, and is estimated as the proportion of simulated datasets for which the true treatment effect is contained within the nominal $(100\times(1 - \alpha))\%$ interval estimate of the estimated treatment effect. In this article, $\alpha=0.05$ is the nominal significance level. The empirical standard error is the standard deviation of the treatment effect estimates across the 2,000 simulated datasets. Therefore, it measures precision or long-run variability, and evaluates aim 4. The mean square error is equivalent to the average of the squared bias plus the variance across the 2,000 simulated datasets. Therefore, it is a summary value of overall accuracy (efficiency), that accounts for both bias (aim 1) and variability (aim 4). 

\section{Results}\label{sec5}

\begin{figure}[!htb]
\center{\includegraphics[height=0.89\textheight, width=\textwidth]{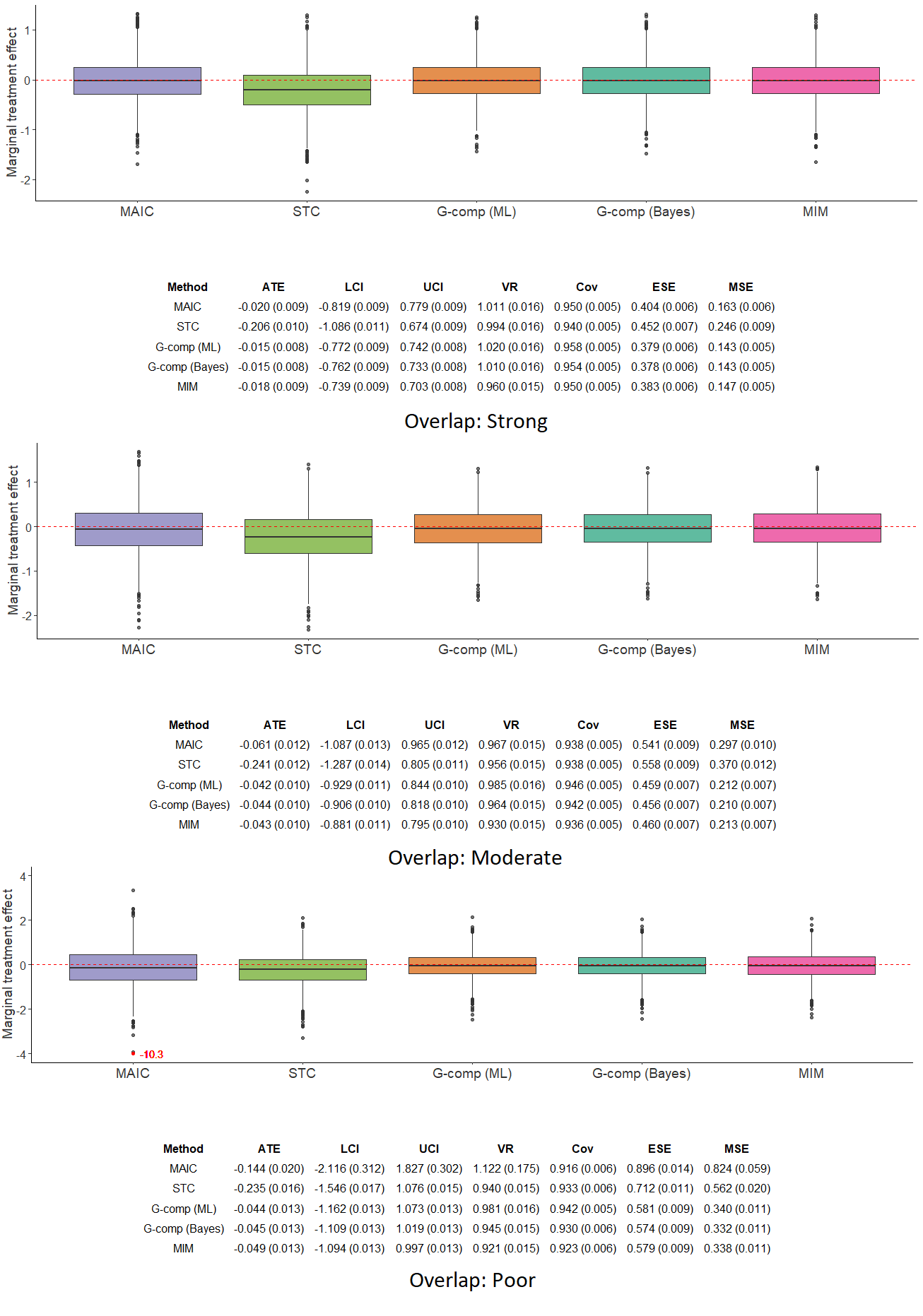}}
\caption{Point estimates and performance metrics across all methods for each simulation scenario with $N=200$. The model standard error for the MAIC outlier in the poor overlap scenario has an inordinate influence on the variability ratio; removing it reduces the variability ratio to 0.980 (0.019). Note that the version of STC evaluated does not actually target a marginal effect. 
\label{fig3}}
\end{figure}

\begin{figure}[!htb]
\center{\includegraphics[height=0.93\textheight, width=\textwidth]{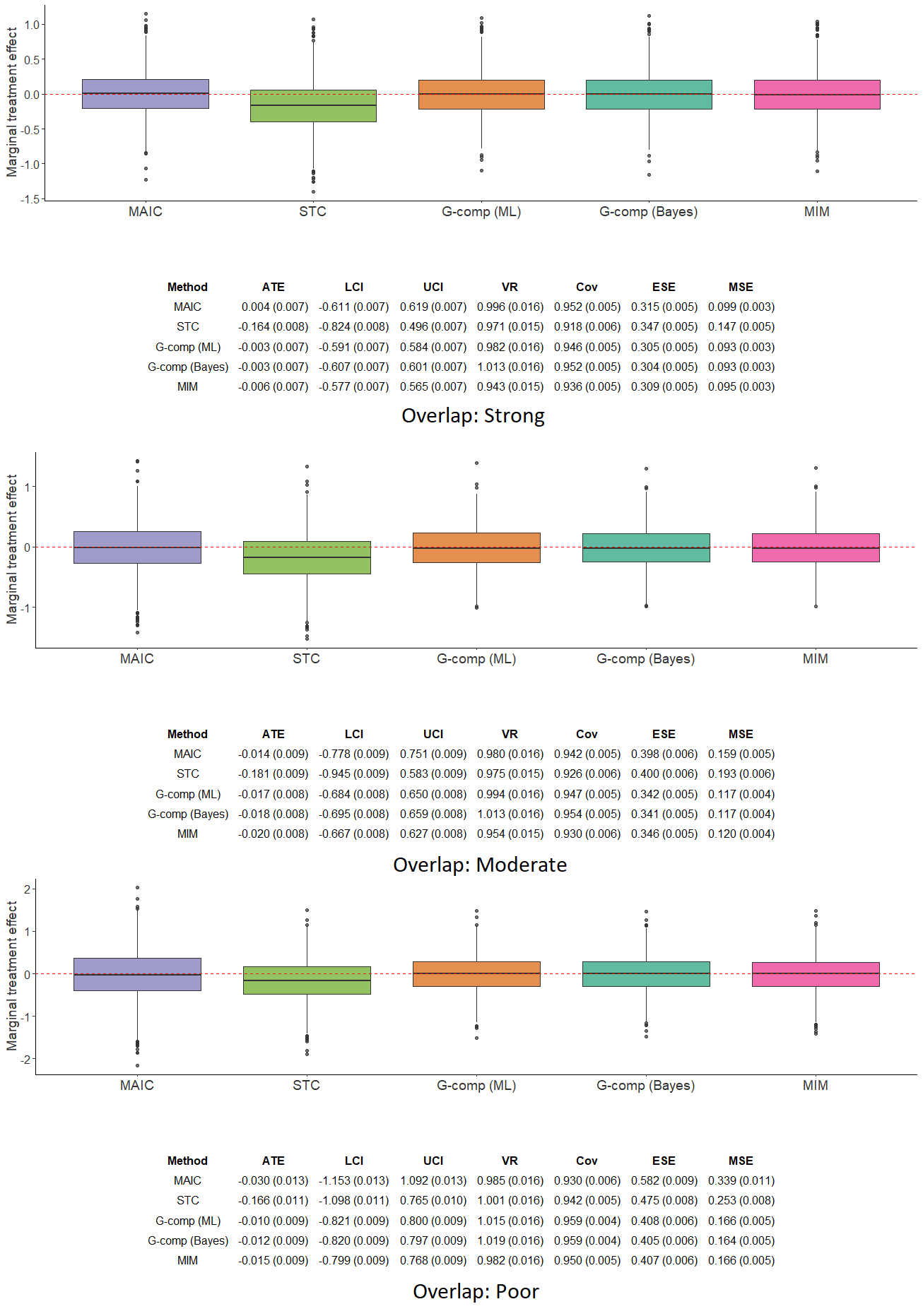}}
\caption{Point estimates and performance metrics across all methods for each simulation scenario with $N=400$.\label{fig4}}
\end{figure}

\begin{figure}[!htb]
\center{\includegraphics[height=0.93\textheight, width=\textwidth]{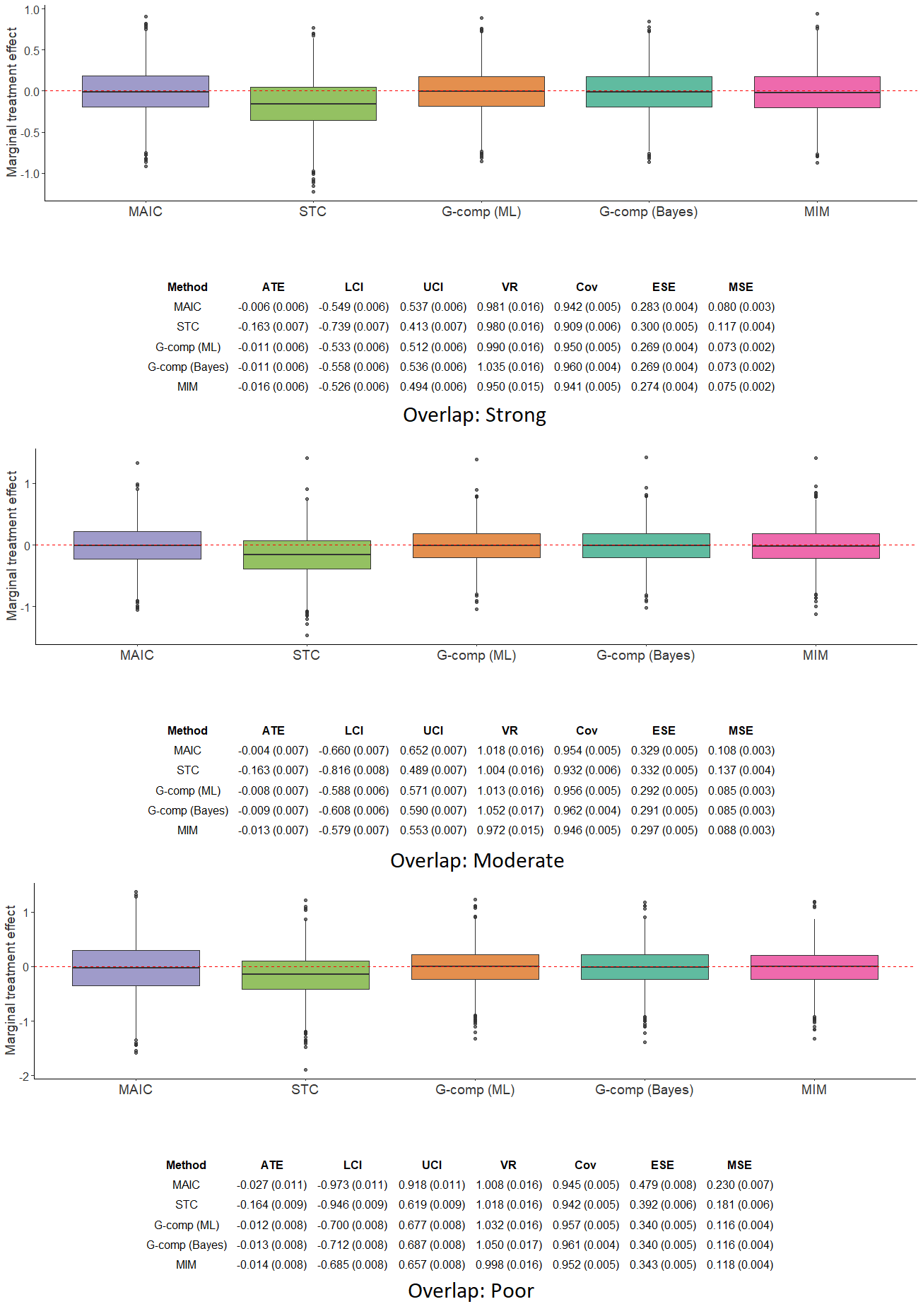}}
\caption{Point estimates and performance metrics across all methods for each simulation scenario with $N=600$.\label{fig5}}
\end{figure}

Performance metrics for all simulation scenarios are displayed in Figure \ref{fig3}, Figure \ref{fig4} and Figure \ref{fig5}. Figure \ref{fig3} displays the results for the three data-generating mechanisms under $N=200$. Figure \ref{fig4} presents the results for the three scenarios with $N=400$. Figure \ref{fig5} depicts the results for the three scenarios with $N=600$. From top to bottom, each figure considers the scenario with strong overlap first, followed by the moderate and poor overlap scenarios. For each scenario, there is a box plot of the point estimates of the $A$ vs.~$B$ marginal treatment effect across the 2,000 simulated datasets. Below, is a summary tabulation of the performance measures for each method. Each performance measure is followed by its Monte Carlo standard error, presented in parentheses, which quantifies the simulation uncertainty. 

In the figures, ATE is the average marginal treatment effect estimate for $A$ vs.~$B$ across the simulated datasets (this is equal to the bias as the true effect is zero). LCI is the average lower bound of the 95\% interval estimate. UCI is the average upper bound of the 95\% interval estimate. VR, ESE and MSE are the variability ratio, empirical standard error and mean square error, respectively. Cov is the empirical coverage rate of the 95\% interval estimates. G-comp (ML) stands for the maximum-likelihood version of parametric G-computation and G-comp (Bayes) denotes its Bayesian counterpart using MCMC estimation. 

In MIM, no simulation replicates produce negative variances and ad hoc truncation is not required. Weight estimation cannot be performed for 4 of the 18,000 replicates in MAIC, where there are no feasible weighting solutions. This issue occurs in the most extreme scenario, corresponding to $N=200$ and poor covariate overlap. Feasible weighting solutions do not exist due to separation problems, i.e., there is a total lack of covariate overlap. Because MAIC is incapable of producing an estimate in these cases, the affected replicates are discarded altogether (the scenario in question analyzes 1,996 simulated datasets for MAIC). This phenomenon has also been observed in a recent simulation study,\cite{phillippo2020assessing} where MAIC cannot generate an estimate in scenarios with small sample sizes and poor overlap.    

\paragraph{Unbiasedness of treatment effect estimates}

The impact of the bias largely depends on the uncertainty in the estimated treatment effect, quantified by the empirical standard error. We compute standardized biases (bias as a percentage of the empirical standard error). With $N=200$, MAIC has standardized biases of magnitude 11.3\% and 16.1\% under moderate and poor covariate overlap, respectively. Otherwise, the magnitude of the standardized bias is below 10\%. Similarly, under $N=200$, the maximum-likelihood version of parametric G-computation has standardized biases of magnitude 13.3\% and 24.8\% in the scenarios with moderate and poor overlap, respectively. In all other scenarios, the standardized bias has magnitude below 10\%. For Bayesian parametric G-computation, standardized biases never have a magnitude above 10\% and troublesome biases are not produced in any of the simulation scenarios. The maximum absolute value of the standardized bias is 9.7\% in a scenario with $N=200$ and moderate covariate overlap. In MIM, no standardized biases are larger than 10\% in either direction, and the maximum absolute value is 9.4\% in a simulation scenario with $N=200$ and moderate overlap. 

To evaluate whether the bias in MAIC and parametric G-computation has any practical significance, we investigate whether the coverage rates are degraded by it. Coverage is not affected for maximum-likelihood parametric G-computation, where empirical coverage rates for all simulation scenarios are very close to the nominal coverage rate, 0.95 for 95\% interval estimates. In the case of MAIC, there is discernible undercoverage in the scenario with $N=200$ and poor covariate overlap (empirical coverage rate of 0.916). This is the scenario with the lowest effective sample size after weighting. Hence, the results are probably related to small-sample bias\cite{greenland2016sparse} in the weighted logistic regression.\cite{vittinghoff2007relaxing} This bias for MAIC was not observed in the more extreme scenarios of a recent simulation study,\cite{remiro2020methods} which considered survival outcomes and the Cox proportional hazards regression as the outcome model. In absolute terms, the bias of MAIC is greater than that of MIM and both versions of parametric G-computation where the number of patients in the $AC$ trial is small ($N=200$) or covariate overlap is poor. In fact, when both of these apply, the bias of MAIC is important (-0.144). Otherwise, with $N=400$ and greater, and moderate or strong overlap, the aforementioned methods produce similarly low levels of bias. 

STC generates problematic negative biases in all nine scenarios considered in this simulation study, with a standardized bias of magnitude greater than 30\% in all cases. This systematic bias arises from the non-collapsibility of the log-odds ratio. STC consistently produces the highest bias of all methods, and the magnitude of this bias appears to increase under the smallest sample size ($N=200$).

\clearpage

\clearpage

\clearpage

\clearpage

\paragraph{Unbiasedness of variance estimates}

In MAIC, the variability ratio of treatment effect estimates is close to one under all simulation scenarios except one. That is the scenario with $N=200$ and poor covariate overlap, where the variability ratio is 1.122. This high value is attributed to the undue influence of an outlier (as seen in the box plot of point estimates) on the average model standard error. Once the outlier is removed, the variability ratio decreases to 0.98, just outside from being within Monte Carlo error of one but not statistically significantly different. This suggests very little bias in the standard error estimates in this scenario, i.e., that the model standard errors tend to coincide with the empirical standard error. In a previous simulation study,\cite{remiro2020methods} robust sandwich standard errors underestimated the variability of estimates in MAIC under small sample sizes and poor covariate overlap. The non-parametric bootstrap seems to provide more conservative variance estimation in these extreme settings. 

In STC, variability ratios are generally close to one with $N=400$ and $N=600$. Any bias in the estimated variances appears to be negligible, although there is a slight decrease in the variability ratios when the $AC$ sample size is small ($N = 200$). Recall that this metric assumes that the correct estimand and corresponding variance are being targeted. However, in our application of STC, both model standard errors and empirical standard errors are taken over an incompatible indirect treatment comparison.  

In maximum-likelihood parametric G-computation, variability ratios are generally very close to one. In Bayesian parametric G-computation, variability ratios are generally close to one but are slightly above it in some scenarios with $N=600$ (1.05 and 1.052 with moderate and poor covariate overlap, respectively). This suggests some overestimation of the empirical standard error by the model standard errors. On the other hand, MIM displays some underestimation of variability by the model standard errors. This is more pronounced under the smallest sample size, with variability ratios of 0.93 and 0.921 for moderate and poor overlap, respectively. The underestimation is likely due to the normality assumptions used to derive the model standard errors --- the posterior distribution of $\Delta_{10}^{(2)}$ is assumed to be normal in the derivation of the combining rules, and low sample sizes may break the normality assumption.

\paragraph{Randomization validity}

From a frequentist viewpoint,\cite{neyman1934two} 95\% interval estimates are randomization-valid if these are guaranteed to include the true treatment effect 95\% of the time. Namely, the empirical coverage rate should be approximately equal to the nominal coverage rate, in this case 0.95 for 95\% interval estimates, to obtain appropriate type I error rates for testing a ``no effect'' null hypothesis. Theoretically, the empirical coverage rate is statistically significantly different to 0.95 if, roughly, it is less than 0.94 or more than 0.96, assuming 2,000 independent simulations per scenario. These values differ by approximately two standard errors from the nominal coverage rate. Poor coverage rates are a decomposition of both the bias and the standard
error used to compute the Wald-type interval estimates. In the simulation scenarios, none of the methods lead to overly conservative inferences but there are some issues with undercoverage. 

Empirical coverage rates for MAIC are significantly different from the advertised nominal coverage rate in three scenarios. In the three, the coverage rate is below 0.94 (empirical coverage rates of 0.938, 0.93 and 0.916). The last two of these occur in scenarios with poor covariate overlap, with the latter corresponding to the smallest effective sample size after weighting ($N=200$). This is the scenario which integrates the two most important determinants of small-sample bias, in which MAIC has exhibited discernible bias. In this case, undercoverage is bias-induced. On the other hand, in our previous simulation study,\cite{remiro2020methods} undercoverage was induced by the robust sandwich variance underestimating standard errors.    

In the conventional version of STC, coverage rates are degraded by the bias induced by the non-collapsibility of the log-odds ratio. Almost invariably, there is undercoverage. Interestingly, the empirical coverage does not markedly deteriorate --- coverage percentages never fall below 90\%, i.e., never at least double the nominal rate of error. In general, both versions of parametric G-computation exhibit appropriate coverage. Only one scenario provides rates below 0.94 (Bayesian G-computation with $N=200$ and poor overlap, with an empirical coverage rate of 0.93). No scenarios have empirical coverage above 0.96. Coverage rates for the maximum-likelihood implementation are always appropriate, with most empirical coverage percentages within Monte Carlo error of 95\%. 

In MIM, coverage rates generally exhibit some underestimation of the advertised nominal coverage rate. Empirical coverage rates are significantly below the nominal rate in four scenarios (empirical coverage rates of 0.936, 0.936, 0.93 and 0.923). Again, the most inappropriate of these (0.923) occurs where there is poor covariate overlap and the $AC$ sample size is low ($N=200$). In this scenario, it is not bias that degrades the coverage rate for MIM. Poor coverage is induced by the standard errors used to construct the Wald-type interval estimates, which underestimate variability.  

\paragraph{Precision and efficiency}

MIM and both versions of parametric G-computation have reduced empirical standard errors compared to MAIC across all scenarios. Interestingly, conventional STC is even less precise than MAIC in most scenarios (all the scenarios with moderate or strong overlap, where reductions in effective sample size after weighting are tolerable). Several trends are revealed upon comparison of the ESEs, and upon visual inspection of the spread of the point estimates in the box plots. As expected, the ESE increases for all methods (i.e., estimates are less precise) as the number of subjects in the $AC$ trial is lower. The decrease in precision is more substantial for MAIC than for the outcome regression methods.  

The degree of covariate overlap has an important influence on the ESE and population adjustment methods incur losses of precision when covariate overlap is poor. Again, this loss of precision is more substantial for MAIC than for the outcome regression approaches. Where overlap is poor, there exists a subpopulation in $BC$ that does not overlap with the $AC$ population. Therefore, inferences in this subpopulation rely largely on extrapolation. Outcome regression approaches require greater extrapolation when the covariate overlap is weaker, thereby incurring a loss of precision. 

Where covariate overlap is strong, MIM and both versions of parametric G-computation display very similar ESEs than MAIC. As mentioned earlier, conventional STC offers even lower precision than MAIC in these cases. To illustrate this, consider the scenario with $N=200$ and moderate overlap, where MAIC is expected to have a low effective sample size after weighting and perform comparatively worse than outcome regression. Even in this scenario, MAIC appears to be more precise (empirical standard error of 0.541) than conventional STC (empirical standard error of 0.558). As overlap decreases, precision is reduced more markedly for MAIC compared to the outcome regression methods. Under poor overlap, MAIC considerably increases the ESE compared to the conventional STC. 

In MAIC, extrapolation is not even possible. Where covariate overlap is poor, the observations in the $AC$ IPD that are not covered by the ranges of the selected covariates in $BC$ are assigned weights that are very close to zero. The relatively small number of individuals in the overlapping region of the covariate space are assigned inflated weights, dominating the reweighted sample. These extreme weights lead to large reductions in effective sample size and affect very negatively the precision of estimates. 

Similar to the trends observed for the ESE, the MSE is also very sensitive to the value of $N$ and to the level of covariate overlap. The MSE decreases for all methods as $N$ and the level of overlap increase. The accuracy of MAIC and the marginalized outcome regression methods is comparable when the $AC$ sample size is high or covariate overlap is strong. As the $AC$ sample size and overlap decrease, the relative accuracy of MAIC with respect to MIM and both approaches to parametric G-computation is markedly reduced. Accuracy for the conventional version of STC is comparatively poor and this is driven by bias. 

Where covariate overlap is strong or moderate, the marginalized outcome regression methods have the highest accuracy, followed by MAIC and STC. Where overlap is poor, the marginalized outcome regression methods are considerably more accurate than MAIC, with much smaller mean square errors. MAIC also provides less accurate estimates than STC in terms of mean square error. The variability of estimates under MAIC increases considerably in these scenarios. The precision is sufficiently poor to offset the loss of bias with respect to STC. 

\section{Discussion}\label{sec6}

\subsection{Summary of results}\label{subsec61}

The marginalized outcome regression methods and MAIC can yield unbiased estimates of the marginal $A$ vs.~$B$ treatment effect in the $BC$ population. Conventional STC targets a conditional treatment effect for $A$ vs.~$C$ that is incompatible in the indirect comparison. Bias is produced because the log-odds ratio is a non-collapsible measure of effect. Across all scenarios, MIM and both versions of parametric G-computation largely eliminate the bias induced by effect modifier imbalances. There is some negative bias in MAIC and the marginalized outcome regression methods where the sample size $N$ is small. In the case of MAIC, this is problematic where covariate overlap is poor. MAIC did not display these biases using Cox proportional hazards regression as the outcome model in the time-to-event setting.\cite{remiro2020methods} The difference in results is likely due to logistic regression being more prone to small-sample bias than Cox regression.\cite{vittinghoff2007relaxing} 

As for precision, the marginalized outcome regression approaches have reduced variability compared to MAIC. The superior precision is demonstrated by their lower empirical standard errors across all scenarios. Because the methods are generally unbiased, precision is the driver of comparative accuracy. The simulation study confirms that, under correct model specification, parametric G-computation and MIM have lower mean square errors than weighting and are therefore more efficient. The differences in performance are exacerbated where covariate overlap is poor and sample sizes are low. In these cases, the effective sample size after weighting is small, and this leads to inflated variances and wider interval estimates for MAIC. Specific bias-variance trade-offs will likely depend on the outcome model of interest.

The performance measures for Bayesian G-computation and MIM are very similar, as they use the same MCMC procedure to fit the outcome model and to predict outcomes. In terms of bias, precision and efficiency, there is no particular reason to believe that the performance metrics for Bayesian G-computation or MIM are superior one to the other. In terms of variance estimation and coverage, the performance measures for both parametric G-computation approaches are superior to those of MIM. MIM exhibits undercoverage in some scenarios due to the model standard errors underestimating variability. Generally speaking, coverage rates for the interval estimates are more appropriate for the parametric G-computation methods than for MIM. Where the outcome regression is a generalized linear model, parametric G-computation is easier to implement and has lesser potential complications. In any case, MIM could be useful with different outcome model types. Further research on this method is required, and will focus on developing alternative variance estimators that avoid negative variances and are more conservative.

For the conventional STC, outcome regression may have decreased precision relative to MAIC, as dictated by the empirical standard errors. On the other hand, the marginalized outcome regression methods are more precise than both MAIC and conventional STC. From a frequentist perspective, the standard error of the estimator of a conditional log-odds ratio for $A$ vs.~$C$, targeted by conventional STC, is larger than the standard error of a regression-adjusted estimate of the marginal log-odds ratio for $A$ vs.~$C$, targeted by G-computation and MIM. This precision comparison likely lacks relevance, because one is comparing estimators that target different estimands. Nevertheless, it supports previous findings on non-collapsible measures of effect when adjusting for prognostic covariates.\cite{moore2009covariate, daniel2020making} When we marginalize and compare estimators targeting like-for-like marginal estimands, we find that outcome regression is no longer detrimental for precision and efficiency compared to weighting.  

The robust sandwich variance estimator for MAIC has underestimated variability and produced narrow interval estimates under small effective sample sizes.\cite{remiro2020methods} This simulation study demonstrates that the bootstrap procedure provides more conservative variance estimation in the more extreme settings. This implies that the bootstrap approach should be preferred for statistical inference where there are violations of the overlap assumption and small sample sizes. 

\subsection{Extended discussion}\label{subsec62}

\paragraph{Extrapolation capabilities and precision considerations}

We expect the conclusions in the previous paragraphs to be robust. The number of simulated datasets per scenario (2,000) is large enough so that the outlined performance differences are not due to chance. Nevertheless, we now clarify some aspects of the conclusions that are more nuanced. 

In real applications, the effective sample sizes and percentage reductions in effective sample size may be lower and higher, respectively, than those considered in this simulation study.\cite{phillippo2019population} In these situations, covariate overlap is poor and this leads to a high loss of precision in MAIC. The marginalized outcome regression methods should be considered because they are substantially more statistically efficient. This is particularly the case where the outcome model is a logistic regression, more prone to small-sample bias,\cite{phillippo2020assessing, vittinghoff2007relaxing} imprecision\cite{annesi1989efficiency} and inefficiency\cite{annesi1989efficiency} than other models, e.g.~the Cox regression. In addition, where sample sizes are small and the number of covariates is large, feasible weighting solutions may not exist for MAIC due to separation problems,\cite{jackson2020alternative} as observed in one of the scenarios of this simulation study ($N=200$ with poor overlap) and, notably, in another recent simulation study.\cite{phillippo2020assessing} An advantage of outcome regression is that it can be applied in these settings. MAIC cannot extrapolate beyond the covariate space observed in the IPD. Therefore, it cannot overcome the failure of assumptions that is the lack of covariate overlap and is incapable of producing an estimate.

Moreover, we note that MAIC requires accounting for all effect modifiers (balanced and imbalanced), as excluding balanced covariates from the weighting procedure does not ensure balance after the weighting. On the other hand, outcome regression methods do not necessarily require the inclusion of the effect modifiers that are in balance, for instance when the outcome model is a linear regression. This may mitigate losses of precision further, particularly where the number of potential effect modifiers is large. 

With limited overlap, outcome regression methods can use the linearity assumption to extrapolate beyond the $AC$ population, provided the true relationship between the covariates and the outcome is adequately captured. We view this as a desirable attribute because poor overlap, with small effective sample sizes and large percentage reductions in effective sample size, is a pervasive issue in health technology appraisals.\cite{phillippo2019population} Nevertheless, where overlap is more considerable, one may wish to restrict inferences to the region of overlap and avoid relying on a model for extrapolation outside this region.\cite{ho2007matching,rubin1997estimating} 

Note that the model extrapolation uncertainty is not reflected in the interval estimates for the outcome regression approaches and that some consider weighting approaches to give a ``more honest reflection of the overall uncertainty''.\cite{vansteelandt2011invited} The gain in efficiency produced by outcome regression must be counterbalanced against the potential for model misspecification bias. Weighting methods are often perceived to rely on less demanding parametric assumptions, yet model misspecification is an issue for both methods as we discuss later in this section. 

It is worth noting that we have used the standard MAIC formulation proposed by Signorovitch et al.\cite{signorovitch2010comparative, phillippo2016nice, remiro2020methods,  phillippo2020equivalence} and that our conclusions are based on this approach. Nevertheless, MAIC is a rapidly developing methodology with novel implementations. An alternative formulation based on entropy balancing has been presented.\cite{petto2019alternative,belger2015inclusion, hainmueller2012entropy,phillippo2020equivalence} This approach is similar to the original version with a subtle modification to the weight estimation procedure. While it has some interesting computational properties, Phillippo et al.\cite{phillippo2020equivalence} have recently shown that the standard method of moments and entropy balancing produce weights that are mathematically equivalent (up to optimization error or a normalizing constant). Drawing from Zubizarreta,\cite{zubizarreta2015stable} Jackson et al.\cite{jackson2020alternative} propose a distinct weight estimation procedure that satisfies the conventional method of moments and maximizes the effective sample size. A larger effective sample size translates into minimizing the variance of the weights, with more stable weights producing a gain in precision at the expense of introducing some bias. 

A potential extension to MAIC could involve estimating the treatment mechanism in the $AC$ trial as well as the trial assignment mechanism. In a randomized trial, the treatment assignment mechanism is known --- the true conditional probability of treatment among the randomized subjects is known and, in expectation, independent of the covariates. Nevertheless, modeling this probability, e.g.~using a parametric model, is beneficial to control for random (chance) imbalances in baseline covariates between study arms. In other contexts, this has improved the precision and efficiency of propensity score estimators.\cite{lunceford2004stratification, robins1995semiparametric, hahn1998role, williamson2014variance, robins1994estimation, li2021generalizing} 

% Outcome regression methods, on the other hand, cannot leverage this information to obtain more precise estimates. 

\paragraph{Bayesian modularity}

The marginalized outcome regression methods, particularly the Bayesian approaches, can be readily adapted to address missing values in the $AC$ IPD. As seen in subsection \ref{subsec361}, Bayesian G-computation and the synthesis stage of MIM follow very closely the principles of multiple imputation, which is also, arguably, a fundamentally Bayesian operation. Missing covariates and outcomes in the IPD could be imputed in each MCMC iteration, accounting naturally for the uncertainty in the missing data. Addressing ``missingness'' in the $BC$ study is not possible task without access to the patient-level data. 

Throughout the text, we have made certain assumptions about the covariate distribution in the $BC$ population. We have treated the covariate moments $\boldsymbol{\theta}$ and the correlation information $\boldsymbol{\rho}$ as fixed. The Bayesian frameworks could be extended to account for this additional layer of uncertainty, in the specification of $\boldsymbol{\theta}$ and $\boldsymbol{\rho}$ and also in the selected marginal distribution forms for $BC$. Bayesian regression approaches can also account for other issues such as measurement error in the IPD.\cite{keil2014autism} Bayesian model averaging can be incorporated to capture structural or model uncertainty.\cite{madigan1994model} By drawing outcome predictions under various models, complex relationships in the patient-level data may be reproduced more accurately, offering some protection against parametric model misspecification. 

In the Bayesian procedures, both ``hard'' (e.g.~the results of a meta-analysis) and ``soft'' (e.g.~clinical rationale from experts) evidence can be used to form the prior distributions for the conditional prognostic and interaction effects. The specification of the parametric outcome model requires ``dichotomizing'' whether a variable is an effect modifier or not, i.e., in statistical terms, specifying whether interactions with treatment do or do not exist. Bayesian shrinkage methods allow interactions to be ``half in, half out'' of the model.\cite{dixon1991bayesian, spiegelhalter1994bayesian, simon1997bayesian} For instance, one can specify skeptical or regularization prior distributions for the interaction effects, over all potential candidate effect modifiers. In the words of Simon and Freedman,\cite{simon1997bayesian} this ``encourages the quantification of prior belief about the size of interactions that may exist. Rather than forcing the investigator to adopt one of two extreme positions regarding interactions, it provides for the specification of intermediate positions''. 

\paragraph{Limitations}

Care must be taken where sample sizes are small in population-adjusted indirect comparisons. Low sample sizes cause substantial issues for the accuracy of MAIC due to unstable weights. Also, MIM assumes that the posterior distribution of $\Delta_{10}^{(2)}$ is approximately normal, and low sample sizes may break this normality assumption. As the sponsor company is directly responsible for setting the value of $N$, the $AC$ trial should be as large as possible to maximize precision and accuracy. The sample size requirements for indirect comparisons, and more generally for economic evaluation, are considerably larger than those required to demonstrate an effect for the main clinical outcome in a single RCT. However, trials are usually powered for the main clinical comparison, even if there is a prospective indirect, potentially adjusted, comparison down the line. Ideally, if the manufacturer intends to use standard or population-adjusted indirect comparisons for reimbursement purposes, its clinical study should be powered for the relevant methods.

Note that sponsors tend to run multiple RCTs instead of one larger RCT for marketing authorization applications. If there are many different IPD RCTs, it is necessary to fit the covariate-adjusted regression to each patient-level dataset and marginalize against the $BC$ pseudo-population in G-computation and MIM. Similarly, one would apply MAIC to each study individually, reweighting each patient-level dataset against the $BC$ study report. Then, a meta-analysis of effect measure estimates can be performed in the same population using the marginalized or weighted results from the IPD studies and the original effect estimate published in the ALD study.

The population adjustment methods outlined in this article are only applicable to pairwise indirect comparisons, and not easily generalizable to larger network structures of treatments and studies. This is because the methods have been developed in the two-study scenario seen in this paper, very common in HTA submissions, where there is one $AC$ study with IPD and another $BC$ study with ALD. In this very sparse network, indirect comparisons are vulnerable to bias induced by effect modifier imbalances. In larger networks, multiple pairwise comparisons do not necessarily generate a consistent set of relative effect estimates for all treatments. This is because the comparisons must be undertaken in the ALD populations.  

Another issue is that the ALD population(s) may not correspond precisely to the target population for the decision. Marginal estimands in different populations may not match if there are differences in the distribution of effect modifiers. This is a problem of external validity: if populations are non-exchangeable, an internally valid estimate for the marginal estimand in one population is not necessarily unbiased for the marginal estimand in the other(s).\cite{westreich2019target, imai2008misunderstandings} To address this, one suggestion would be for the decision-maker to define a target population for a specific disease into which all manufacturers should conduct their indirect comparisons. The outcome regression approaches discussed in this article could be applied to produce marginal effects in any target population. The target could be represented by the joint covariate distribution of a registry, cohort study or some other observational dataset, and one would marginalize over this distribution. Similarly, MAIC can reweight the IPD with respect to a different population than that of the $BC$ study.

Recently, a novel population adjustment method named multilevel network meta-regression (ML-NMR) has been introduced.\cite{phillippo2020multilevel, phillippo2019calibration} ML-NMR generalizes IPD network meta-regression\cite{berlin2002individual} to include aggregate-level data, reducing to this method when IPD are available for all studies. ML-NMR is a timely addition; it is applicable in treatment networks of any size with the aforementioned two-study scenario as a special case. This is important because a recent review\cite{phillippo2019population} finds that 56\% of NICE technology appraisals include larger networks, where the methods discussed in this article cannot be readily applied. 

ML-NMR is an outcome regression approach, with the outcome model of interest being identical to that of parametric G-computation and MIM. While the methods share the same assumptions in the two-study scenario, ML-NMR generalizes the regression to handle larger networks. Like Bayesian G-computation and MIM, ML-NMR has been developed under a Bayesian framework and estimates the outcome model using MCMC. It also makes parametric assumptions to characterize the marginal covariate distributions in $BC$ and reconstructs the joint covariate distribution using a copula. The methods average over the $BC$ population in different ways; Bayesian G-computation and MIM simulate individual-level covariates from their approximate joint distribution and ML-NMR uses numerical integration over the approximate joint distribution (quasi-Monte Carlo methods). 

In its original publication,\cite{phillippo2020multilevel, phillippo2019calibration} ML-NMR targets a conditional treatment effect (avoiding the compatibility issues of conventional STC), because the effect estimate is derived from the treatment coefficient of a covariate-adjusted multivariable regression. However, ML-NMR can directly calculate marginalization integrals akin to those required for Bayesian G-computation and MIM (Equations \ref{eqn14} and \ref{eqn15}). Phillippo et al. have recently demonstrated that ML-NMR can be adapted to target marginal treatment effects.\cite{phillippo2021target} We previously mentioned that pairwise population-adjusted indirect comparisons target marginal estimands that are specific to the $BC$ study. These may not be directly relevant for HTA decision-making. On the other hand, ML-NMR can potentially estimate marginal effects in any target population, presenting novel and exciting opportunities for evidence synthesis. 

\paragraph{Method assumptions}

Population-adjusted indirect comparisons mostly depend on the same assumptions (detailed in \hyperref[SA]{Supplementary Appendix A}) including: (i) internal validity of the $AC$ and $BC$ trials, (ii) consistency under parallel studies, (iii) accounting for all effect modifiers of treatment $A$ vs.~$C$ in the adjustment (i.e., the conditional constancy of the $A$ vs.~$C$ marginal treatment effect or the conditional ignorability, unconfoundedness or exchangeability of trial assignment/selection for such treatment effect), (iv) that there is overlap between the covariate distributions in $AC$ and $BC$ (more specifically, that the ranges of the selected covariates in the $AC$ trial cover some of their respective ranges in the $BC$ population), (v) that the joint covariate distribution of the $BC$ population has been correctly specified, (vi) and parametric modeling assumptions. 

% A final assumption, the shared effect modifier assumption (described in subsection \ref{subsec43}), is required to transport the marginal treatment effect estimate from the $BC$ population to any given target population. Otherwise, one has to assume that the $BC$ population is the target for the analysis to be valid.

Assumptions (i) and (ii) are made by any indirect treatment comparison or meta-analysis. The other, largely untestable, assumptions are unique to population-adjusted analyses and their violation may lead to bias. The most crucial assumptions underlying population-adjusted indirect comparisons relate to the correct specification of the trial assignment logistic regression (in the case of MAIC), and of the covariate-adjusted outcome regression (in the case of conventional STC, parametric G-computation and MIM). 

In practice, there will be model misspecification if there is incomplete information on effect modifiers for one or both of the trials. Conditional exchangeability (``no omitted effect modifiers'') is a fundamental assumption for all methods. However, it is not directly testable with the available data due to the lack of additional individual-level outcome information for the $BC$ study.\cite{hartman2015sample} In collaboration with clinical experts, the most plausible effect modifiers should be selected for the base-case analysis. Nevertheless, the effect modifier status of covariates is difficult to ascertain, particularly for novel treatments with limited prior empirical evidence and clinical domain knowledge.\cite{remiro2020principled} Therefore, we will never be completely certain that all effect modifiers have been accounted for, or of the validity of the population adjustment. 

Consequently, sensitivity analyses are warranted under alternative model specifications to explore the dependence of inferences on the model and the robustness of results.\cite{crosensitivity, nguyen2017sensitivity, dahabreh2019extending} In the context of ``generalizability'', Nguyen et al.\cite{nguyen2017sensitivity} have recently developed an approach for sensitivity analysis. This is applicable where potential effect modifiers are measured only in the $AC$ trial but not in the $BC$ study, given some assumptions about the missing effect modifiers. Dahabreh\cite{dahabreh2020extending} proposes a bias function strategy for sensitivity analyses, which does not require individual-level information on unobserved effect modifiers. Further research should adapt this recent work to our ``limited patient-level data'' setup.  

Parametric modeling assumptions will not hold under incorrect model specification, e.g.~in the outcome regression methods, if only linear relationships are considered and the selected covariates have non-linear interactions with treatment on the linear predictor scale. This simulation study only considers a best-case scenario with correct parametric model specification. To predict the outcomes, we use the logistic regression model implied by the data-generating mechanism. Similarly, the model for estimating the weights in MAIC is the best-case model because the right subset of covariates has been selected as effect modifiers and the balancing property holds for the weights with respect to the effect modifier means, as mentioned in subsection \ref{subsec441}.\footnote{The MAIC implementation is optimal in terms of precision and accuracy because the trial assignment model only balances the two covariates that interact with treatment. Nevertheless, these are not the only two covariates that are associated with trial assignment. Consider balancing the full set of covariates that predict trial assignment (a total of four covariates, including the two predictors with only main effects in the data-generating outcome model). Variance would be increased without improving the potential for bias reduction in the $BC$ population. The behavior of MAIC would be more unstable because of weaker overlap. More extreme weights would be produced, and finite-sample or ``chance'' overlap violations would be more likely, particularly with small $AC$ sample sizes.} Also, effect modification has been correctly specified as linear, but scale conflicts would arise if effect modification status, which is scale-specific, had been justified on the wrong scale, e.g.~if the true treatment effect modification were non-linear or multiplicative, e.g.~age in cardiovascular disease treatments.

In real applications, these modeling assumptions are difficult to hold because, unlike in simulations, the correct specification is unknown, particularly where there are a large number of covariates and complex relationships exist between them. The simulation study presented in this article demonstrates proof-of-concept for the outcome regression methods and for MAIC, but does not investigate how robust the methods are to failures in assumptions. Future simulation studies should explore performance in scenarios where assumptions are violated, in order to draw more accurate conclusions with respect to practical applications and limitations. 

The general-purpose nature of the methods presented in this article may provide some degree of robustness against model misspecification because the covariate-adjusted outcome model does not necessarily need to be parametric. Non-parametric regression techniques or other data-adaptive estimation approaches can be used to detect (higher-order) interactions, product terms and non-linear relationships, offering more flexible functions to predict the conditional outcome expectations. These may enhance the likelihood of correct model specification with respect to parametric regressions, but are prone to overfitting, particularly with small sample sizes. They can also minimize ``data snooping'' problems (e.g.~the analyst selecting the model specification or the effect modifiers on the basis of statistically significant treatment effects), specially when there are no clear hypotheses about effect modification ex ante. 

\paragraph{Specification of the $BC$ population}

Population-adjusted indirect comparisons make certain assumptions to approximate the joint distribution of covariates in the $BC$ trial, but these assumptions differ slightly. In MAIC, as stated in the NICE Decision Support Unit technical support document,\cite{phillippo2016nice} ``when covariate correlations are not available from the ($BC$) population, and therefore cannot be balanced by inclusion in the weighting model, they are assumed to be equal to the correlations amongst covariates in the pseudo-population formed by weighting the ($AC$) population.'' In the conventional version of STC, the correlations between the $BC$ covariates are assumed to be equal to the correlations between covariates in the $AC$ trial. 

In the marginalization methods proposed in this article (parametric G-computation and MIM), more explicit and stringent distributional assumptions are made in the ``covariate simulation'' step. The methods assume the joint distribution of the $BC$ covariates is specified correctly, by the combination of the specified marginal distributions and correlation structure. In the simulation study, we have assumed that the pairwise correlations of the covariates and the parametric forms of their marginal distributions are identical across trials. It is important to assess the robustness of the methods to failures in these distributional assumptions. 

Note that the covariate distributional assumptions could be relaxed or verified empirically if trial publications included more complete summary statistics, e.g.~information on the covariates' correlation structure or their observed marginal distributions, as opposed to simple summary tables of means/proportions and standard deviations. This information would allow us to approximate the full joint distribution of the $BC$ covariates more accurately and reduce the risk of misspecifying the $BC$ population. We have decided to mimic the $AC$ pairwise correlations as, in principle, the relationships between covariates should be similar across trials. 

% By defining relationships between covariates, we elicit prior beliefs irrespective of the outcomes of each treatment. 

% Nevertheless, we suspect that these assumptions are not problematic in most cases.

% It is worth noting that this set of assumptions only induces bias in the methods if higher-order interactions (involving two or more covariates) are unaccounted for or misspecified. If the inclusion of these interactions is not explored in the weighting model for MAIC or in the covariate-adjusted regression for the outcome modeling approaches, the specification of pairwise correlations will not make a difference in terms of bias, as observed in a recent simulation study that investigates this set of assumptions.\cite{phillippo2020multilevel} 

\paragraph{Concluding remarks}

The traditional regression adjustment approach in population-adjusted indirect comparisons targets a conditional treatment effect for $A$ vs.~$C$. We have showed empirically that this effect is incompatible in the indirect treatment comparison, producing biased estimation where the measure of effect is non-collapsible. In addition, this effect is not of interest in our scenario because we seek marginal effects for policy decisions at the population level. We have proposed several approaches for marginalizing the conditional estimates produced by covariate-adjusted regressions. The procedures are applicable to a wide range of outcome models and target marginal treatment effects for $A$ vs.~$C$ that have no compatibility issues in the indirect treatment comparison. 

We have demonstrated that the novel marginalized outcome regression approaches achieve greater precision than MAIC and are unbiased under no failures of assumptions. Hence, the development of these approaches is appealing and impactful. As observed in the simulation study, these methodologies are more efficient than weighting, providing more stable estimators. We can now capitalize on the advantages offered by outcome regression with respect to weighting in our scenario, e.g.~extrapolation capabilities and increased statistical precision. Furthermore, we have shown that the marginalized regression-adjusted estimates provide greater statistical precision than the conditional estimates produced by the conventional version of STC. While this precision comparison is irrelevant, because it is made for estimators of different estimands, it supports previous research on non-collapsible measures of effect.\cite{moore2009covariate, daniel2020making} 

Marginal and conditional effects are regularly conflated in the literature on population-adjusted indirect comparisons, with many simulation studies comparing the bias, precision and efficiency of estimators of different effect measures. The implications of this conflation must be acknowledged in order to provide meaningful comparisons of methods. We have built on previous research conducted by the original authors of STC, who have also suggested the use of a preliminary ``covariate simulation'' step.\cite{ishak2015simulated, ishak2015simulation} Nevertheless, up until now, there was no consensus on how to marginalize the conditional effect estimates. For instance, in a previous simulation study,\cite{remiro2020methods} we discouraged the ``covariate simulation'' approach when attempting to marginalize on the linear predictor scale. Averaging on the linear predictor scale, i.e., computing the conditional linear prediction under each treatment for every simulated subject and averaging the linear predictions across all subjects, then calculating the difference between the average predictions, reduces to the conventional version of STC (i.e., to ``plugging in'' the mean $BC$ covariate values). It is equivalent to averaging ``predictions at the mean''\cite{bartlett2018covariate} or estimating the ``mean at mean covariates''\cite{qu2015estimation} (as discussed in subsection \ref{subsec341}), hence producing conditional effect estimates for $A$ vs.~$C$, as opposed to marginal estimates. We hope to have established some clarity. 

The presented marginalization methods have been developed in a very specific context, common in HTA, where access to patient-level data is limited and an indirect comparison is required. However, their principles are applicable to estimate marginal treatment effects in any situation. For instance, in scenarios which require marginalizing out regression-adjusted estimates over the study sample in which they have been computed. Alternatively, the frameworks can be used to transport the results of a randomized experiment to any other external target population; not necessarily that of the $BC$ trial. In both cases, the required assumptions are weaker than those required for population-adjusted indirect comparisons. 

Finally, we have assumed that the $AC$ study is a randomized trial. Our approach to parametric G-computation could be extended to the situation where the $AC$ trial is an observational study by including all confounders of the treatment-outcome relationship in the outcome model. In this scenario, one must overcome the limited internal validity of the study design. Because treatment assignment is non-random, additional assumptions would be required, e.g. conditional exchangeability within the study arms (``no unmeasured confounding'') and the associated overlap/positivity condition.\cite{faria2015nice, robins2009estimation} These assumptions are similar to those discussed in subsection \ref{subsec22} but would be expected to hold across treatment arms in the IPD study in addition to across study populations.

\section*{Acknowledgments}

The authors thank Andrea Gabrio for discussions that contributed to the quality of the manuscript. The authors acknowledge his advice and expertise in missing data. In addition, the authors thank the peer reviewers of a previous article of theirs.\cite{remiro2020methods} Their comments were extremely insightful and helped improved the underlying motivation of this article. The authors are hugely thankful to Tim Morris, whose comments on G-computation and marginalization helped motivate the article. Finally, the authors are grateful to David Phillippo, who has provided very valuable feedback to our work, and helped in substantially improving our research. This article is based on research supported by Antonio Remiro-Az\'ocar's PhD scholarship from the Engineering and Physical Sciences Research Council of the United Kingdom. Anna Heath is supported by the Canada Research Chair in Statistical Trial Design and funded by the Discovery Grant Program of the Natural Sciences and Engineering Research Council of Canada (RGPIN-2021-03366). An earlier and more rudimentary version of this research\cite{azocar2019pns311} was presented in ISPOR Europe 2019, which took place 2-6 November 2019 in Copenhagen, Denmark. 

\subsection*{Financial disclosure}

Funding agreements ensure the authors’ independence in developing the methodology, designing the simulation study, interpreting the results, writing, and publishing the article.

\subsection*{Conflict of interest}

The authors declare no potential conflict of interests.

\subsection*{Data Availability Statement}

The files required to generate the data, run the simulations, and reproduce the results are available at \url{http://github.com/remiroazocar/marginalized_indirect_comparisons_simstudy}.

\clearpage

\section*{Supplementary Appendix A: Method assumptions}\label{SA}
\addcontentsline{toc}{section}{Supplementary Appendix A: Method assumptions}

Indirect comparisons of treatments seek to mimic the analysis that would be conducted in a head-to-head RCT and to recover the causal effect of treatment on the clinical outcome of interest. In our particular case, the term ``causal'' also alludes to the need to control for effect modification, an inherently causal concept. Hence, we base our discussion of assumptions on the ideas underlying the Neyman-Rubin Model for causal treatment effects. This was originally suggested by Neyman\cite{neyman1923application} in experiments with randomization-based inference, with extensions to observational studies later introduced by Rubin.\cite{imbens2015causal, rubin1978bayesian, rubin2005causal} The central concept of this general framework is that of a potential outcomes approach to causal inference. Note that this discussion is non-technical and detailed theory and notation based on potential outcomes are not presented.

MAIC and the outcome regression approaches share the following assumptions, required to make valid causal inferences in the $BC$ population: (1) internal validity; (2) consistency under parallel studies; (3) conditional strong ignorability of trial assignment for the $A$ vs.~$C$ treatment effect (this requires both the conditional constancy of relative effects and overlap/positivity across the covariate distributions); (4) correct specification of the $BC$ population; and (5) (typically parametric) modeling assumptions. The first two assumptions are made by any indirect treatment comparison or meta-analysis. Assumptions that are not specific to indirect treatment comparisons, e.g.~those that are specific to the type of regression model used, such as proportional hazards or non-informative censoring for a Cox regression, are not discussed.

Note that the following assumptions can only guarantee a valid indirect comparison if the within-trial relative effects target compatible estimands of the same type. The majority of RCTs publish an estimate for $B$ vs.~$C$ that targets a marginal treatment effect (any published conditional treatment effect is likely incompatible with that for $A$ vs.~$C$). Therefore, population adjustment methods should target a marginal treatment effect for $A$ vs.~$C$. If a comparison of conditional treatment effects is performed, these would have to be adjusted across identical sets of covariates, using the same model specification.

Those studying the generalizability of treatment effects often make a distinction between sample-average and population-average marginal effects.\cite{stuart2011use, cole2010generalizing, kern2016assessing, hartman2015sample} Typically, another implicit assumption made by population-adjusted indirect comparisons is that the marginal treatment effects estimated in the $BC$ sample, as described by its published covariate moments in the case of the $A$ vs.~$C$ treatment effect, coincide with those that would be estimated in the target population of the trial. Namely, either the study sample on which inferences are made is the study target population, or it is a simple random sample (i.e., representative) of such population, ignoring sampling variability. 

Furthermore, when referring to ``effect modifiers'', we describe the covariates modifying the treatment effect measure for $A$ vs.~$C$ in the linear predictor scale. We select the effect modifiers of treatment $A$ with respect to $C$ (as opposed to the effect modifiers of treatment $B$ with respect to $C$), because we have to adjust for these to perform the indirect comparison in the $BC$ population, implicitly assumed to be the target population. If we had IPD for the $BC$ study and ALD for the $AC$ study, we would have to account for the covariates that modify the effect of treatment $B$ vs.~$C$, in order to perform the comparison in the $AC$ population.

\subsection*{Internal validity}

The first set of assumptions relates to the internal validity of the $AC$ and $BC$ trials. The trials are internally valid under the following structural assumptions, which are necessary for causal inference:

\begin{itemize}
\item Stable unit treatment value assignment (SUTVA). This assumption implies that: (1) the treatment of a given subject does not affect the potential outcomes of other individuals (non-interference);\cite{rubin1980randomization, hudgens2008toward} and (2) there is only one version of each treatment (treatment-variation irrelevance),\cite{vanderweele2009concerning} implying that the treatment is comparable across units.\cite{vanderweele2013causal} The first condition is questionable, for example, in a vaccine trial, where the outcome of an individual (i.e., developing the flu) depends on the vaccination status of others because of herd immunity. The second condition is questionable if there are differences among versions of treatment, e.g.~in the delivery mechanism, that are relevant to the outcome of interest. 
\item Strongly ignorable treatment assignment. Ignorability implies that treatment assignment is independent of the potential outcomes.\cite{rosenbaum1983central} Ignorability can be conditional on the observed baseline covariates or unconditional. Conditional ignorability is strong when there is positivity or overlap,\cite{hernan2006estimating} i.e., any subject has a positive probability of being assigned to either treatment group given the baseline covariates. 
\end{itemize}

The SUTVA assumption is met by appropriate study design.\cite{rubin2005causal} By design, the conditions of positivity\cite{cole2008constructing} and ignorability,\cite{greenland2009identifiability}, whether this is plain or conditional on baseline characteristics, are met by randomized trials. The random allocation of treatment ensures that, on expectation, there are no systematic differences in the distribution of (measured and unmeasured) baseline covariates between treatment groups, i.e., there is covariate balance.\cite{austin2013performance, greenland1990randomization} Note that balance is a large sample property. In small samples, one may still observe modest residual differences in baseline characteristics. As formulated by Senn,\cite{senn1994testing} in a RCT, over all the randomizations the groups are balanced, but for a particular randomization they may be unbalanced.

Therefore, the internal validity assumptions are met if the $AC$ and $BC$ studies are appropriately designed trials with appropriate randomization and reasonably large sample sizes. Finally, we have assumed that internal validity in each trial is not compromised by other issues, such that there is negligible measurement error or missing data, the absence of non-compliance, etc.

\subsection*{Consistency under parallel studies}

Consistency under parallel studies\cite{hartman2015sample} is the cross-trial version of the second condition of SUTVA (treatment-variation irrelevance). This assumption implies that potential outcomes for an individual under a given treatment are homogeneous regardless of the study assigned to the individual. For instance, treatment $C$ should be administered in the same setting in both trials, or differences in the nature of treatment, e.g.~in the clinical protocol or delivery mechanism, should not change its effect. In there are non-negligible differences in the versions of treatment, for instance, if treatment $C$ is accompanied by adherence counseling in one of the trials, while such counseling is absent in the other, this assumption could be invalid. 

Consistency under parallel studies means that population adjustment methods cannot adjust for cross-trial differences related to the nature of treatments, e.g.~treatment administration, switching, dosing formulation, titration, or co-treatments. Differences of this type are perfectly confounded with treatment,\cite{phillippo2016nice} and MAIC and the outcome regression methods can only adjust for differences in trial population characteristics. This assumption is required to perform a valid indirect comparison across studies.

\subsection*{Conditionally strong ignorability of trial assignment}

Strongly ignorable trial assignment (specifically, assignment to the $AC$ trial), conditional on the selected covariates, is the primary assumption underlying population-adjusted indirect comparisons and is required for unbiased estimation of $\Delta_{10}^{(2)}$. This is akin to the strongly ignorable sample or trial assignment assumption\cite{hartman2015sample} commonly used in the generalizability, transportability or external validity literature.\cite{hartman2015sample, cole2010generalizing, stuart2011use,kern2016assessing} This literature seeks to calibrate relative treatment effects obtained from a RCT into a, more diverse, target population. In MAIC and the discussed outcome regression methods, the indirect comparison is performed in the $BC$ population, and the $A$ vs.~$C$ treatment effect is transported to the $BC$ population. (Conditionally) strong ignorability consists of two assumptions: (conditional) ignorability and overlap (or positivity). Note that, even though strong ignorability has been proposed in the context of propensity score modeling, it is also a crucial assumption for the causal interpretation of outcome regression results in the $BC$ population.

\subsubsection*{Conditional ignorability}

There are many ways to articulate this assumption. One can consider that trial assignment/selection is conditionally ignorable, unconfounded or exchangeable for the $A$ vs.~$C$ treatment effect (the potential $A$ vs.~$C$ relative outcomes), i.e., conditionally independent of the treatment effect, given the selected effect modifiers. This means that after adjusting for these effect modifiers, treatment effect heterogeneity and trial assignment are conditionally independent. The NICE technical support document\cite{phillippo2016nice, phillippo2018methods} describes this assumption as the conditional constancy of relative effects across populations (namely, given the selected effect-modifying covariates, the $A$ vs.~$C$ treatment effect is constant across populations).

MAIC will only meet conditional ignorability if \textit{all} (observed or unobserved) effect modifiers are accounted for, regardless of whether these are balanced before the weighting. Excluding balanced covariates from the weighting procedure does not ensure balance after the weighting. The outcome regression methods meet conditional ignorability if all \textit{imbalanced} effect modifiers are accounted for in the covariate-adjusted regression model (in the case of multiple imputation marginalization, that is the first-stage regression).

This is a demanding assumption in practice, which is also untestable. On one hand, it is tied to the measure used to define the treatment effects and effect modifiers. Most crucially, ignorability is hard to meet because it requires complete information on all treatment effect modifiers to be measured and available across trials $AC$ and $BC$, and for all effect modifiers to be accounted for by the analyst. Firstly, it is conceivable that information on some effect modifiers is unavailable or unpublished in one or both studies. Secondly, the analyst may select the effect modifiers incorrectly. It is generally difficult to ascertain the effect modifier status of variables, particularly for new treatments with limited prior empirical evidence and clinical domain knowledge. We can never eliminate the possibility that this assumption is broken, as we cannot guarantee that there are no unobserved or unmeasured effect modifiers. Nevertheless, the careful selection of effect modifiers\cite{remiro2020principled} from the observed baseline covariates is within the investigator's control and can provide some protection. Overspecification of effect modifiers should not bias the comparison but may inflate standard errors and lead to a subsequent loss of precision.\cite{phillippo2016nice} 

Effect modifier status is often determined by carrying out subgroup analyses in the IPD, or by examining statistical covariate-treatment interactions in outcome regressions fitted to the IPD.\cite{rothman1980concepts, dahabreh2016using, dahabreh2017heterogeneity} The latter is a preferred approach and the former is typically  discouraged.\cite{remiro2020principled} Non-parametric tree-based regressions have recently been used for this purpose. These are appealing because they are data-driven and can detect interactions without pre-specifying which candidate variables to include in the model.\cite{steingrimsson2019subgroup, yang2021causal} Nevertheless, all statistical approaches are hindered by the lack of power of individual RCTs to identify interactions.\cite{remiro2020methods}

\subsubsection*{Overlap}

Conditional ignorability of trial assignment is strong if there is positivity or overlap, i.e., if every subject in the $BC$ population has a positive probability of being assigned to the $AC$ trial given the covariates accounted for in the adjustment mechanism. This implies that the ranges of the covariates in the $BC$ population are covered by their respective ranges in the $AC$ trial. This assumption may pose a problem if the inclusion/exclusion criteria of $AC$ and $BC$ are inconsistent. For instance, consider a situation where age is selected as an effect modifier and the age ranges of trial $AC$ and trial $BC$ are 60-70 and 40-70, respectively. There exists a subpopulation (age 40-60) in $BC$ that does not overlap with the $AC$ population. Hence, the $AC$ study provides no evidence about the treatment effect and treatment effect modification in the excluded age group, and the $A$ vs.~$C$ treatment effect estimate may be biased in the full comparator trial population (ages 40-70). 

In such cases, reweighting methods like MAIC cannot extrapolate beyond the observed covariate space in the $AC$ IPD, as there are no subjects to reweight. Where overlap is insufficient, outcome regression methods can extrapolate beyond the $AC$ population, using the linearity assumption or other appropriate assumptions about the input space. However, valid extrapolation requires accurately capturing the true relationship between the covariates and the outcome. 
Conversely, the exclusion of patients enrolled in $AC$ from the $BC$ population, e.g.~if the $AC$ population is more diverse, does not necessarily violate the overlap assumption. This is because these methods deliver estimates in the $BC$ population. Hence, adjustment in this scenario is an interpolation as opposed to an extrapolation of the observed $AC$ data. In this scenario, $\hat{\Delta}_{10}^{(2)}$ may be unbiased because the $BC$ population is covered within that of $AC$. In MAIC, the excluded subpopulation will receive very low weights (low odds of enrolment in $BC$ vs.~$AC$), while the included subpopulation receives high weights and dominates the reweighted sample. These extreme weights lead to large reductions in effective sample size and to the deterioration of precision and efficiency. Removing observations from the $AC$ patient-level data, so that inclusion/exclusion criteria are consistent, explicitly lowers the $AC$ sample size and may degrade precision further. Of course, when there is no interpolation or extrapolation overlap whatsoever, MAIC cannot generate population-adjusted estimate for the treatment effect, as a feasible weighting solution does not exist due to separation problems.\cite{jackson2020alternative}  

If IPD were available for the $BC$ study, the overlap assumption could be easily checked by visualizing the ranges of the selected covariates and their empirical distributions. However, this is challenging in our setup without further distributional assumptions due to patient-level data limitations for $BC$. 

\subsection*{Specification of the joint covariate distribution in $BC$}

Population-adjusted indirect comparisons make certain assumptions to approximate the joint distribution of covariates in the $BC$ trial. The restriction of limited IPD makes it unlikely that such joint distribution is available. Summary statistics for the marginal distributions are typically published instead. Where no correlation information is available for the $BC$ study, MAIC and the conventional centered version of STC seem to assume that the joint $BC$ covariate distribution is the product of the published marginal distributions. The implicit assumptions are, in fact, more nuanced and differ slightly between methods. 

In MAIC, as stated in the NICE Decision Support Unit,\cite{phillippo2016nice} ``when covariate correlations are not available from the ($BC$) population, and therefore cannot be balanced by inclusion in the weighting model, they are assumed to be equal to the correlations amongst covariates in the pseudo-population formed by weighting the ($AC$) population.'' In typical usage, MAIC only balances the marginal distributions of the selected baseline covariates, not the multidimensional joint covariate distributions, due to the lack of published correlation data for $BC$. In the typical usage of STC (i.e., the ``plug-in'' approach to the method), the assumption differs slightly. The correlations between the $BC$ covariates are assumed to be equal to the correlations between covariates in the $AC$ study.

In the novel outcome regression methods discussed in this article (parametric G-computation and MIM), more explicit distributional assumptions are made to characterize the $BC$ population in the ``covariate simulation'' step. The methods assume the joint distribution of the $BC$ covariates is specified correctly, by the combination of the specified marginal distributions and correlation structure. In the simulation study, pseudo-populations are constructed under certain parametric assumptions. We have assumed that the pairwise correlations of the covariates and the parametric forms of their marginal distributions are identical across trials, because the correlation structure observed in the $AC$ IPD is used in the ``covariate simulation'' step. These assumptions cannot be verified empirically as we have no information on the covariates' correlation structure and true marginal distributions. Information on correlations or on the joint distribution of covariates for $BC$ is rarely published, but could be requested. We have decided to mimic the $AC$ pairwise correlations as, in principle, the relationships between covariates should be similar across trials. 

% Nevertheless, we suspect that these assumptions are not problematic in most cases. 

% Furthermore, suspected cross-trial differences in dependency structures may be due to reasons which discourage population adjustment in the first place, such as differences in study design. The leeway in this modeling decision is less vulnerable to the investigator bias that characterizes other steps of population adjustment. By defining relationships between covariates, we elicit prior beliefs irrespective of the outcomes of each treatment. On the other hand, decisions such as the selection of effect modifiers are more difficult to ascertain and directly define covariate-outcome relationships. 

% In an anchored comparison, only effect-modifying covariates need balancing, so the assumption can be relaxed to only include effect modifiers. This set of assumptions only induces bias if higher-order interactions (involving two or more covariates) are unaccounted for or misspecified. If the inclusion of these interactions is not explored in the weighting model for MAIC or in the covariate-adjusted regression for the outcome modeling approaches, the specification of pairwise correlations will not make a difference in terms of bias, as observed in a recent simulation study that investigates this set of assumptions.\cite{phillippo2020multilevel} 

\subsection*{Modeling assumptions}

Indirect treatment comparisons are typically conducted on the linear predictor scale,\cite{phillippo2016nice} upon which the treatment effect is assumed to be additive for all indirect comparisons. In the main text, the anchored population-adjusted indirect comparisons have additionally assumed that the effect modifiers have been defined on the linear predictor scale and are additive on this scale, but the linearity assumption is not always appropriate. Hence, all population adjustment methods are subject to scale conflicts or to bias if effect modification status, which is scale-specific, has been justified on the wrong scale, e.g.~when treatment effect modification is specified as linear but is non-linear or multiplicative, e.g.~age in cardiovascular disease treatments. 

This form of model misspecification is more evident in the outcome modeling approaches, where an explicit outcome regression is formulated. The parametric model depends on functional form assumptions that will be violated if the relationship between the covariates and the outcome is not captured correctly, in which case the methods may be biased. Even though the logistic regression model for the weights in MAIC does not make reference to the outcome, the method is also susceptible to model misspecification bias, albeit in a more implicit form. The model for estimating the weights is approximately correct in the simulation study because the right subset of covariates has been selected as effect modifiers and the balancing property holds for the weights, as mentioned in subsection \ref{subsec441}. In practice, the model will be incorrectly specified if this is not the case, potentially leading to a biased estimate. Note that, in practice, we find that it may be more difficult to specify a correct parametric model for the outcome than an approximately correct parametric model for the trial assignment weights.  

% \subsection*{Shared effect modification}

% The methods discussed in this article only produce an estimate of the treatment effect for $A$ vs.~$B$ that is valid in the $BC$ population, implicitly assumed to be the target population. This may differ from the target population for the decision. An additional assumption, the shared effect modifier assumption,\cite{phillippo2016nice} can be made to transport the $A$ vs.~$B$ treatment effect to any other target population. Otherwise, the analysis is only valid if the target population of the comparison is the $BC$ population.

% This assumption is untestable with the available data and implies that: (1) active treatments $A$ and $B$ have the same set of treatment effect modifiers (with respect to $C$); and (2) the interaction effects of each effect modifier are identical for both treatments. Then, the treatment effect estimate can be generalized to any given target population as effect modifiers are guaranteed to cancel out (relative effects for $A$ vs. $B$ are conditionally constant across all populations). 

% Shared effect modification is hard to meet in practice if the competing interventions do not belong to the same class, and have dissimilar mechanisms of action or clinical properties. In that case, there is little to suggest that treatments $A$ and $B$ have the same set of effect-modifying covariates and that these interact with active treatment in the same way in both trials. 

\subsection*{Concluding remarks}

In practice, some of the assumptions above may be hard to meet. If these are violated, the resulting treatment effect may be biased. Hence, it is important to assess the robustness of the methods to failures of assumptions and under different degrees of model misspecification in future simulation studies.

\clearpage

\section*{Supplementary Appendix B: Example code}\label{SB}
\addcontentsline{toc}{section}{Supplementary Appendix B: Example code}

Example \texttt{R} code implementing MAIC, the conventional STC, maximum-likelihood parametric G-computation, Bayesian parametric G-computation and MIM on a simulated dataset is provided below. The code and data are available at \url{https://github.com/remiroazocar/marginalized_indirect_comparisons_simstudy} in the \texttt{Example} subdirectory. Full code for implementing the simulation study is available in the online repository. 

The simulation study and the provided example use binary outcomes and a logistic regression outcome model. Nevertheless, all methods are general-purpose frameworks that, under a generalized linear modeling formulation, can be easily adapted to different outcome models, outcome types, and scalar measures of treatment effect. The code below can be altered by changing the link function in the outcome model. For instance: (1) for a normal linear regression, by setting \texttt{family=gaussian} in the arguments to the \texttt{glm} (or \texttt{stanglm}) function, such that the link is the identity function (for the weighted outcome model, in the case of MAIC, and for the first- and second-stage regressions, in the case of MIM); (2) for a Gamma regression, set \texttt{family=Gamma}, and, for parametric G-computation, transform the predicted marginal outcome means to the linear predictor scale using the ``negative inverse'' link (\(g(\mu)=-\mu^{-1}\), for outcome mean $\mu$); (3) for a Poisson regression, set \texttt{family=poisson}, and, for parametric G-computation, transform the marginal outcome means to the linear predictor scale using the log link (\(g(\mu)=\ln(\mu)\)); and (4) for an inverse Gaussian regression, set \texttt{family=inverse.gaussian}, and, for parametric G-computation, transform the marginal outcome means to the linear predictor scale using the ``inverse squared'' link (\(g(\mu)=\mu^{-2}\)).

\subsection*{MAIC}

\begin{lstlisting}
library("boot") # for non-parametric bootstrap

AC.IPD <- read.csv("Example/AC_IPD.csv") # load AC patient-level data
BC.ALD <- read.csv("Example/BC_ALD.csv") # load BC aggregate-level data

set.seed(555) # set seed for reproducibility

# objective function to be minimized for standard method of moments
Q <- function(alpha, X.EM) {
  return(sum(exp(X.EM %*% alpha)))
}

# function to be bootstrapped
maic.boot <- function(data, indices) {
  dat <- data[indices,] # AC bootstrap sample
  N <- nrow(dat) # number of subjects in sample
  x.EM <- dat[,c("X1","X2")] # AC effect modifiers 
  # BC effect modifier means, assumed fixed
  theta <- BC.ALD[c("mean.X1", "mean.X2")] 
  K.EM <- ncol(x.EM) # number of effect modifiers 
  # center the AC effect modifiers on the BC means
  x.EM$X1 <- x.EM$X1 - theta$mean.X1
  x.EM$X2 <- x.EM$X2 - theta$mean.X2
  # MAIC weight estimation using method of moments
  alpha <- rep(1,K.EM) # arbitrary starting point for the optimizer
  # objective function minimized using BFGS
  Q.min <- optim(fn=Q, X.EM=as.matrix(x.EM), par=alpha, method="BFGS")
  # finite solution is the logistic regression parameters
  hat.alpha <- Q.min$par 
  log.hat.w <- rep(0, N)
  for (k in 1:K.EM) {
    log.hat.w <- log.hat.w + hat.alpha[k]*x.EM[,k]
  }
  hat.w <- exp(log.hat.w) # estimated weights
  # fit weighted logistic regression model using glm
  outcome.fit <- glm(y~trt, family="quasibinomial", weights=hat.w, 
                     data=dat)
  # fitted treatment coefficient is marginal effect for A vs. C
  hat.Delta.AC <- coef(outcome.fit)["trt"] 
  return(hat.Delta.AC)
}

# non-parametric bootstrap with 1000 resamples
boot.object <- boot::boot(data=AC.IPD, statistic=maic.boot, R=1000)
# bootstrap mean of marginal A vs. C treatment effect estimate
hat.Delta.AC <- mean(boot.object$t)
# bootstrap variance of A vs. C treatment effect estimate   
hat.var.Delta.AC <- var(boot.object$t)
# B vs. C marginal treatment effect from reported event counts
hat.Delta.BC <- with(BC.ALD, log(y.B.sum*(N.C-y.C.sum)/
                                   (y.C.sum*(N.B-y.B.sum))))
# B vs. C marginal effect variance using the delta method
hat.var.Delta.BC <- with(BC.ALD, 1/y.C.sum+1/(N.C-y.C.sum)+
                           1/y.B.sum+1/(N.B-y.B.sum))
hat.Delta.AB <- hat.Delta.AC - hat.Delta.BC # A vs. B
hat.var.Delta.AB <- hat.var.Delta.AC + hat.var.Delta.BC
# construct Wald-type normal distribution-based confidence interval
uci.Delta.AB <- hat.Delta.AB + qnorm(0.975)*sqrt(hat.var.Delta.AB)
lci.Delta.AB <- hat.Delta.AB + qnorm(0.025)*sqrt(hat.var.Delta.AB)
\end{lstlisting}

\subsection*{Conventional STC}

\begin{lstlisting}
AC.IPD <- read.csv("Example/AC_IPD.csv") # load AC patient-level data
BC.ALD <- read.csv("Example/BC_ALD.csv") # load BC aggregate-level data

# fit regression model of outcome on treatment and covariates
# IPD effect modifiers centered at the mean BC values 
# purely prognostic variables are included but not centered
outcome.model <- glm(y~X3+X4+trt*I(X1-BC.ALD$mean.X1)+
                       trt*I(X2-BC.ALD$mean.X2),
                     data=AC.IPD, family=binomial)
# fitted treatment coefficient is relative A vs. C conditional effect 
hat.Delta.AC <- coef(outcome.model)["trt"] 
# estimated variance for A vs. C from model fit
hat.var.Delta.AC <- vcov(outcome.model)["trt", "trt"] 
# B vs. C marginal treatment effect estimated from reported event counts
hat.Delta.BC <- with(BC.ALD, log(y.B.sum*(N.C-y.C.sum)/
                                   (y.C.sum*(N.B-y.B.sum))))
# B vs. C marginal treatment effect variance using the delta method 
hat.var.Delta.BC <- with(BC.ALD, 1/y.C.sum+1/(N.C-y.C.sum)+
                           1/y.B.sum+1/(N.B-y.B.sum))
hat.Delta.AB <- hat.Delta.AC - hat.Delta.BC # A vs. B
hat.var.Delta.AB <- hat.var.Delta.AC + hat.var.Delta.BC
# construct Wald-type normal distribution-based confidence interval
uci.Delta.AB <- hat.Delta.AB + qnorm(0.975)*sqrt(hat.var.Delta.AB)
lci.Delta.AB <- hat.Delta.AB + qnorm(0.025)*sqrt(hat.var.Delta.AB)
\end{lstlisting}

\subsection*{Maximum-likelihood parametric G-computation}

\begin{lstlisting}
library("copula") # for simulating BC covariates from Gaussian copula
library("boot") # for non-parametric bootstrap

AC.IPD <- read.csv("Example/AC_IPD.csv") # load AC patient-level data
BC.ALD <- read.csv("Example/BC_ALD.csv") # load BC aggregate-level data

set.seed(555) # set seed for reproducibility

# matrix of pairwise correlations between IPD covariates  
rho <- cor(AC.IPD[,c("X1","X2","X3","X4")]) 
#  covariate simulation for BC trial using copula package
cop <- normalCopula(param=c(rho[1,2],rho[1,3],rho[1,4],rho[2,3],
                            rho[2,4],rho[3,4]), 
                    dim=4, dispstr="un") # AC IPD pairwise correlations
# sample covariates from approximate joint distribution using copula
mvd <- mvdc(copula=cop, margins=c("norm", "norm", # Gaussian marginals
                                  "norm", "norm"), 
            # BC covariate means and standard deviations
            paramMargins=list(list(mean=BC.ALD$mean.X1, sd=BC.ALD$sd.X1),
                              list(mean=BC.ALD$mean.X2, sd=BC.ALD$sd.X2),       
                              list(mean=BC.ALD$mean.X3, sd=BC.ALD$sd.X3),
                              list(mean=BC.ALD$mean.X4, sd=BC.ALD$sd.X4)))
# simulated BC pseudo-population of size 1000
x_star <- as.data.frame(rMvdc(1000, mvd))
colnames(x_star) <- c("X1", "X2", "X3", "X4")
# this function will be bootstrapped
gcomp.ml <- function(data, indices) {
  dat = data[indices,]
  # outcome logistic regression fitted to IPD using maximum likelihood
  outcome.model <- glm(y~X3+X4+trt*X1+trt*X2, data=dat, family=binomial)
  # counterfactual datasets
  data.trtA <- data.trtC <- x_star
  # intervene on treatment while keeping set covariates fixed
  data.trtA$trt <- 1 # dataset where everyone receives treatment A
  data.trtC$trt <- 0 # dataset where all observations receive C
  # predict counterfactual event probs, conditional on treatment/covariates
  hat.mu.A.i <- predict(outcome.model, type="response", newdata=data.trtA)
  hat.mu.C.i <- predict(outcome.model, type="response", newdata=data.trtC)
  hat.mu.A <- mean(hat.mu.A.i) # (marginal) mean probability prediction under A
  hat.mu.C <- mean(hat.mu.C.i) # (marginal) mean probability prediction under C
  # marginal A vs. C log-odds ratio (mean difference in expected log-odds)  
  # estimated by transforming from probability to linear predictor scale 
  hat.Delta.AC <- log(hat.mu.A/(1-hat.mu.A)) - log(hat.mu.C/(1-hat.mu.C))    
  # hat.Delta.AC <- qlogis(hat.mu.A) - qlogis(hat.mu.C) 
  return(hat.Delta.AC)
}  
# non-parametric bootstrap with 1000 resamples
boot.object <- boot::boot(data=AC.IPD, statistic=gcomp.ml, R=1000)
# bootstrap mean of marginal A vs. C treatment effect estimate
hat.Delta.AC <- mean(boot.object$t)
# bootstrap variance of A vs. C treatment effect estimate   
hat.var.Delta.AC <- var(boot.object$t)
# marginal log-odds ratio for B vs. C from reported event counts
hat.Delta.BC <- with(BC.ALD,log(y.B.sum*(N.C-y.C.sum)/
                                  (y.C.sum*(N.B-y.B.sum))))
# variance of B vs. C using delta method
hat.var.Delta.BC <- with(BC.ALD,1/y.C.sum+1/(N.C-y.C.sum)+
                           1/y.B.sum+1/(N.B-y.B.sum))
# marginal treatment effect for A vs. B
hat.Delta.AB <- hat.Delta.AC - hat.Delta.BC 
# variance for A vs. B
hat.var.Delta.AB <- hat.var.Delta.AC + hat.var.Delta.BC 
# construct Wald-type normal distribution-based confidence interval
uci.Delta.AB <- hat.Delta.AB + qnorm(0.975)*sqrt(hat.var.Delta.AB)
lci.Delta.AB <- hat.Delta.AB + qnorm(0.025)*sqrt(hat.var.Delta.AB)
\end{lstlisting}

\subsection*{Bayesian parametric G-computation}

\begin{lstlisting}
library("copula") # for simulating BC covariates from Gaussian copula
# for outcome regression and drawing outcomes from posterior predictive dist.
library("rstanarm") 

AC.IPD <- read.csv("Example/AC_IPD.csv") # load AC patient-level data
BC.ALD <- read.csv("Example/BC_ALD.csv") # load BC aggregate-level data

set.seed(555) # set seed for reproducibility

# matrix of pairwise correlations between IPD covariates  
rho <- cor(AC.IPD[,c("X1","X2","X3","X4")]) 
#  covariate simulation for BC trial using copula package
cop <- normalCopula(param=c(rho[1,2],rho[1,3],rho[1,4],rho[2,3],
                            rho[2,4],rho[3,4]), 
                    dim=4, dispstr="un") # AC IPD pairwise correlations
# sample covariates from approximate joint distribution using copula
mvd <- mvdc(copula=cop, margins=c("norm", "norm", # Gaussian marginals
                                  "norm", "norm"), 
            # BC covariate means and standard deviations
            paramMargins=list(list(mean=BC.ALD$mean.X1, sd=BC.ALD$sd.X1),
                              list(mean=BC.ALD$mean.X2, sd=BC.ALD$sd.X2),       
                              list(mean=BC.ALD$mean.X3, sd=BC.ALD$sd.X3),
                              list(mean=BC.ALD$mean.X4, sd=BC.ALD$sd.X4)))
# simulated BC pseudo-population of size 1000
x_star <- as.data.frame(rMvdc(1000, mvd))
colnames(x_star) <- c("X1", "X2", "X3", "X4")  
# outcome logistic regression fitted to IPD using MCMC (Stan)  
outcome.model <- stan_glm(y~X3+X4+trt*X1+trt*X2, data=AC.IPD, 
                          family=binomial, algorithm="sampling",
                          iter=4000, warmup=2000, chains=2) 
# counterfactual datasets
data.trtA <- data.trtC <- x_star
# intervene on treatment while keeping set covariates fixed
data.trtA$trt <- 1 # dataset where everyone receives treatment A
data.trtC$trt <- 0 # dataset where all observations receive C  
# draw binary responses from posterior predictive distribution
# matrix of posterior predictive draws under A
y.star.A <- posterior_predict(outcome.model, newdata=data.trtA) 
# matrix of posterior predictive draws under C
y.star.C <- posterior_predict(outcome.model, newdata=data.trtC)
# compute marginal log-odds ratio for A vs. C for each MCMC sample
# by transforming from probability to linear predictor scale  
hat.delta.AC <- qlogis(rowMeans(y.star.A)) - qlogis(rowMeans(y.star.C)) 
hat.Delta.AC <- mean(hat.delta.AC) # average over samples
hat.var.Delta.AC <- var(hat.delta.AC) # sample variance
# B vs. C from reported aggregate event counts in contingency table
hat.Delta.BC <- with(BC.ALD, log(y.B.sum*(N.C-y.C.sum)/
                                   (y.C.sum*(N.B-y.B.sum))))
# B vs. C variance using the delta method 
hat.var.Delta.BC <- with(BC.ALD, 1/y.C.sum+1/(N.C-y.C.sum)+
                           1/y.B.sum+1/(N.B-y.B.sum))
# marginal treatment effect for A vs. B
hat.Delta.AB <- hat.Delta.AC - hat.Delta.BC 
# A vs. B variance
hat.var.Delta.AB <- hat.var.Delta.AC + hat.var.Delta.BC 
# construct Wald-type normal distribution-based confidence interval
uci.Delta.AB <- hat.Delta.AB + qnorm(0.975)*sqrt(hat.var.Delta.AB)
lci.Delta.AB <- hat.Delta.AB + qnorm(0.025)*sqrt(hat.var.Delta.AB) 
\end{lstlisting}

\subsection*{MIM}

\begin{lstlisting}
library("copula") # for simulating BC covariates from Gaussian copula
library("rstanarm") # for MCMC posterior sampling in data synthesis stage 

AC.IPD <- read.csv("Example/AC_IPD.csv") # load AC patient-level data
BC.ALD <- read.csv("Example/BC_ALD.csv") # load BC aggregate-level data

set.seed(555) # set seed for reproducibility

# hyper-parameter settings
M <- 1000 # number of syntheses used in analysis stage
N_star <- 1000 # size of syntheses or simulated BC pseudo-populations
alloc <- 2/3 # 2:1 A:C allocation ratio in synthesis
# MCMC info
n.chains <- 2 # number of Markov chains for MCMC
warmup <- 2000 # discarded warmup/burn-in iterations per chain
iters <- 4000 # total iterations per chain (including warmup)

## SYNTHESIS STAGE (as per Bayesian G-computation) ##
# matrix of pairwise correlations between IPD covariates  
rho <- cor(AC.IPD[,c("X1","X2","X3","X4")]) 
#  covariate simulation for BC trial using copula package
cop <- normalCopula(param=c(rho[1,2],rho[1,3],rho[1,4],rho[2,3],
                            rho[2,4],rho[3,4]), 
                    dim=4, dispstr="un") # AC IPD pairwise correlations
# sample covariates from approximate joint distribution using copula
mvd <- mvdc(copula=cop, margins=c("norm", "norm", # Gaussian marginals
                                  "norm", "norm"), 
            # BC covariate means and standard deviations
            paramMargins=list(list(mean=BC.ALD$mean.X1, sd=BC.ALD$sd.X1),
                              list(mean=BC.ALD$mean.X2, sd=BC.ALD$sd.X2),       
                              list(mean=BC.ALD$mean.X3, sd=BC.ALD$sd.X3),
                              list(mean=BC.ALD$mean.X4, sd=BC.ALD$sd.X4)))
# simulated BC pseudo-population of size N_star
x_star <- as.data.frame(rMvdc(N_star, mvd))
colnames(x_star) <- c("X1", "X2", "X3", "X4")  
# first-stage logistic regression fitted to IPD using MCMC (Stan)
outcome.model <- stan_glm(y~X3+X4+trt*X1+trt*X2,
                          data=AC.IPD, family=binomial, 
                          algorithm="sampling", iter=iters, 
                          warmup=warmup, chains=n.chains, 
                          # thin to use M independent samples in analysis
                          thin=(n.chains*(iters-warmup))/M) 
# tratment assignment in synthesis 
N_active <- round(N_star*alloc) # number of patients in synthesis under A
N_control <- N_star - N_active # number of patients in synthesis under C
trt_star <- c(rep(1,N_active), rep(0,N_control)) 
x_star$trt <- trt_star
# draw binary outcomes from posterior predictive distribution
y_star <- posterior_predict(outcome.model, newdata=x_star) 

## ANALYSIS stage ##
# second-stage regression (marginal structural model) on each synthesis
reg2.fits <- lapply(1:M, function(m) glm(y_star[m,]~trt_star, 
                                         family=binomial))
# treatment coefficient is marginal effect for A vs. C in m-th synthesis
hats_delta_AC <- unlist(lapply(reg2.fits, 
                               function(fit) coef(fit)["trt_star"][[1]]))         
# estimated point estimates of the variance for A vs. C
hats_v <- unlist(lapply(reg2.fits, 
                        function(fit) vcov(fit)["trt_star", "trt_star"]))
# quantities originally defined by Rubin (1987) for multiple imputation
bar_delta_AC <- mean(hats_delta_AC) # average of point estimates 
bar_v <- mean(hats_v) # within variance (average of point estimates of the variance)
# between variance (sample variance of point estimates)
b <- var(hats_delta_AC) 
# pooling + indirect comparison (combining rules)
# average of point estimates is the marginal effect for A vs. C
hat.Delta.AC <- bar_delta_AC
# variance combining rule for A vs. C
hat.var.Delta.AC <- (1+(1/M))*b-bar_v 
# B vs. C from reported aggregate event counts in contingency table
hat.Delta.BC <- with(BC.ALD, log(y.B.sum*(N.C-y.C.sum)/
                                   (y.C.sum*(N.B-y.B.sum))))
# B vs. C variance using the delta method 
hat.var.Delta.BC <- with(BC.ALD, 1/y.C.sum+1/(N.C-y.C.sum)+
                           1/y.B.sum+1/(N.B-y.B.sum))
# marginal treatment effect for A vs. B
hat.Delta.AB <- hat.Delta.AC - hat.Delta.BC 
# A vs. B variance
hat.var.Delta.AB <- hat.var.Delta.AC + hat.var.Delta.BC  
# construct Wald-type normal distribution-based confidence interval
uci.Delta.AB <- hat.Delta.AB + qnorm(0.975)*sqrt(hat.var.Delta.AB)
lci.Delta.AB <- hat.Delta.AB + qnorm(0.025)*sqrt(hat.var.Delta.AB) 
\end{lstlisting}

\clearpage

% \nocite{*}% Show all bib entries - both cited and uncited; comment this line to view only cited bib entries;
\bibliographystyle{unsrt}% We choose the "plain" reference style

\addcontentsline{toc}{section}{References}

\bibliography{references}

\begin{thebibliography}{100}

\bibitem{vreman2020decision}
Rick~A Vreman, Huseyin Naci, Wim~G Goettsch, Aukje~K Mantel-Teeuwisse,
  Sebastian~G Schneeweiss, Hubert~GM Leufkens, and Aaron~S Kesselheim.
\newblock Decision making under uncertainty: comparing regulatory and health
  technology assessment reviews of medicines in the united states and europe.
\newblock {\em Clinical Pharmacology \& Therapeutics}, 108(2):350--357, 2020.

\bibitem{temple2000placebo}
Robert Temple and Susan~S Ellenberg.
\newblock Placebo-controlled trials and active-control trials in the evaluation
  of new treatments. part 1: ethical and scientific issues.
\newblock {\em Annals of internal medicine}, 133(6):455--463, 2000.

\bibitem{paul2001fourth}
John~E Paul and Paul Trueman.
\newblock ‘fourth hurdle reviews’, nice, and database applications.
\newblock {\em Pharmacoepidemiology and drug safety}, 10(5):429--438, 2001.

\bibitem{sutton2008use}
Alex Sutton, AE~Ades, Nicola Cooper, and Keith Abrams.
\newblock Use of indirect and mixed treatment comparisons for technology
  assessment.
\newblock {\em Pharmacoeconomics}, 26(9):753--767, 2008.

\bibitem{dias2013evidence}
Sofia Dias, Alex~J Sutton, AE~Ades, and Nicky~J Welton.
\newblock Evidence synthesis for decision making 2: a generalized linear
  modeling framework for pairwise and network meta-analysis of randomized
  controlled trials.
\newblock {\em Medical Decision Making}, 33(5):607--617, 2013.

\bibitem{bucher1997results}
Heiner~C Bucher, Gordon~H Guyatt, Lauren~E Griffith, and Stephen~D Walter.
\newblock The results of direct and indirect treatment comparisons in
  meta-analysis of randomized controlled trials.
\newblock {\em Journal of clinical epidemiology}, 50(6):683--691, 1997.

\bibitem{phillippo2018methods}
David~M Phillippo, Anthony~E Ades, Sofia Dias, Stephen Palmer, Keith~R Abrams,
  and Nicky~J Welton.
\newblock Methods for population-adjusted indirect comparisons in health
  technology appraisal.
\newblock {\em Medical Decision Making}, 38(2):200--211, 2018.

\bibitem{signorovitch2010comparative}
James~E Signorovitch, Eric~Q Wu, P~Yu Andrew, Charles~M Gerrits, Evan Kantor,
  Yanjun Bao, Shiraz~R Gupta, and Parvez~M Mulani.
\newblock Comparative effectiveness without head-to-head trials.
\newblock {\em Pharmacoeconomics}, 28(10):935--945, 2010.

\bibitem{caro2010no}
J~Jaime Caro and K~Jack Ishak.
\newblock No head-to-head trial? simulate the missing arms.
\newblock {\em Pharmacoeconomics}, 28(10):957--967, 2010.

\bibitem{miettinen1972standardization}
Olli~S Miettinen.
\newblock Standardization of risk ratios.
\newblock {\em American Journal of Epidemiology}, 96(6):383--388, 1972.

\bibitem{stuart2011use}
Elizabeth~A Stuart, Stephen~R Cole, Catherine~P Bradshaw, and Philip~J Leaf.
\newblock The use of propensity scores to assess the generalizability of
  results from randomized trials.
\newblock {\em Journal of the Royal Statistical Society: Series A (Statistics
  in Society)}, 174(2):369--386, 2011.

\bibitem{phillippo2020multilevel}
David~M Phillippo, Sofia Dias, AE~Ades, Mark Belger, Alan Brnabic, Alexander
  Schacht, Daniel Saure, Zbigniew Kadziola, and Nicky~J Welton.
\newblock Multilevel network meta-regression for population-adjusted treatment
  comparisons.
\newblock {\em Journal of the Royal Statistical Society: Series A (Statistics
  in Society)}, 2020.

\bibitem{phillippo2019calibration}
David~M Phillippo.
\newblock {\em Calibration of treatment effects in network meta-analysis using
  individual patient data}.
\newblock PhD thesis, University of Bristol, Bristol, UK, 2019.

\bibitem{phillippo2016nice}
David Phillippo, Tony Ades, Sofia Dias, Stephen Palmer, Keith~R Abrams, and
  Nicky Welton.
\newblock Nice dsu technical support document 18: methods for
  population-adjusted indirect comparisons in submissions to nice.
\newblock 2016.

\bibitem{baio2012bayesian}
Gianluca Baio.
\newblock {\em Bayesian methods in health economics}.
\newblock CRC Press, 2012.

\bibitem{claxton2005probabilistic}
Karl Claxton, Mark Sculpher, Chris McCabe, Andrew Briggs, Ron Akehurst, Martin
  Buxton, John Brazier, and Tony O'Hagan.
\newblock Probabilistic sensitivity analysis for nice technology assessment:
  not an optional extra.
\newblock {\em Health economics}, 14(4):339--347, 2005.

\bibitem{remiro2020methods}
Antonio Remiro-Az{\'o}car, Anna Heath, and Gianluca Baio.
\newblock Methods for population adjustment with limited access to individual
  patient data: A review and simulation study.
\newblock {\em arXiv preprint arXiv:2004.14800}, 2020.

\bibitem{cheng2019statistical}
David Cheng, Rajeev Ayyagari, and James Signorovitch.
\newblock The statistical performance of matching-adjusted indirect
  comparisons.
\newblock {\em arXiv preprint arXiv:1910.06449}, 2019.

\bibitem{hatswell2020effects}
Anthony~James Hatswell, Nick Freemantle, and Gianluca Baio.
\newblock The effects of model misspecification in unanchored matching-adjusted
  indirect comparison (maic): Results of a simulation study.
\newblock {\em Value in Health}, 2020.

\bibitem{phillippo2020assessing}
David~M Phillippo, Sofia Dias, AE~Ades, and Nicky~J Welton.
\newblock Assessing the performance of population adjustment methods for
  anchored indirect comparisons: A simulation study.
\newblock {\em Statistics in Medicine}, 39(30):4885--4911, 2020.

\bibitem{stuart2010matching}
Elizabeth~A Stuart.
\newblock Matching methods for causal inference: A review and a look forward.
\newblock {\em Statistical science: a review journal of the Institute of
  Mathematical Statistics}, 25(1):1, 2010.

\bibitem{lee2011weight}
Brian~K Lee, Justin Lessler, and Elizabeth~A Stuart.
\newblock Weight trimming and propensity score weighting.
\newblock {\em PloS one}, 6(3):e18174, 2011.

\bibitem{hirano2001estimation}
Keisuke Hirano and Guido~W Imbens.
\newblock Estimation of causal effects using propensity score weighting: An
  application to data on right heart catheterization.
\newblock {\em Health Services and Outcomes research methodology},
  2(3-4):259--278, 2001.

\bibitem{jackson2020alternative}
Dan Jackson, Kirsty Rhodes, and Mario Ouwens.
\newblock Alternative weighting schemes when performing matching-adjusted
  indirect comparisons.
\newblock {\em Research Synthesis Methods}, 2020.

\bibitem{van2011targeted}
Mark~J Van~der Laan and Sherri Rose.
\newblock {\em Targeted learning: causal inference for observational and
  experimental data}.
\newblock Springer Science \& Business Media, 2011.

\bibitem{neugebauer2005prefer}
Romain Neugebauer and Mark van~der Laan.
\newblock Why prefer double robust estimators in causal inference?
\newblock {\em Journal of statistical planning and inference},
  129(1-2):405--426, 2005.

\bibitem{robins1992estimating}
James~M Robins, Steven~D Mark, and Whitney~K Newey.
\newblock Estimating exposure effects by modelling the expectation of exposure
  conditional on confounders.
\newblock {\em Biometrics}, pages 479--495, 1992.

\bibitem{phillippo2019population}
David~M Phillippo, Sofia Dias, Ahmed Elsada, AE~Ades, and Nicky~J Welton.
\newblock Population adjustment methods for indirect comparisons: A review of
  national institute for health and care excellence technology appraisals.
\newblock {\em International journal of technology assessment in health care},
  pages 1--8, 2019.

\bibitem{hauck1998should}
Walter~W Hauck, Sharon Anderson, and Sue~M Marcus.
\newblock Should we adjust for covariates in nonlinear regression analyses of
  randomized trials?
\newblock {\em Controlled clinical trials}, 19(3):249--256, 1998.

\bibitem{daniel2020making}
Rhian Daniel, Jingjing Zhang, and Daniel Farewell.
\newblock Making apples from oranges: Comparing noncollapsible effect
  estimators and their standard errors after adjustment for different covariate
  sets.
\newblock {\em Biometrical Journal}, 2020.

\bibitem{robins1986new}
James Robins.
\newblock A new approach to causal inference in mortality studies with a
  sustained exposure period—application to control of the healthy worker
  survivor effect.
\newblock {\em Mathematical modelling}, 7(9-12):1393--1512, 1986.

\bibitem{robins1987graphical}
James Robins.
\newblock A graphical approach to the identification and estimation of causal
  parameters in mortality studies with sustained exposure periods.
\newblock {\em Journal of chronic diseases}, 40:139S--161S, 1987.

\bibitem{moore2009covariate}
Kelly~L Moore and Mark~J van~der Laan.
\newblock Covariate adjustment in randomized trials with binary outcomes:
  targeted maximum likelihood estimation.
\newblock {\em Statistics in medicine}, 28(1):39--64, 2009.

\bibitem{austin2010absolute}
Peter~C Austin.
\newblock Absolute risk reductions, relative risks, relative risk reductions,
  and numbers needed to treat can be obtained from a logistic regression model.
\newblock {\em Journal of clinical epidemiology}, 63(1):2--6, 2010.

\bibitem{rosenblum2010simple}
Michael Rosenblum and Mark~J Van Der~Laan.
\newblock Simple, efficient estimators of treatment effects in randomized
  trials using generalized linear models to leverage baseline variables.
\newblock {\em The international journal of biostatistics}, 6(1), 2010.

\bibitem{zhang2008estimating}
Zhiwei Zhang.
\newblock Estimating a marginal causal odds ratio subject to confounding.
\newblock {\em Communications in Statistics-Theory and methods},
  38(3):309--321, 2008.

\bibitem{vo2019novel}
Tat-Thang Vo, Raphael Porcher, Anna Chaimani, and Stijn Vansteelandt.
\newblock A novel approach for identifying and addressing case-mix
  heterogeneity in individual participant data meta-analysis.
\newblock {\em Research synthesis methods}, 10(4):582--596, 2019.

\bibitem{vo2021assessing}
Tat-Thang Vo, Rapha{\"e}l Porcher, and Stijn Vansteelandt.
\newblock Assessing the impact of case-mix heterogeneity in individual
  participant data meta-analysis: Novel use of i 2 statistic and prediction
  interval.
\newblock {\em Research Methods in Medicine \& Health Sciences}, 2(1):12--30,
  2021.

\bibitem{rubin2004multiple}
Donald~B Rubin.
\newblock {\em Multiple imputation for nonresponse in surveys}.
\newblock John Wiley \& Sons, 1987.

\bibitem{glenny2005indirect}
AM~Glenny, DG~Altman, F~Song, C~Sakarovitch, JJ~Deeks, R~D’amico, M~Bradburn,
  and AJ~Eastwood.
\newblock Indirect comparisons of competing interventions.
\newblock 2005.

\bibitem{remiro2021target}
Antonio Remiro-Az{\'o}car.
\newblock Target estimands for population-adjusted indirect comparisons.
\newblock {\em arXiv preprint arXiv:2112.08023}, 2021.

\bibitem{austin2011introduction}
Peter~C Austin.
\newblock An introduction to propensity score methods for reducing the effects
  of confounding in observational studies.
\newblock {\em Multivariate behavioral research}, 46(3):399--424, 2011.

\bibitem{imbens2004nonparametric}
Guido~W Imbens.
\newblock Nonparametric estimation of average treatment effects under
  exogeneity: A review.
\newblock {\em Review of Economics and statistics}, 86(1):4--29, 2004.

\bibitem{hernan2020causal}
Miguel~A Hern{\'a}n and James~M Robins.
\newblock Causal inference: what if, 2020.

\bibitem{remiro2021marginalization}
Antonio Remiro-Az{\'o}car, Anna Heath, and Gianluca Baio.
\newblock The marginalization of regression-adjusted estimates is necessary for
  reimbursement decisions at the population level.

\bibitem{vanderweele2009concerning}
Tyler~J VanderWeele.
\newblock Concerning the consistency assumption in causal inference.
\newblock {\em Epidemiology}, 20(6):880--883, 2009.

\bibitem{rothman1980concepts}
Kenneth~J Rothman, Sander Greenland, and Alexander~M Walker.
\newblock Concepts of interaction.
\newblock {\em American journal of epidemiology}, 112(4):467--470, 1980.

\bibitem{manski2019meta}
Charles~F Manski.
\newblock Meta-analysis for medical decisions.
\newblock 2019.

\bibitem{song2003validity}
Fujian Song, Douglas~G Altman, Anne-Marie Glenny, and Jonathan~J Deeks.
\newblock Validity of indirect comparison for estimating efficacy of competing
  interventions: empirical evidence from published meta-analyses.
\newblock {\em Bmj}, 326(7387):472, 2003.

\bibitem{remiro2020conflating}
Antonio Remiro-Az{\'o}car, Anna Heath, and Gianluca Baio.
\newblock Conflating marginal and conditional treatment effects: Comments
  on'assessing the performance of population adjustment methods for anchored
  indirect comparisons: A simulation study'.
\newblock {\em arXiv preprint arXiv:2011.06334}, 2020.

\bibitem{cole2010generalizing}
Stephen~R Cole and Elizabeth~A Stuart.
\newblock Generalizing evidence from randomized clinical trials to target
  populations: the actg 320 trial.
\newblock {\em American journal of epidemiology}, 172(1):107--115, 2010.

\bibitem{kern2016assessing}
Holger~L Kern, Elizabeth~A Stuart, Jennifer Hill, and Donald~P Green.
\newblock Assessing methods for generalizing experimental impact estimates to
  target populations.
\newblock {\em Journal of research on educational effectiveness},
  9(1):103--127, 2016.

\bibitem{hartman2015sample}
Erin Hartman, Richard Grieve, Roland Ramsahai, and Jasjeet~S Sekhon.
\newblock From sample average treatment effect to population average treatment
  effect on the treated: combining experimental with observational studies to
  estimate population treatment effects.
\newblock {\em Journal of the Royal Statistical Society: Series A (Statistics
  in Society)}, 178(3):757--778, 2015.

\bibitem{rosenbaum1987model}
Paul~R Rosenbaum.
\newblock Model-based direct adjustment.
\newblock {\em Journal of the American Statistical Association},
  82(398):387--394, 1987.

\bibitem{van2003unified}
Mark~J Van~der Laan, MJ~Laan, and James~M Robins.
\newblock {\em Unified methods for censored longitudinal data and causality}.
\newblock Springer Science \& Business Media, 2003.

\bibitem{ertefaie2010comparing}
Ashkan Ertefaie and David~A Stephens.
\newblock Comparing approaches to causal inference for longitudinal data:
  inverse probability weighting versus propensity scores.
\newblock {\em The international journal of biostatistics}, 6(2), 2010.

\bibitem{daniel2013methods}
Rhian~M Daniel, SN~Cousens, BL~De~Stavola, Michael~G Kenward, and JAC Sterne.
\newblock Methods for dealing with time-dependent confounding.
\newblock {\em Statistics in medicine}, 32(9):1584--1618, 2013.

\bibitem{lunceford2004stratification}
Jared~K Lunceford and Marie Davidian.
\newblock Stratification and weighting via the propensity score in estimation
  of causal treatment effects: a comparative study.
\newblock {\em Statistics in medicine}, 23(19):2937--2960, 2004.

\bibitem{shenoy2015elderly}
Premnath Shenoy and Anand Harugeri.
\newblock Elderly patients’ participation in clinical trials.
\newblock {\em Perspectives in clinical research}, 6(4):184, 2015.

\bibitem{khan2020participation}
Safi~U Khan, Muhammad~Zia Khan, Charumathi~Raghu Subramanian, Haris Riaz,
  Muhammad~U Khan, Ahmad~Naeem Lone, Muhammad~Shahzeb Khan, Eve-Marie Benson,
  Mohamad Alkhouli, Michael~J Blaha, et~al.
\newblock Participation of women and older participants in randomized clinical
  trials of lipid-lowering therapies: a systematic review.
\newblock {\em JAMA network open}, 3(5):e205202--e205202, 2020.

\bibitem{bang2005doubly}
Heejung Bang and James~M Robins.
\newblock Doubly robust estimation in missing data and causal inference models.
\newblock {\em Biometrics}, 61(4):962--973, 2005.

\bibitem{vansteelandt2011invited}
Stijn Vansteelandt and Niels Keiding.
\newblock Invited commentary: G-computation--lost in translation?
\newblock {\em American journal of epidemiology}, 173(7):739--742, 2011.

\bibitem{dahabreh2019generalizing}
Issa~J Dahabreh, Sarah~E Robertson, Eric~J Tchetgen, Elizabeth~A Stuart, and
  Miguel~A Hern{\'a}n.
\newblock Generalizing causal inferences from individuals in randomized trials
  to all trial-eligible individuals.
\newblock {\em Biometrics}, 75(2):685--694, 2019.

\bibitem{kang2007demystifying}
Joseph~DY Kang, Joseph~L Schafer, et~al.
\newblock Demystifying double robustness: A comparison of alternative
  strategies for estimating a population mean from incomplete data.
\newblock {\em Statistical science}, 22(4):523--539, 2007.

\bibitem{tan2007comment}
Zhiqiang Tan.
\newblock Comment: Understanding or, ps and dr.
\newblock {\em Statistical Science}, 22(4):560--568, 2007.

\bibitem{pearl2014external}
Judea Pearl and Elias Bareinboim.
\newblock External validity: From do-calculus to transportability across
  populations.
\newblock {\em Statistical Science}, pages 579--595, 2014.

\bibitem{remiro2020principled}
Antonio Remiro-Az{\'o}car, Anna Heath, and Gianluca Baio.
\newblock Principled selection of effect modifiers: Comments
  on'matching-adjusted indirect comparisons: Application to time-to-event
  data'.
\newblock {\em arXiv preprint arXiv:2012.05127}, 2020.

\bibitem{nelsen2007introduction}
Roger~B Nelsen.
\newblock {\em An introduction to copulas}.
\newblock Springer Science \& Business Media, 2007.

\bibitem{sklar1959fonctions}
M~Sklar.
\newblock Fonctions de repartition an dimensions et leurs marges.
\newblock {\em Publ. Inst. Statist. Univ. Paris}, 8:229--231, 1959.

\bibitem{royston2004multiple}
Patrick Royston.
\newblock Multiple imputation of missing values.
\newblock {\em The Stata Journal}, 4(3):227--241, 2004.

\bibitem{buuren2010mice}
S~van Buuren and Karin Groothuis-Oudshoorn.
\newblock mice: Multivariate imputation by chained equations in r.
\newblock {\em Journal of statistical software}, pages 1--68, 2010.

\bibitem{kruschke2014doing}
John Kruschke.
\newblock {\em Doing Bayesian data analysis: A tutorial with R, JAGS, and
  Stan}.
\newblock Academic Press, 2014.

\bibitem{chan2014increasing}
An-Wen Chan, Fujian Song, Andrew Vickers, Tom Jefferson, Kay Dickersin, Peter~C
  G{\o}tzsche, Harlan~M Krumholz, Davina Ghersi, and H~Bart Van Der~Worp.
\newblock Increasing value and reducing waste: addressing inaccessible
  research.
\newblock {\em The Lancet}, 383(9913):257--266, 2014.

\bibitem{voigt2017eu}
Paul Voigt and Axel Von~dem Bussche.
\newblock The eu general data protection regulation (gdpr).
\newblock {\em A Practical Guide, 1st Ed., Cham: Springer International
  Publishing}, 2017.

\bibitem{nowok2016synthpop}
Beata Nowok, Gillian~M Raab, Chris Dibben, et~al.
\newblock synthpop: Bespoke creation of synthetic data in r.
\newblock {\em J Stat Softw}, 74(11):1--26, 2016.

\bibitem{bonofigliorecovery}
Federico Bonofiglio, Martin Schumacher, and Harald Binder.
\newblock Recovery of original individual person data (ipd) inferences from
  empirical ipd summaries only: Applications to distributed computing under
  disclosure constraints.
\newblock {\em Statistics in Medicine}.

\bibitem{ishak2015simulation}
K~Jack Ishak, Irina Proskorovsky, and Agnes Benedict.
\newblock Simulation and matching-based approaches for indirect comparison of
  treatments.
\newblock {\em Pharmacoeconomics}, 33(6):537--549, 2015.

\bibitem{petto2019alternative}
Helmut Petto, Zbigniew Kadziola, Alan Brnabic, Daniel Saure, and Mark Belger.
\newblock Alternative weighting approaches for anchored matching-adjusted
  indirect comparisons via a common comparator.
\newblock {\em Value in Health}, 22(1):85--91, 2019.

\bibitem{belger2015inclusion}
M~Belger, A~Brnabic, Z~Kadziola, H~Petto, and D~Faries.
\newblock Inclusion of multiple studies in matching adjusted indirect
  comparisons (maic).
\newblock {\em Value in Health}, 18(3):A33, 2015.

\bibitem{aouni2020matching}
Jihane Aouni, Nadia Gaudel-Dedieu, and Bernard Sebastien.
\newblock Matching-adjusted indirect comparisons: Application to time-to-event
  data.
\newblock {\em Statistics in Medicine}, 2020.

\bibitem{grimm2019nivolumab}
Sabine~E Grimm, Nigel Armstrong, Bram~LT Ramaekers, Xavier Pouwels, Shona Lang,
  Svenja Petersohn, Rob Riemsma, Gillian Worthy, Lisa Stirk, Janine Ross,
  et~al.
\newblock Nivolumab for treating metastatic or unresectable urothelial cancer:
  an evidence review group perspective of a nice single technology appraisal.
\newblock {\em Pharmacoeconomics}, 37(5):655--667, 2019.

\bibitem{ren2019pembrolizumab}
Shijie Ren, Hazel Squires, Emma Hock, Eva Kaltenthaler, Andrew Rawdin, and
  Constantine Alifrangis.
\newblock Pembrolizumab for locally advanced or metastatic urothelial cancer
  where cisplatin is unsuitable: An evidence review group perspective of a nice
  single technology appraisal.
\newblock {\em PharmacoEconomics}, 37(9):1073--1080, 2019.

\bibitem{janes2010quantifying}
Holly Janes, Francesca Dominici, and Scott Zeger.
\newblock On quantifying the magnitude of confounding.
\newblock {\em Biostatistics}, 11(3):572--582, 2010.

\bibitem{greenland1987interpretation}
Sander Greenland.
\newblock Interpretation and choice of effect measures in epidemiologic
  analyses.
\newblock {\em American journal of epidemiology}, 125(5):761--768, 1987.

\bibitem{greenland1999confounding}
Sander Greenland, James~M Robins, and Judea Pearl.
\newblock Confounding and collapsibility in causal inference.
\newblock {\em Statistical science}, pages 29--46, 1999.

\bibitem{austin2014use}
Peter~C Austin.
\newblock The use of propensity score methods with survival or time-to-event
  outcomes: reporting measures of effect similar to those used in randomized
  experiments.
\newblock {\em Statistics in medicine}, 33(7):1242--1258, 2014.

\bibitem{keil2018bayesian}
Alexander~P Keil, Eric~J Daza, Stephanie~M Engel, Jessie~P Buckley, and
  Jessie~K Edwards.
\newblock A bayesian approach to the g-formula.
\newblock {\em Statistical methods in medical research}, 27(10):3183--3204,
  2018.

\bibitem{snowden2011implementation}
Jonathan~M Snowden, Sherri Rose, and Kathleen~M Mortimer.
\newblock Implementation of g-computation on a simulated data set:
  demonstration of a causal inference technique.
\newblock {\em American journal of epidemiology}, 173(7):731--738, 2011.

\bibitem{wang2017g}
Aolin Wang, Roch~A Nianogo, and Onyebuchi~A Arah.
\newblock G-computation of average treatment effects on the treated and the
  untreated.
\newblock {\em BMC medical research methodology}, 17(1):1--5, 2017.

\bibitem{imbens2015causal}
Guido~W Imbens and Donald~B Rubin.
\newblock {\em Causal inference in statistics, social, and biomedical
  sciences}.
\newblock Cambridge University Press, 2015.

\bibitem{phillippo2020target}
David~M Phillippo, Sofia Dias, AE~Ades, and Nicky~J Welton.
\newblock Target estimands for efficient decision making: Response to comments
  on ``assessing the performance of population adjustment methods for anchored
  indirect comparisons: A simulation study''.
\newblock {\em Statistics in Medicine}, 2021.

\bibitem{stitelman2011targeted}
Ori~M Stitelman, C~William Wester, Victor De~Gruttola, and Mark~J van~der Laan.
\newblock Targeted maximum likelihood estimation of effect modification
  parameters in survival analysis.
\newblock {\em The international journal of biostatistics}, 7(1), 2011.

\bibitem{varadhan2016cross}
Ravi Varadhan, Nicholas~C Henderson, and Carlos~O Weiss.
\newblock Cross-design synthesis for extending the applicability of trial
  evidence when treatment effect is heterogeneous: Part i. methodology.
\newblock {\em Communications in Statistics: Case Studies, Data Analysis and
  Applications}, 2(3-4):112--126, 2016.

\bibitem{zhang2016new}
Zhiwei Zhang, Lei Nie, Guoxing Soon, and Zonghui Hu.
\newblock New methods for treatment effect calibration, with applications to
  non-inferiority trials.
\newblock {\em Biometrics}, 72(1):20--29, 2016.

\bibitem{gabrio2019full}
Andrea Gabrio, Alexina~J Mason, and Gianluca Baio.
\newblock A full bayesian model to handle structural ones and missingness in
  economic evaluations from individual-level data.
\newblock {\em Statistics in medicine}, 38(8):1399--1420, 2019.

\bibitem{bartlett2018covariate}
Jonathan~W Bartlett.
\newblock Covariate adjustment and estimation of mean response in randomised
  trials.
\newblock {\em Pharmaceutical statistics}, 17(5):648--666, 2018.

\bibitem{qu2015estimation}
Yongming Qu and Junxiang Luo.
\newblock Estimation of group means when adjusting for covariates in
  generalized linear models.
\newblock {\em Pharmaceutical statistics}, 14(1):56--62, 2015.

\bibitem{lane1982analysis}
Peter~W Lane and John~A Nelder.
\newblock Analysis of covariance and standardization as instances of
  prediction.
\newblock {\em Biometrics}, pages 613--621, 1982.

\bibitem{dahabreh2020extending}
Issa~J Dahabreh, Sarah~E Robertson, Jon~A Steingrimsson, Elizabeth~A Stuart,
  and Miguel~A Hernan.
\newblock Extending inferences from a randomized trial to a new target
  population.
\newblock {\em Statistics in medicine}, 39(14):1999--2014, 2020.

\bibitem{efron1986bootstrap}
Bradley Efron and Robert Tibshirani.
\newblock Bootstrap methods for standard errors, confidence intervals, and
  other measures of statistical accuracy.
\newblock {\em Statistical science}, pages 54--75, 1986.

\bibitem{rubin1987logit}
Donald~B Rubin and Nathaniel Schenker.
\newblock Logit-based interval estimation for binomial data using the jeffreys
  prior.
\newblock {\em Sociological methodology}, pages 131--144, 1987.

\bibitem{aalen1997markov}
Odd~O Aalen, Vernon~T Farewell, Daniela De~Angelis, Nicholas~E Day, and
  O~N{\"o}el~Gill.
\newblock A markov model for hiv disease progression including the effect of
  hiv diagnosis and treatment: application to aids prediction in england and
  wales.
\newblock {\em Statistics in medicine}, 16(19):2191--2210, 1997.

\bibitem{keil2014autism}
Alexander~P Keil, Julie~L Daniels, and Irva Hertz-Picciotto.
\newblock Autism spectrum disorder, flea and tick medication, and adjustments
  for exposure misclassification: the charge (childhood autism risks from
  genetics and environment) case--control study.
\newblock {\em Environmental Health}, 13(1):1--10, 2014.

\bibitem{josefsson2021bayesian}
Maria Josefsson and Michael~J Daniels.
\newblock Bayesian semi-parametric g-computation for causal inference in a
  cohort study with mnar dropout and death.
\newblock {\em Journal of the Royal Statistical Society: Series C (Applied
  Statistics)}, 70(2):398--414, 2021.

\bibitem{rubin1978bayesian}
Donald~B Rubin.
\newblock Bayesian inference for causal effects: The role of randomization.
\newblock {\em The Annals of statistics}, pages 34--58, 1978.

\bibitem{saarela2015predictive}
Olli Saarela, Elja Arjas, David~A Stephens, and Erica~EM Moodie.
\newblock Predictive bayesian inference and dynamic treatment regimes.
\newblock {\em Biometrical Journal}, 57(6):941--958, 2015.

\bibitem{lunn2012bugs}
David Lunn, Chris Jackson, Nicky Best, Andrew Thomas, and David Spiegelhalter.
\newblock {\em The BUGS book: A practical introduction to Bayesian analysis}.
\newblock CRC press, 2012.

\bibitem{plummer2003jags}
Martyn Plummer et~al.
\newblock Jags: A program for analysis of bayesian graphical models using gibbs
  sampling.
\newblock In {\em Proceedings of the 3rd international workshop on distributed
  statistical computing}, volume 124, pages 1--10. Vienna, Austria., 2003.

\bibitem{carpenter2017stan}
Bob Carpenter, Andrew Gelman, Matthew~D Hoffman, Daniel Lee, Ben Goodrich,
  Michael Betancourt, Marcus Brubaker, Jiqiang Guo, Peter Li, and Allen
  Riddell.
\newblock Stan: A probabilistic programming language.
\newblock {\em Journal of statistical software}, 76(1), 2017.

\bibitem{rue2009approximate}
H{\aa}vard Rue, Sara Martino, and Nicolas Chopin.
\newblock Approximate bayesian inference for latent gaussian models by using
  integrated nested laplace approximations.
\newblock {\em Journal of the royal statistical society: Series b (statistical
  methodology)}, 71(2):319--392, 2009.

\bibitem{latimer2013survival}
Nicholas~R Latimer.
\newblock Survival analysis for economic evaluations alongside clinical
  trials—extrapolation with patient-level data: inconsistencies, limitations,
  and a practical guide.
\newblock {\em Medical Decision Making}, 33(6):743--754, 2013.

\bibitem{bagust2014survival}
Adrian Bagust and Sophie Beale.
\newblock Survival analysis and extrapolation modeling of time-to-event
  clinical trial data for economic evaluation: an alternative approach.
\newblock {\em Medical Decision Making}, 34(3):343--351, 2014.

\bibitem{grieve2013extrapolation}
Richard Grieve, Neil Hawkins, and Mark Pennington.
\newblock Extrapolation of survival data in cost-effectiveness analyses:
  improving the current state of play, 2013.

\bibitem{vickers2019evaluation}
Adrian Vickers.
\newblock An evaluation of survival curve extrapolation techniques using
  long-term observational cancer data.
\newblock {\em Medical Decision Making}, 39(8):926--938, 2019.

\bibitem{baio2020survhe}
Gianluca Baio.
\newblock survhe: Survival analysis for health economic evaluation and
  cost-effectiveness modeling.
\newblock {\em Journal of Statistical Software}, 95(1):1--47, 2020.

\bibitem{guyot2012enhanced}
Patricia Guyot, AE~Ades, Mario~JNM Ouwens, and Nicky~J Welton.
\newblock Enhanced secondary analysis of survival data: reconstructing the data
  from published kaplan-meier survival curves.
\newblock {\em BMC medical research methodology}, 12(1):9, 2012.

\bibitem{meng1994multiple}
Xiao-Li Meng.
\newblock Multiple-imputation inferences with uncongenial sources of input.
\newblock {\em Statistical Science}, pages 538--558, 1994.

\bibitem{leon2003semiparametric}
Selene Leon, Anastasios~A Tsiatis, and Marie Davidian.
\newblock Semiparametric estimation of treatment effect in a pretest-posttest
  study.
\newblock {\em Biometrics}, 59(4):1046--1055, 2003.

\bibitem{schafer1997analysis}
Joseph~L Schafer.
\newblock {\em Analysis of incomplete multivariate data}.
\newblock Chapman and Hall/CRC, 1997.

\bibitem{sato2003marginal}
Tosiya Sato and Yutaka Matsuyama.
\newblock Marginal structural models as a tool for standardization.
\newblock {\em Epidemiology}, pages 680--686, 2003.

\bibitem{raghunathan2003multiple}
Trivellore~E Raghunathan, Jerome~P Reiter, and Donald~B Rubin.
\newblock Multiple imputation for statistical disclosure limitation.
\newblock {\em Journal of official statistics}, 19(1):1, 2003.

\bibitem{rubin1993statistical}
Donald~B Rubin.
\newblock Statistical disclosure limitation.
\newblock {\em Journal of official Statistics}, 9(2):461--468, 1993.

\bibitem{reiter2002satisfying}
Jerome~P Reiter.
\newblock Satisfying disclosure restrictions with synthetic data sets.
\newblock {\em Journal of Official Statistics}, 18(4):531, 2002.

\bibitem{reiter2005releasing}
Jerome~P Reiter.
\newblock Releasing multiply imputed, synthetic public use microdata: An
  illustration and empirical study.
\newblock {\em Journal of the Royal Statistical Society: Series A (Statistics
  in Society)}, 168(1):185--205, 2005.

\bibitem{si2011comparison}
Yajuan Si and Jerome~P Reiter.
\newblock A comparison of posterior simulation and inference by combining rules
  for multiple imputation.
\newblock {\em Journal of Statistical Theory and Practice}, 5(2):335--347,
  2011.

\bibitem{reiter2007multiple}
Jerome~P Reiter and Trivellore~E Raghunathan.
\newblock The multiple adaptations of multiple imputation.
\newblock {\em Journal of the American Statistical Association},
  102(480):1462--1471, 2007.

\bibitem{raab2016practical}
Gillian~M Raab, Beata Nowok, and Chris Dibben.
\newblock Practical data synthesis for large samples.
\newblock {\em Journal of Privacy and Confidentiality}, 7(3):67--97, 2016.

\bibitem{reiter2003inference}
Jerome~P Reiter.
\newblock Inference for partially synthetic, public use microdata sets.
\newblock {\em Survey Methodology}, 29(2):181--188, 2003.

\bibitem{raghunathan2020synthetic}
Trivellore~E Raghunathan.
\newblock Synthetic data.
\newblock {\em Annual Review of Statistics and Its Application}, 8, 2020.

\bibitem{morris2019using}
Tim~P Morris, Ian~R White, and Michael~J Crowther.
\newblock Using simulation studies to evaluate statistical methods.
\newblock {\em Statistics in medicine}, 38(11):2074--2102, 2019.

\bibitem{team2013r}
R~Core Team et~al.
\newblock R: A language and environment for statistical computing.
\newblock 2013.

\bibitem{ripley2009stochastic}
Brian~D Ripley.
\newblock {\em Stochastic simulation}, volume 316.
\newblock John Wiley \& Sons, 2009.

\bibitem{skipka2016methodological}
Guido Skipka, Beate Wieseler, Thomas Kaiser, Stefanie Thomas, Ralf Bender,
  J{\"u}rgen Windeler, and Stefan Lange.
\newblock Methodological approach to determine minor, considerable, and major
  treatment effects in the early benefit assessment of new drugs.
\newblock {\em Biometrical Journal}, 58(1):43--58, 2016.

\bibitem{stanley2007design}
Kenneth Stanley.
\newblock Design of randomized controlled trials.
\newblock {\em Circulation}, 115(9):1164--1169, 2007.

\bibitem{flury1986standard}
Bernhard~K Flury and Hans Riedwyl.
\newblock Standard distance in univariate and multivariate analysis.
\newblock {\em The American Statistician}, 40(3):249--251, 1986.

\bibitem{cohen2013statistical}
Jacob Cohen.
\newblock {\em Statistical power analysis for the behavioral sciences}.
\newblock Academic press, 2013.

\bibitem{austin2013performance}
Peter~C Austin.
\newblock The performance of different propensity score methods for estimating
  marginal hazard ratios.
\newblock {\em Statistics in medicine}, 32(16):2837--2849, 2013.

\bibitem{austin2016variance}
Peter~C Austin.
\newblock Variance estimation when using inverse probability of treatment
  weighting (iptw) with survival analysis.
\newblock {\em Statistics in medicine}, 35(30):5642--5655, 2016.

\bibitem{lesko2017bias}
Catherine~R Lesko and Bryan Lau.
\newblock Bias due to confounders for the exposure-competing risk relationship.
\newblock {\em Epidemiology (Cambridge, Mass.)}, 28(1):20, 2017.

\bibitem{phillippo2020equivalence}
David~M Phillippo, Sofia Dias, AE~Ades, and Nicky~J Welton.
\newblock Equivalence of entropy balancing and the method of moments for
  matching-adjusted indirect comparison.
\newblock {\em Research Synthesis Methods}, 2020.

\bibitem{efron1994introduction}
Bradley Efron and Robert~J Tibshirani.
\newblock {\em An introduction to the bootstrap}.
\newblock CRC press, 1994.

\bibitem{sikirica2013comparative}
Vanja Sikirica, Robert~L Findling, James Signorovitch, M~Haim Erder, Ryan
  Dammerman, Paul Hodgkins, Mei Lu, Jipan Xie, and Eric~Q Wu.
\newblock Comparative efficacy of guanfacine extended release versus
  atomoxetine for the treatment of attention-deficit/hyperactivity disorder in
  children and adolescents: applying matching-adjusted indirect comparison
  methodology.
\newblock {\em CNS drugs}, 27(11):943--953, 2013.

\bibitem{white1980heteroskedasticity}
Halbert White et~al.
\newblock A heteroskedasticity-consistent covariance matrix estimator and a
  direct test for heteroskedasticity.
\newblock {\em econometrica}, 48(4):817--838, 1980.

\bibitem{huber1967behavior}
Peter~J Huber et~al.
\newblock The behavior of maximum likelihood estimates under nonstandard
  conditions.
\newblock In {\em Proceedings of the fifth Berkeley symposium on mathematical
  statistics and probability}, volume~1, pages 221--233. University of
  California Press, 1967.

\bibitem{hill2006interval}
Jennifer Hill and Jerome~P Reiter.
\newblock Interval estimation for treatment effects using propensity score
  matching.
\newblock {\em Statistics in medicine}, 25(13):2230--2256, 2006.

\bibitem{kauermann2001note}
G{\"o}ran Kauermann and Raymond~J Carroll.
\newblock A note on the efficiency of sandwich covariance matrix estimation.
\newblock {\em Journal of the American Statistical Association},
  96(456):1387--1396, 2001.

\bibitem{fay2001small}
Michael~P Fay and Barry~I Graubard.
\newblock Small-sample adjustments for wald-type tests using sandwich
  estimators.
\newblock {\em Biometrics}, 57(4):1198--1206, 2001.

\bibitem{windmeijer2005finite}
Frank Windmeijer.
\newblock A finite sample correction for the variance of linear efficient
  two-step gmm estimators.
\newblock {\em Journal of econometrics}, 126(1):25--51, 2005.

\bibitem{dehejia1999causal}
Rajeev~H Dehejia and Sadek Wahba.
\newblock Causal effects in nonexperimental studies: Reevaluating the
  evaluation of training programs.
\newblock {\em Journal of the American statistical Association},
  94(448):1053--1062, 1999.

\bibitem{waernbaum2010propensity}
Ingeborg Waernbaum.
\newblock Propensity score model specification for estimation of average
  treatment effects.
\newblock {\em Journal of Statistical Planning and Inference},
  140(7):1948--1956, 2010.

\bibitem{zhao2008sensitivity}
Zhong Zhao.
\newblock Sensitivity of propensity score methods to the specifications.
\newblock {\em Economics Letters}, 98(3):309--319, 2008.

\bibitem{rubin2000combining}
Donald~B Rubin and Neal Thomas.
\newblock Combining propensity score matching with additional adjustments for
  prognostic covariates.
\newblock {\em Journal of the American Statistical Association},
  95(450):573--585, 2000.

\bibitem{goodrich2018rstanarm}
Ben Goodrich, Jonah Gabry, Imad Ali, and Sam Brilleman.
\newblock rstanarm: Bayesian applied regression modeling via stan.
\newblock {\em R package version}, 2(4):1758, 2018.

\bibitem{team2016rstan}
Stan~Developent Team et~al.
\newblock Rstan: the r interface to stan.
\newblock {\em R package version}, 2(1):522, 2016.

\bibitem{gelman2008weakly}
Andrew Gelman, Aleks Jakulin, Maria~Grazia Pittau, Yu-Sung Su, et~al.
\newblock A weakly informative default prior distribution for logistic and
  other regression models.
\newblock {\em The annals of applied statistics}, 2(4):1360--1383, 2008.

\bibitem{gelman2013bayesian}
Andrew Gelman, John~B Carlin, Hal~S Stern, David~B Dunson, Aki Vehtari, and
  Donald~B Rubin.
\newblock {\em Bayesian data analysis}.
\newblock CRC press, 2013.

\bibitem{bland2000odds}
J~Martin Bland and Douglas~G Altman.
\newblock The odds ratio.
\newblock {\em Bmj}, 320(7247):1468, 2000.

\bibitem{leyrat2014propensity}
Cl{\'e}mence Leyrat, Agn{\`e}s Caille, Allan Donner, and Bruno Giraudeau.
\newblock Propensity score methods for estimating relative risks in cluster
  randomized trials with low-incidence binary outcomes and selection bias.
\newblock {\em Statistics in medicine}, 33(20):3556--3575, 2014.

\bibitem{greenland2016sparse}
Sander Greenland, Mohammad~Ali Mansournia, and Douglas~G Altman.
\newblock Sparse data bias: a problem hiding in plain sight.
\newblock {\em bmj}, 352, 2016.

\bibitem{vittinghoff2007relaxing}
Eric Vittinghoff and Charles~E McCulloch.
\newblock Relaxing the rule of ten events per variable in logistic and cox
  regression.
\newblock {\em American journal of epidemiology}, 165(6):710--718, 2007.

\bibitem{neyman1934two}
Jerzy Neyman.
\newblock On the two different aspects of the representative method: the method
  of stratified sampling and the method of purposive selection.
\newblock {\em Journal of the Royal Statistical Society}, 97(4):558--625, 1934.

\bibitem{annesi1989efficiency}
Isabella Annesi, Thierry Moreau, and Joseph Lellouch.
\newblock Efficiency of the logistic regression and cox proportional hazards
  models in longitudinal studies.
\newblock {\em Statistics in medicine}, 8(12):1515--1521, 1989.

\bibitem{ho2007matching}
Daniel~E Ho, Kosuke Imai, Gary King, and Elizabeth~A Stuart.
\newblock Matching as nonparametric preprocessing for reducing model dependence
  in parametric causal inference.
\newblock {\em Political analysis}, 15(3):199--236, 2007.

\bibitem{rubin1997estimating}
Donald~B Rubin.
\newblock Estimating causal effects from large data sets using propensity
  scores.
\newblock {\em Annals of internal medicine}, 127(8\_Part\_2):757--763, 1997.

\bibitem{hainmueller2012entropy}
Jens Hainmueller.
\newblock Entropy balancing for causal effects: A multivariate reweighting
  method to produce balanced samples in observational studies.
\newblock {\em Political Analysis}, 20(1):25--46, 2012.

\bibitem{zubizarreta2015stable}
Jos{\'e}~R Zubizarreta.
\newblock Stable weights that balance covariates for estimation with incomplete
  outcome data.
\newblock {\em Journal of the American Statistical Association},
  110(511):910--922, 2015.

\bibitem{robins1995semiparametric}
James~M Robins and Andrea Rotnitzky.
\newblock Semiparametric efficiency in multivariate regression models with
  missing data.
\newblock {\em Journal of the American Statistical Association},
  90(429):122--129, 1995.

\bibitem{hahn1998role}
Jinyong Hahn.
\newblock On the role of the propensity score in efficient semiparametric
  estimation of average treatment effects.
\newblock {\em Econometrica}, pages 315--331, 1998.

\bibitem{williamson2014variance}
Elizabeth~J Williamson, Andrew Forbes, and Ian~R White.
\newblock Variance reduction in randomised trials by inverse probability
  weighting using the propensity score.
\newblock {\em Statistics in medicine}, 33(5):721--737, 2014.

\bibitem{robins1994estimation}
James~M Robins, Andrea Rotnitzky, and Lue~Ping Zhao.
\newblock Estimation of regression coefficients when some regressors are not
  always observed.
\newblock {\em Journal of the American statistical Association},
  89(427):846--866, 1994.

\bibitem{li2021generalizing}
Fan Li, Ashley~L Buchanan, and Stephen~R Cole.
\newblock Generalizing trial evidence to target populations in non-nested
  designs: Applications to aids clinical trials.
\newblock {\em arXiv preprint arXiv:2103.04907}, 2021.

\bibitem{madigan1994model}
David Madigan and Adrian~E Raftery.
\newblock Model selection and accounting for model uncertainty in graphical
  models using occam's window.
\newblock {\em Journal of the American Statistical Association},
  89(428):1535--1546, 1994.

\bibitem{dixon1991bayesian}
Dennis~O Dixon and Richard Simon.
\newblock Bayesian subset analysis.
\newblock {\em Biometrics}, pages 871--881, 1991.

\bibitem{spiegelhalter1994bayesian}
David~J Spiegelhalter, Laurence~S Freedman, and Mahesh~KB Parmar.
\newblock Bayesian approaches to randomized trials.
\newblock {\em Journal of the Royal Statistical Society: Series A (Statistics
  in Society)}, 157(3):357--387, 1994.

\bibitem{simon1997bayesian}
Richard Simon and Laurence~S Freedman.
\newblock Bayesian design and analysis of two x two factorial clinical trials.
\newblock {\em Biometrics}, pages 456--464, 1997.

\bibitem{westreich2019target}
Daniel Westreich, Jessie~K Edwards, Catherine~R Lesko, Stephen~R Cole, and
  Elizabeth~A Stuart.
\newblock Target validity and the hierarchy of study designs.
\newblock {\em American journal of epidemiology}, 188(2):438--443, 2019.

\bibitem{imai2008misunderstandings}
Kosuke Imai, Gary King, and Elizabeth~A Stuart.
\newblock Misunderstandings between experimentalists and observationalists
  about causal inference.
\newblock {\em Journal of the royal statistical society: series A (statistics
  in society)}, 171(2):481--502, 2008.

\bibitem{berlin2002individual}
Jesse~A Berlin, Jill Santanna, Christopher~H Schmid, Lynda~A Szczech, and
  Harold~I Feldman.
\newblock Individual patient-versus group-level data meta-regressions for the
  investigation of treatment effect modifiers: ecological bias rears its ugly
  head.
\newblock {\em Statistics in medicine}, 21(3):371--387, 2002.

\bibitem{phillippo2021target}
David~M Phillippo, Sofia Dias, Anthony~E Ades, and Nicky~J Welton.
\newblock Target estimands for efficient decision making: Response to comments
  on “assessing the performance of population adjustment methods for anchored
  indirect comparisons: A simulation study”.
\newblock {\em Statistics in Medicine}, 40(11):2759--2763, 2021.

\bibitem{crosensitivity}
Suzie Cro, Tim~P Morris, Michael~G Kenward, and James~R Carpenter.
\newblock Sensitivity analysis for clinical trials with missing continuous
  outcome data using controlled multiple imputation: A practical guide.
\newblock {\em Statistics in Medicine}.

\bibitem{nguyen2017sensitivity}
Trang~Quynh Nguyen, Cyrus Ebnesajjad, Stephen~R Cole, and Elizabeth~A Stuart.
\newblock Sensitivity analysis for an unobserved moderator in
  rct-to-target-population generalization of treatment effects.
\newblock {\em The Annals of Applied Statistics}, pages 225--247, 2017.

\bibitem{dahabreh2019extending}
Issa~J Dahabreh.
\newblock {\em Extending Causal Inferences From a Randomized Trial to Another
  Population}.
\newblock PhD thesis, 2019.

\bibitem{ishak2015simulated}
KJ~Ishak, M~Rael, H~Phatak, C~Masseria, and T~Lanitis.
\newblock Simulated treatment comparison of time-to-event (and other
  non-linear) outcomes.
\newblock {\em Value in Health}, 18(7):A719, 2015.

\bibitem{faria2015nice}
R~Faria, M~Hernandez~Alava, A~Manca, and A~Wailoo.
\newblock Nice dsu technical support document 17: the use of observational data
  to inform estimates of treatment effectiveness for technology appraisal:
  methods for comparative individual patient data.
\newblock {\em Sheffield: NICE Decision Support Unit}, 2015.

\bibitem{robins2009estimation}
James~M Robins and Miguel~A Hern{\'a}n.
\newblock Estimation of the causal effects of time-varying exposures.
\newblock {\em Longitudinal data analysis}, 553:599, 2009.

\bibitem{azocar2019pns311}
A~Remiro Azocar, Gianluca Baio, and Anna Heath.
\newblock Pns311 predictive-adjusted indirect comparison (paic): A novel method
  for population-adjusted indirect comparison.
\newblock {\em Value in Health}, 22:S816, 2019.

\bibitem{neyman1923application}
Jerzy~S Neyman.
\newblock On the application of probability theory to agricultural experiments.
  essay on principles. section 9.(translated and edited by dm dabrowska and tp
  speed, statistical science (1990), 5, 465-480).
\newblock {\em Annals of Agricultural Sciences}, 10:1--51, 1923.

\bibitem{rubin2005causal}
Donald~B Rubin.
\newblock Causal inference using potential outcomes: Design, modeling,
  decisions.
\newblock {\em Journal of the American Statistical Association},
  100(469):322--331, 2005.

\bibitem{rubin1980randomization}
Donald~B Rubin.
\newblock Randomization analysis of experimental data: The fisher randomization
  test comment.
\newblock {\em Journal of the American Statistical Association},
  75(371):591--593, 1980.

\bibitem{hudgens2008toward}
Michael~G Hudgens and M~Elizabeth Halloran.
\newblock Toward causal inference with interference.
\newblock {\em Journal of the American Statistical Association},
  103(482):832--842, 2008.

\bibitem{vanderweele2013causal}
Tyler~J VanderWeele and Miguel~A Hernan.
\newblock Causal inference under multiple versions of treatment.
\newblock {\em Journal of causal inference}, 1(1):1, 2013.

\bibitem{rosenbaum1983central}
Paul~R Rosenbaum and Donald~B Rubin.
\newblock The central role of the propensity score in observational studies for
  causal effects.
\newblock {\em Biometrika}, 70(1):41--55, 1983.

\bibitem{hernan2006estimating}
Miguel~A Hern{\'a}n and James~M Robins.
\newblock Estimating causal effects from epidemiological data.
\newblock {\em Journal of Epidemiology \& Community Health}, 60(7):578--586,
  2006.

\bibitem{cole2008constructing}
Stephen~R Cole and Miguel~A Hern{\'a}n.
\newblock Constructing inverse probability weights for marginal structural
  models.
\newblock {\em American journal of epidemiology}, 168(6):656--664, 2008.

\bibitem{greenland2009identifiability}
Sander Greenland and James~M Robins.
\newblock Identifiability, exchangeability and confounding revisited.
\newblock {\em Epidemiologic Perspectives \& Innovations}, 6(1):4, 2009.

\bibitem{greenland1990randomization}
Sander Greenland.
\newblock Randomization, statistics, and causal inference.
\newblock {\em Epidemiology}, pages 421--429, 1990.

\bibitem{senn1994testing}
Stephen Senn.
\newblock Testing for baseline balance in clinical trials.
\newblock {\em Statistics in medicine}, 13(17):1715--1726, 1994.

\bibitem{dahabreh2016using}
Issa~J Dahabreh, Rodney Hayward, and David~M Kent.
\newblock Using group data to treat individuals: understanding heterogeneous
  treatment effects in the age of precision medicine and patient-centred
  evidence.
\newblock {\em International journal of epidemiology}, 45(6):2184--2193, 2016.

\bibitem{dahabreh2017heterogeneity}
Issa~J Dahabreh, Thomas~A Trikalinos, David~M Kent, and Christopher~H Schmid.
\newblock Heterogeneity of treatment effects.
\newblock {\em Methods in comparative effectiveness research}, page 227, 2017.

\bibitem{steingrimsson2019subgroup}
Jon~Arni Steingrimsson and Jiabei Yang.
\newblock Subgroup identification using covariate-adjusted interaction trees.
\newblock {\em Statistics in medicine}, 38(21):3974--3984, 2019.

\bibitem{yang2021causal}
Jiabei Yang, Issa~J Dahabreh, and Jon~A Steingrimsson.
\newblock Causal interaction trees: Finding subgroups with heterogeneous
  treatment effects in observational data.
\newblock {\em Biometrics}, 2021.

\end{thebibliography}

\end{document}